\documentclass[aps,prd,twocolumn,floatfix,noshowpacs,tightenlines,noshowkeys,superscriptaddress,amsmath,amssymb,nofootinbib]{revtex4-2}

\usepackage{txfonts}

\usepackage{graphicx}
\usepackage{xcolor}
\usepackage{dcolumn}
\usepackage{hyperref}
\usepackage{amssymb}
\usepackage{xspace}
\usepackage{amsmath}
\usepackage{soul}
\usepackage[mathlines]{lineno}
\usepackage{aas_macros}
\usepackage{orcid}

\newcommand{\new}[1]{{\color{black}#1}}

\newcommand{\Ngw}{\ensuremath{N_{\rm GW}}\xspace}
\newcommand{\Ngal}{\ensuremath{N_{\rm gal}}\xspace}
\newcommand{\Ncbc}{\ensuremath{N_{\rm CBC}}\xspace}
\newcommand{\Nexp}{\ensuremath{N_{\rm exp}}\xspace}
\newcommand{\Pdet}{\ensuremath{P_{\rm det}}\xspace}
\newcommand{\ppop}{\ensuremath{p_{\rm pop}}\xspace}
\newcommand{\Rgal}{\ensuremath{R^*_{\rm gal,0}}\xspace}
\newcommand{\hu}{\xspace \ensuremath{{\rm km \, s^{-1} \, Mpc^{-1}}\xspace}}

\newcommand{\mmin}{\xspace m_{\rm min}\xspace}
\newcommand{\mmax}{\xspace m_{\rm max}\xspace}
\newcommand{\Msol}{\xspace\rm{M}_{\odot}\xspace}

\newcommand{\Gpcyr}{\ensuremath{\rm Gpc^{-3} yr^{-1}}\xspace}

\newcommand{\Tobs}{\ensuremath{T_{\rm obs}}\xspace}

\newcommand{\nn}{\nonumber}

\newcommand{\de}{{\rm d}}
\def\ie{{\emph{i.e.~}}}
\def\eg{{\emph{e.g.~}}}

\newcommand{\gammamax}{12.0\xspace}
\newcommand{\gammamin}{0.0\xspace}
\newcommand{\kappamax}{6.0\xspace}
\newcommand{\kappamin}{0.0\xspace}
\newcommand{\zpmax}{4.0\xspace}
\newcommand{\zpmin}{0.0\xspace}

\newcommand{\Rnotmax}{100.0\xspace}
\newcommand{\Rnotmin}{0.0\xspace}

\newcommand{\PLGalphamax}{12.0\xspace}
\newcommand{\PLGalphamin}{1.5\xspace}
\newcommand{\PLGbetamax}{12.0\xspace}
\newcommand{\PLGbetamin}{-4.0\xspace}
\newcommand{\PLGmmaxmax}{200.0\xspace}
\newcommand{\PLGmmaxmin}{50.0\xspace}
\newcommand{\PLGmminmax}{10.0\xspace}
\newcommand{\PLGmminmin}{2.0\xspace}
\newcommand{\PLGmugmax}{50.0\xspace}
\newcommand{\PLGmugmin}{20.0\xspace}
\newcommand{\PLGsigmagmax}{10.0\xspace}
\newcommand{\PLGsigmagmin}{0.4\xspace}
\newcommand{\PLGlambdapeakmax}{1.0\xspace}
\newcommand{\PLGlambdapeakmin}{0.0\xspace}
\newcommand{\PLGdeltammax}{10.0\xspace}
\newcommand{\PLGdeltammin}{0.0\xspace}

\begin{document}
\preprint{LIGO P2300139}

\title{Joint Population and cosmological properties inference with gravitational waves standard sirens and galaxy surveys}

\author{Simone Mastrogiovanni \orcidlink{0000-0003-1606-4183}}
\email{simone.mastrogiovanni@roma1.infn.it}
\affiliation{INFN, Sezione di Roma, I-00185 Roma, Italy}

\author{Danny Laghi \orcidlink{0000-0001-7462-3794}}
\affiliation{Laboratoire des 2 Infinis - Toulouse (L2IT-IN2P3), Universit\'e de Toulouse, CNRS, UPS, F-31062 Toulouse Cedex 9, France}

\author{Rachel Gray \orcidlink{0000-0002-5556-9873}}
\affiliation{SUPA, University of Glasgow, Glasgow, G12 8QQ, United Kingdom}
\affiliation{Department of Physics \& Astronomy, Queen Mary University of London, Mile End Road, London, E1 4NS, United Kingdom}

\author{Giada Caneva Santoro \orcidlink{0000-0002-0642-5507}}
\affiliation{Institut de Física d’Altes Energies (IFAE), Barcelona Institute of Science and Technology, Barcelona, Spain}

\author{Archisman Ghosh \orcidlink{0000-0003-0423-3533}}
\affiliation{Department of Physics and Astronomy, Ghent University, Proeftuinstraat 86, 9000 Ghent, Belgium}

\author{Christos Karathanasis \orcidlink{0000-0002-0642-5507}}
\affiliation{Institut de Física d’Altes Energies (IFAE), Barcelona Institute of Science and Technology, Barcelona, Spain}

\author{Konstantin Leyde \orcidlink{0000-0002-0642-5507}}
\affiliation{Universit\'e Paris Cit\'e, CNRS, Astroparticule et Cosmologie, F-75013 Paris, France}

\author{Dani\`ele A. Steer \orcidlink{0000-0002-8781-1273}}
\affiliation{Universit\'e Paris Cit\'e, CNRS, Astroparticule et Cosmologie, F-75013 Paris, France}

\author{Stéphane Perriès \orcidlink{0000-0003-2213-3579}}
\affiliation{Université Lyon, Université Claude Bernard Lyon 1, CNRS, IP2I Lyon/IN2P3,UMR 5822, F-69622 Villeurbanne, France}

\author{Grégoire Pierra \orcidlink{0000-0003-3970-7970}}
\affiliation{Université Lyon, Université Claude Bernard Lyon 1, CNRS, IP2I Lyon/IN2P3,UMR 5822, F-69622 Villeurbanne, France}


\date{\today}

\begin{abstract}
Gravitational wave (GW) sources at cosmological distances can be used to probe the expansion rate of the Universe. GWs directly provide a distance estimation of the source but no direct information on its redshift. The optimal scenario to obtain a redshift is through the direct identification of an electromagnetic (EM) counterpart and its host galaxy.
With almost 100 GW sources detected without EM counterparts (dark sirens), it is becoming crucial to have statistical techniques able to perform cosmological studies in the absence of EM emission.  
Currently, only two techniques for dark sirens are used on GW observations: the spectral siren method, which is based on the source-frame mass distribution to estimate conjointly cosmology and the source's merger rate, and the galaxy survey method, which uses galaxy surveys to assign a probabilistic redshift to the source while fitting cosmology. 
It has been recognized, however, that these two methods are two sides of the same coin. In this paper, we present a novel approach to unify these two methods. We apply this approach to several observed GW events using the \textsc{glade+} galaxy catalog discussing limiting cases. We provide estimates of the Hubble constant, modified gravity propagation effects, and population properties for binary black holes. We also estimate the binary black hole merger rate per galaxy to be $10^{-6}-10^{-5} {\rm yr^{-1}}$ depending on the galaxy catalog hypotheses. 

\end{abstract}

\maketitle


\section{Introduction}

Compact binary coalescences (CBCs) detected via gravitational waves (GWs) are rapidly becoming a central tool to study cosmology. GWs directly provide the luminosity distance of the source, but they are unable to provide the cosmological redshift estimation that is required to measure the cosmic expansion. An estimation of the source redshift could be obtained from the observation of an electromagnetic counterpart (EM) as occurred in the case of the binary neutron star merger GW170817 \citep{ligobns, MMApaper} observed by the LIGO \citep{LIGOScientific:2014pky} and Virgo \citep{VIRGO:2014yos} GW detectors. 
However, most of the sources currently detected are binary black holes (BBHs) \cite{gwtc3} which are not accompanied by an EM counterpart. In order to exploit this population of ``dark sirens'' for cosmological studies several methods have been proposed. In this study, we focus on two methods which, however, as we will see, are not independent: the galaxy catalog method and the spectral siren method. 
Additional methods to assign an implicit redshift from the GW alone include the ``cross-correlation method'', which explores the spatial clustering between GW sources and galaxies \citep{PhysRevD.93.083511,Mukherjee:2019wcg, Mukherjee:2020hyn,Bera:2020jhx, Diaz:2021pem}, and the ``Equation of State method'', which uses the measurement of tidal deformability of neutron stars and knowledge of their Equation of State \citep{2021PhRvD.104h3528C,Ghosh:2022muc}. For a review on these methods, see \cite{Bambi:2020tsh,2022LRR....25....6M}.

The {\it galaxy catalog} method, also referred to as the ``statistical method'', consists in using galaxy catalogs to statistically assign a host galaxy, hence a redshift, to the GW event \citep{schutz,2005ApJ...629...15H,Dalal:2006qt}. This method has been applied in recent studies \citep{PhysRevD.86.043011,chen17,fishbach,2020PhRvD.101l2001G,Gray2022,PhysRevD.105.023523} using various galaxy catalogs. One of the main bottlenecks of this method is the fact that at high redshift galaxy catalogs are not complete and therefore could not bring any relevant information on the source redshift. The galaxy catalog method also makes assumptions about the source frame mass distribution that we comment on more in detail later.
Another method used in the literature is the {\it spectral siren} method \citep{1993ApJ...411L...5C,Taylor_2012,Farr_2019,2020arXiv200602211M,mastrogiovanni_2021,Mukherjee:2021rtw,2021PhLB..82236665E,2022JCAP...09..012L,Ezquiaga_2022, Karathanasis:2022rtr} which aims at obtaining a redshift estimate for the GW sources from the relation between source-frame masses and detector-frame masses. In fact, it is possible to obtain a constraint on cosmology by jointly fitting for the population distribution of source-frame masses. The spectral siren method comes with the price of assuming a functional form and some priors for the source-frame mass distribution. In particular, wrong modeling of the population distribution of dark sirens (in particular source-frame masses) can introduce a bias in the estimation of the cosmological parameters~\cite{mastrogiovanni_2021}. \new{In analogy to the mass method, information on the possible redshift of GW events can also come from the CBC merger rate as a function of redshift. If the CBC merger rate presents local over-densities, such as a global merger rate redshift peak, then one can perform cosmological studies even in the absence of electromagnetic counterparts \cite{2019JCAP...04..033D,2021PhRvD.104d3507Y}.}

As discussed in \cite{mastrogiovanni_2021}, the galaxy catalog method is in fact a particular extension of the spectral siren method in which additional information from galaxy surveys is encoded. 
The galaxy catalog method requires the definition of a source-frame mass distribution, or more generally a population model for the CBCs. Assumptions about the population of sources typically enter the galaxy catalog method in two ways: \textit{(i)} they provide a systematic preference among host candidates at different redshift and \textit{(ii)} they provide a redshift preference for the \new{detected} source through the \textit{completeness correction}. The completeness correction accounts for the missing galaxies in the catalog. 
In fact, in the case that the galaxy catalog is 0\% complete (it contains no observed galaxy), the catalog method reduces to the spectral siren method. 
The tight connection between the spectral siren and the galaxy catalog methods has strong implications in interpreting results from current GW detections. For instance, in \cite{gwtc3_H0} it is shown that the constraint on the Hubble constant ($H_0$) from 42 BBHs reported in the third GW transient catalog (GWTC-3) is dominated by the population assumptions for the distribution of BBHs even when a galaxy catalog is employed.

In this paper, we present a new methodology, which we refer to as the {\it ``galaxy density method''}, that is able to unify the galaxy catalog {\it and} the spectral siren methods. 
As a result, it is possible to marginalize over all the uncertainties describing the population model while taking into account information from a galaxy catalog. The core of this new method is to link the CBC merger rate (spectral siren) to the galaxy number density (galaxy catalog). 

The paper is organized as follows.
In Sec.~\ref{sec:2} we review the basics of hierarchical Bayesian inference in presence of selection biases in the context of GW detections. In Sec.~\ref{sec:3} we introduce a parametrization for the CBC merger rate in terms of galaxy number density, and we discuss limiting cases, including how this new approach collapses to the standard spectral siren method. In Sec.~\ref{sec:4}, we show some tests and discuss the behavior of the method with the two best localized standard sirens to date, GW170817 and GW190814. In Sec.~\ref{sec:5} we perform a full population and galaxy catalog-based analysis of the 42 BBH detections from GWTC-3 analyzed in \cite{gwtc3_H0} and we discuss the obtained results. Finally, in Sec.~\ref{sec:6} we draw our conclusions.

\section{The hierarchical likelihood}
\label{sec:2}

The GW detection of \Ngw events in a given observing time $T_{\rm obs}$, collectively represented by some data $\{x\}$, can be described in terms of an inhomogeneous Poissonian process \cite{mandel,2022hgwa.bookE..45V} characterised by some parameters $\Lambda$ by using the following hierarchical likelihood: 
\begin{eqnarray}
    \mathcal{L}(\{x\}|\Lambda) &\propto&  e^{-N_{\rm exp}(\Lambda)} \prod_i^{\Ngw} \Tobs \int \de \theta  \de z \; \mathcal{L}_{\rm GW}(x_i|\theta,z,\Lambda_c)  \times \nn \\ && \times \frac{1}{1+z} \frac{\de \Ncbc}{\de \theta \de z \de t_s}(\Lambda).
    \label{eq:fund}
\end{eqnarray}
Here \Nexp is the number of expected CBC detections in \Tobs, while $z$ and $\theta$ indicate the redshift and a set of parameters characteristic of each GW event (such as spins and luminosity distance), respectively.
The central quantity of the hierarchical inference is the CBC merger rate in source-frame time $t_s$: 
\begin{equation}
    \frac{\de \Ncbc}{\de \theta \de z \de t_s} (\Lambda),
\end{equation}
where $\Lambda=\{\Lambda_p,\Lambda_c\}$ contains population-level parameters $\Lambda_p$ as well as cosmological parameters $\Lambda_c$ such as $H_0$ and $\Omega_{m}$. 
The information about how precisely we are able to measure the GW parameters $\theta$ (such as spins and source masses) from the data is given by the GW likelihood $\mathcal{L}_{\rm GW}(x_i|\theta,z,\Lambda_c)$. The expected number of GW events accounts for the GW selection biases:
\begin{equation}
    \Nexp (\Lambda)= \Tobs \int \de \theta \de z \; \Pdet(z,\theta,\Lambda_c) \frac{\de \Ncbc}{\de z  \de \theta \de t_s}(\Lambda) \frac{1}{1+z} ,
    \label{eq:Nexp}
\end{equation}
 where the detection probability $\Pdet(z,\theta,\Lambda)$ is defined as:
 \begin{equation}
     \Pdet(z,\theta,\Lambda_c) = \int_{x \in {\rm detectable}} \de x \; \mathcal{L}_{\rm GW}(x_i|\theta,z,\Lambda_c).
 \end{equation}
Eq.~\ref{eq:fund} is formally equivalent to:  
\begin{equation}
\begin{split}
     \mathcal{L}(\{x\}|\Lambda) & \propto e^{-\Nexp(\Lambda)} [\Nexp (\Lambda)]^{\Ngw} \times \\ 
     & \times \prod_i^{\Ngw} \frac{\int  \de \theta  \de z \; \mathcal{L}_{\rm GW}(x_i|\theta,z,\Lambda_c) \frac{1}{1+z} \frac{\de \Ncbc}{\de \theta \de z \de t_s}(\Lambda)}{\int  \de \theta  \de z \;  \Pdet(\theta,z,\Lambda_c) \frac{1}{1+z} \frac{\de \Ncbc}{\de \theta \de z \de t_s}(\Lambda)},     
\end{split}
\end{equation}
\new{that can be rewritten as: 
\begin{equation}
\begin{split}
    \mathcal{L}(\{x\}|\Lambda) & \propto e^{-\Nexp(\Lambda)} [\Nexp (\Lambda)]^{\Ngw} \times \\ 
     & \times \prod_i^{\Ngw} \frac{\int  \de \theta  \de z \; \mathcal{L}_{\rm GW}(x_i|\theta,z,\Lambda_c) \pi_{\rm pop}(\theta,z|\Lambda)}{\int  \de \theta  \de z \;  \Pdet(\theta,z,\Lambda_c) \pi_{\rm pop}(\theta,z|\Lambda)},
     \label{eq:form2}
\end{split}
\end{equation}
by defining the population parameter probability as:
\begin{equation}
    \pi_{\rm pop}(\theta,z|\Lambda)=\frac{1}{C(\Lambda)} \frac{1}{1+z} \frac{\de \Ncbc}{\de \theta \de z \de t_s}(\Lambda),
\end{equation}
where $C(\Lambda)$ is a normalization constant. Eq. \ref{eq:form2}  is the form of the hierarchical likelihood typically reported in \cite{LIGOScientific:2020kqk,LIGOScientific:2021psn,mastrogiovanni_2021}.} 

However, from a numerical point of view, Eq.~\ref{eq:fund} has some advantages in comparison to Eq.~\ref{eq:form2} \cite{2022PhRvD.105f4030M}: \textit{(i)} it is more connected to a physical quantity, the CBC merger rate  and \textit{(ii)} it involves the numerical computation of two integrals instead of three. In order to obtain Eq.~\ref{eq:form2} from Eq.~\ref{eq:fund}, recall that the observing time is given by:
\begin{equation}
    T_{\rm obs}=\frac{N_{\rm exp}(\Lambda)}  {\int \de \theta \de z \; \Pdet(z,\theta,\Lambda_c) \frac{\de N_{\rm CBC}}{\de z \de \theta \de t_s}(\Lambda) \frac{1}{1+z}},
\end{equation}
which, after substitution into Eq.~\ref{eq:fund}, gives Eq.~\ref{eq:form2}. 

The two forms of the hierarchical likelihood in Eq.~\ref{eq:fund} and Eq.~\ref{eq:form2} can be written in a ``scale-free'' version that does not depend on \Nexp (but still accounts for GW selection biases). By choosing a prior\footnote{Note that a prior $\pi(\Nexp)\propto 1/\Nexp$ would correspond to a prior $\pi(R_0)\propto 1/R_0$, where $R_0$ is the CBC merger rate per \Gpcyr.} $\pi(\Nexp)\propto 1/\Nexp$, it is possible to marginalize analytically over \Nexp and obtain \cite{2017ApJ...851L..25F}
\begin{equation}
    \mathcal{L}(\{x\}|\Lambda)\propto \prod_i^{N_{\rm GW}} \frac{\int \de \theta \de z \; \mathcal{L}(x_i|\theta,z,\Lambda_c) \frac{dN_{\rm CBC}(\Lambda)}{\de z \de \theta \de t_s} \frac{1}{1+z}}{\int \de \theta \de z \; P_{\rm det}(\theta,z,\Lambda_c)  \frac{\de N_{\rm CBC}(\Lambda)}{\de z \de \theta \de t_s} \frac{1}{1+z}}.
    \label{eq:fund_scalefree}
\end{equation}

The Poissonian process described by the hierarchical likelihoods in Eqs.~\ref{eq:fund}-\ref{eq:form2}-\ref{eq:fund_scalefree} can be used to jointly measure the CBC merger rate and cosmological parameters. The fundamental aspect that opens up this possibility is the fact that GW detectors directly measure 
the luminosity distance $d_L$ to the source and the detector frame masses $\vec{m}_d$.  We recall that, for a flat $\Lambda$CDM model, the luminosity distance  is:
\begin{equation}
    d_L(z)=\frac{c(1+z)}{H_0} \int_0^z \frac{\de z'}{E(z')}, 
\end{equation}
where $E(z)=\sqrt{\Omega_m(1+z)^3+(1-\Omega_m)}$. Furthermore, detector masses are linked to source masses through $\vec{m}_d=\vec{m}_s(1+z)$.
In \cite{1993ApJ...411L...5C,Taylor_2012,Farr_2019,2020arXiv200602211M,mastrogiovanni_2021,Mukherjee:2021rtw,2022JCAP...09..012L,Ezquiaga_2022, Karathanasis:2022rtr} it was shown that by modeling the CBC merger rate in terms of source-frame masses, it is possible to obtain a joint measurement of both the CBC merger rate and the cosmological parameters.  Similarly, in 
\citep{2019JCAP...04..033D,2021PhRvD.104d3507Y} it was demonstrated that if the CBC merger rate shows some features in redshift, such as peaks, this might help to measure cosmological parameters even without any information about the source-frame masses. 
In this work, we will apply Eq.~\ref{eq:form2} and Eq.~\ref{eq:fund_scalefree} either to measure the Hubble constant only, or to jointly measure $H_0$ and the population parameters related to the mass and redshift models used (see App.~\ref{app:moredetails}). In the rest of this paper, we assume a flat $\Lambda$CDM cosmology model with free $H_0$ and fixed $\Omega_m=0.308$, with $\Omega_{\Lambda}=1 - \Omega_m$.

\section{CBC merger rate parametrizations}
\label{sec:3}

In Sec.~\ref{sec:2} we have seen that the relevant GW parameters which are crucial for GW cosmology are the source-frame masses $\vec{m}_s$ and the redshift $z$.  However, when working with galaxy catalogs also the sky position, which we label by $\Omega$, is a central quantity since it can help us to identify possible host galaxies along a given line-of-sight. Therefore, in the remainder of this paper, we will work with a CBC merger rate written as:
\begin{equation}
  \frac{\de N_{\rm CBC}}{\de \vec{m}_s \de \theta d\Omega \de z \de t_s}(\Lambda).
  \label{eq:ratefun}
\end{equation}
Note that Eq.~\ref{eq:ratefun} is a CBC source-frame merger rate per source-frame mass, per GW parameters $\theta$, per sky position $\Omega$, per redshift $z$. To simplify the notation, we will refer to this quantity as the ``CBC merger rate'', unless we want to discuss an explicit parametrization of it. For brevity, we will also drop the $\Lambda$ dependence.

\subsection{The CBC spectral siren parametrization}
\label{sec:3a}
In standard cosmological-population analyses such as \cite{mastrogiovanni_2021,2022PhRvD.105f4030M}, which do not make explicit use of galaxy surveys, the CBC merger rate is always factorized in terms of comoving volume rather than the redshift, since it is easier to model and also easier to motivate physically. The CBC merger rate thus becomes: 
\begin{eqnarray}
    \frac{\de N_{\rm CBC}}{\de \vec{m}_s \de \theta d\Omega \de z \de t_s} &=& \frac{\de N_{\rm CBC}}{\de \vec{m}_s \de \theta \de V_c \de t_s} \frac{\de V_c}{\de z \de \Omega}(\Lambda_c)\nn \\ &=&  R_0  \psi(z;\Lambda)  \ppop (\vec{m}_s,\theta|z,\Lambda) \frac{\de V_c}{\de z \de \Omega}, \nn \\ &&
    \label{eq:rateiso}
\end{eqnarray}
where $R_0$ is the CBC merger rate per comoving volume at redshift $z=0$, $\psi(z;\Lambda)$ is a phenomenological parametrization for the CBC merger rate as a function of redshift such that $\psi(z=0;\Lambda)=1$, \ppop is a probability density function describing the distribution of GW sources in masses and other parameters $\theta$,  and 
\begin{equation}
 \frac{\de V_c}{\de z \de \Omega} = \left[\frac{c}{H_0}\right]^3 \left[\int_0^z \frac{\de z'}{E(z')}\right]^2   
\end{equation}
is the differential of the comoving volume. Typically, it is assumed that \ppop in Eq.~\ref{eq:rateiso} is independent of redshift, \ie the distribution of masses and GW population parameters does not evolve with redshift. \new{This is indeed a simplifying assumption as source mass distributions might depend on redshift \cite{2021ApJ...912...98F, 2022ApJ...931...17V} as well as spin parameters \cite{2022ApJ...932L..19B,2022A&A...665A..59B}. This assumption, if wrong, could introduce a bias in the estimation of the cosmological parameters. However, it has been shown in \cite{Ezquiaga_2022} that for a mild evolution one should still be able to obtain unbiased estimation of $H_0$. }

\subsection{The CBC merger rate informed by galaxy surveys}
\label{sec:3b}
We now proceed to derive the CBC merger rate encoding information from the galaxy catalog. To construct this CBC merger rate, we will make two 
assumptions: \textit{(i)} mergers can only occur inside galaxies and \textit{(ii)} mergers can be more or less frequent according to the absolute magnitude $M$ (or luminosity) and redshift of the galaxy. \new{In this paper, we will mostly use galaxies observed in the K-band, but also other colors could be used, in particular if they trace the binary merger rate in redshift.}
We can write the overall CBC merger rate as the integral over the absolute magnitude of the CBC rate per galaxy absolute magnitude, namely:
\begin{equation}
    \frac{\de N_{\rm CBC}}{\de \vec{m}_s \de \theta d\Omega \de z \de t_s} = \int \de M \, \frac{\de N_{\rm CBC}}{\de \vec{m}_s \de \theta d\Omega \de z \de t_s \de M}.
    \label{eq:eqM}
\end{equation}
Similar to the spectral siren case (see Eq.~\ref{eq:rateiso}), we parametrize the CBC rate per galaxy (rather than comoving volume), namely:
\begin{equation}
    \frac{\de N_{\rm CBC}}{\de \vec{m}_s \de \theta d\Omega \de z \de t_s \de M} = \frac{\de \Ncbc }{\de \Ngal \de \theta \de \vec{m}_s \de t_s}  \frac{\de \Ngal }{\de z \de \Omega \de M}.
    \label{eq:deco}
\end{equation}
In order to write Eq.~\ref{eq:deco}, we have further assumed that the number of galaxies only depends on sky position, absolute magnitude, and redshift.
The first term in Eq.~\ref{eq:deco}  is the CBC rate per galaxy which, similarly to Eq.~\ref{eq:rateiso}, we parametrize as:
\begin{equation}
    \frac{\de \Ncbc}{\de \Ngal \de \vec{m}_s \de \theta \de t_s}= \Rgal \Psi_{\rm L}(z,M;\Lambda_p)\ppop(\vec{m}_s,\theta|z,M,\Lambda_p),   \label{eq:cbcrate}
\end{equation}
where \Rgal is the CBC rate per year at a reference galaxy luminosity (or absolute magnitude) $L_*=L(M_*)$, while $\Psi_{\rm L}(z,M;\Lambda_p)$ is an evolution function for the CBC merger rate per galaxy such that $\Psi_{\rm L}(z=0,M=M_*;\Lambda_p)=1$. In Eq.~\ref{eq:cbcrate} $\ppop$ has been expressed in a general form assuming that the probability of having some GW parameters might be dependent on the galaxy redshift and absolute magnitude. For the analysis presented in this paper, we will not consider this possibility and we will choose $\ppop(\vec{m}_s,\theta|z,M,\Lambda_p)=\ppop(\vec{m}_s,\theta|\Lambda_p)$. Regarding the rate function $\Psi_{\rm L}(z,M;\Lambda_p)$ we use the following parametrization:
\begin{equation}
    \Psi_{\rm L}(z,L(M);\Lambda_p)=\psi(z;\Lambda_p) \|\frac{L}{L_{*}}\|^\epsilon=\psi(z;\Lambda_p)10^{0.4 \epsilon(M_*-M)},
    \label{eq:ratepresc}
\end{equation}
\ie we factorize the rate evolution function into a redshift-dependent part and a factor that depends on the luminosity of the galaxy. Similarly to Eq.~\ref{eq:rateiso}, the redshift part $\psi(z;\Lambda_p)$ is modeled to track the merger rate as a function of redshift, while the luminosity part models the fact that the merger rate might depend on the luminosity of the galaxy (which could be a tracer of the Star Formation Rate~\cite{Singer:2016eax}).
As an example, the parameter $\epsilon>0$ indicates that more luminous galaxies are more likely to host CBCs, while  $\epsilon=0$ indicates that there is no such preference.
\new{In principle, one could  also treat $\epsilon$ as a population parameter and jointly fit also for it. However, considering $\epsilon$ as a free parameter poses some numerical challenges because when trying different values of $\epsilon$, the galaxy catalog must be handled multiple times.}

A good choice for the reference luminosity $L_*$ is the ``knee'' luminosity of the Schechter function. 
The Schechter function describes the number density of galaxies in the comoving volume per luminosity (or absolute magnitude). In terms of absolute magnitude, it is identified by five parameters: a galaxy number density $\phi_*$ per comoving volume, a knee absolute magnitude $M_*$, a faint end $M_{\rm max}$, a bright end $M_{\rm min}$, and a slope parameter $\alpha$. Note that in practice the bright end choice is not important as the Schecter function quickly drops to 0 after $M_*(H_0)$. However, here we write it explicitly for a pedagogic derivation. 
We can thus write the Schechter function as:
\begin{widetext}

\begin{equation}
{\rm Sch}(M;\phi_*,M_*,M_{\rm max},M_{\rm min},\alpha) = 
\begin{cases}
    0.4 \ln(10) \phi_* 10^{0.4(\alpha+1)(M_*-M)} {\rm \exp}[-10^{0.4(M_*-M)}], & M_{\rm min} \leq M \leq M_{\rm max}, \\
    0, & {\rm otherwise}.
\end{cases}
\end{equation}
\end{widetext}
Usually, Schechter parameters are provided for $H_0 = 100 h \,\hu$ with $h=1$, so one typically needs to rescale them. The scaling relations are given in terms of absolute magnitude by:
\begin{eqnarray}
    M_*(h)&=& M_*(h=1)+5 \log_{10}h,
    \label{eq:Mt}\\
    \phi_*(h)&=&\phi_*(h=1) h^3 \label{eq:phit}.
\end{eqnarray}
\new{
Note that in writing down the Schecter function, we have implicitly assumed that it does not evolve in redshift. This is a simplification as we can measure for some colors an evolution of the Schecter function in redshift (see \eg \cite{2001AJ....121.2358B}).  As we will see later, the modeling of the Schecter function is crucial in the case that the galaxy catalog is not complete, as it enters the completeness correction. A wrong modelization of the Schecter function evolution might result in a wrong modelization of the CBC merger rate as a function of redshift (as mergers are expected to track galaxy density). This will translate to bias on $H_0$. In order to understand the magnitude of the bias on $H_0$, and how this interacts with other population parameters, we would require a dedicated mock data challenge which is beyond the scope of this paper.}

We now discuss how to build the number density of galaxies in Eq.~\ref{eq:deco}. 
In the limit in which we are provided with a galaxy survey containing all the galaxies in the Universe, this term can be constructed \new{as a sum of delta functions at the redshift of each galaxy}, sky position, and absolute magnitude. However, galaxy catalogs are not complete and therefore the galaxy density must be corrected to include the missing galaxies. In other words, the overall galaxy density can be calculated as the sum of a galaxy density detected by the catalog and a completeness correction, \ie
\begin{equation}
    \frac{\de \Ngal}{\de z \de \Omega \de M}=\frac{\de N_{\rm gal,cat}}{\de z \de \Omega \de M}+\frac{\de N_{\rm gal,out}}{\de z \de \Omega \de M}.
    \label{eq:dndzdodM}
\end{equation}
The overall CBC merger rate will be then given by plugging the galaxy density distribution, Eq.~\ref{eq:dndzdodM}, and the model of the CBC rate per galaxy, Eq.~\ref{eq:ratepresc}, into Eq.~\ref{eq:cbcrate}, thus obtaining:
\begin{equation}\label{eq:total}
\begin{split}
    &\frac{\de N_{\rm CBC}}{\de \vec{m}_s \de \theta d\Omega \de z \de t_s \de M} 
    = \Rgal \ppop(\vec{m}_s,\theta|\Lambda_p) \psi(z;\Lambda_p) \times \\ 
    & \times 10^{0.4 \epsilon(M_*-M)} \left[\frac{\de N_{\rm gal,cat}}{\de z \de \Omega \de M}+\frac{\de N_{\rm gal,out}}{\de z \de \Omega \de M} \right].
\end{split}
\end{equation}
Note that in order to compute the hierarchical likelihood in Eq.~\ref{eq:fund} we need to marginalize out the absolute magnitude from Eq.~\ref{eq:total}, as GW detectors are not able to measure this quantity. In fact, we will need to calculate:
\begin{eqnarray}
    & \dfrac{\de N_{\rm CBC}}{\de \vec{m}_s \de \theta d\Omega \de z \de t_s} =  \Rgal \ppop(\vec{m}_s,\theta|\Lambda_p) \psi(z;\Lambda_p) \times \nn \\ & \times \int_{M_{\rm min}}^{M_{\rm max}} \de M \, 10^{0.4 \epsilon(M_*-M)} \left[\frac{\de N_{\rm gal,cat}}{\de z \de \Omega \de M}+\frac{\de N_{\rm gal,out}}{\de z \de \Omega \de M} \right].\nn \\ \label{eq:total1}
\end{eqnarray}

\subsubsection{The completeness correction}

The second term in Eq.~\ref{eq:dndzdodM}, which is the completeness correction, or ``out-of-catalog'' term \cite{2020PhRvD.101l2001G,Gray2022}, can be written as the background galaxy density times the probability $p_{\rm miss}$ to miss a galaxy at redshift $z$, absolute magnitude $M$, and sky position $\Omega$, namely:
\begin{equation}
    \frac{\de N_{\rm gal,out}}{\de z \de \Omega \de M}=\frac{\de N_{\rm gal,bg}}{\de z \de \Omega \de M} p_{\rm miss}(z,\Omega,M).
    \label{eq:dnout}
\end{equation}
The background galaxy density can be defined in terms of Schechter function with slope $\alpha$ as: 
\begin{equation}
    \frac{\de N_{\rm gal,bg}}{\de z \de \Omega \de M} = {\rm Sch}(M;\alpha) \frac{\de V_c}{\de z \de \Omega},
\end{equation}
where for brevity we just write ${\rm Sch}(M;\alpha)$ to indicate the Schechter function.
Moving to $p_{\rm miss}$, we recall that usually galaxy catalogs are flux-limited, \ie a galaxy is detected if its apparent magnitude $m$ is brighter than a given threshold $m_{\rm thr}$. Note that $m_{\rm thr}$ could be sky-angle dependent to indicate that the galaxy survey is not able to cover in the same way all the sky, \eg a blind region would be identified by $m_{\rm thr}=-\infty$. Therefore, we can model $p_{\rm miss}$ as an Heaviside step function, $p_{\rm miss} = \Theta[(M>M_{\rm thr}(z,m_{\rm thr}(\Omega),H_0)]$. 

We can now use the out-of-catalog part to calculate its contribution to the CBC merger rate in Eq.~\ref{eq:total1}. In doing so we first note that the integral of the Schechter function multiplied by the luminosity weight can be written as: 
\begin{equation}
\begin{split}
      &\int_{M_{\rm min}}^{M_{\rm max}} \de M \, {\rm Sch}(M;\alpha)  10^{0.4 \epsilon(M_*(H_0)-M)} = \\ 
      &= \int_{M_{\rm min}}^{M_{\rm max}} \de M \, {\rm Sch}(M;\alpha+\epsilon) \\ 
      &= \phi_*(H_0) \Gamma_{\rm inc}(\alpha+\epsilon+1,x_{\rm min},x_{\rm max}),
      \label{eq:schres}
\end{split}
\end{equation}
where $\Gamma_{\rm inc}$ is the incomplete gamma function, while in the last step of the above equation, we have changed variable to $x=10^{0.4(M_*(H_0)-M)}$  and 
$x_{{\text{min/max}}}=10^{0.4(M_*(H_0)-M_{\rm min/max})}$.
Bearing in mind the result obtained in Eq.~\ref{eq:schres} and using Eq.~\ref{eq:dnout} we can define an ``effective'' galaxy density per steradian per redshift which are CBC emitters:
\begin{equation}
\begin{split}
    \frac{d N^{\rm eff}_{\rm gal,out}}{\de z \de \Omega} &= \int_{M_{\rm min}(H_0)}^{M_{\rm max}(H_0)} \de M \, 10^{0.4 \epsilon(M_*-M)} \frac{\de N_{\rm gal,out}}{\de z \de \Omega \de M} \\  
    &= \int_{M_{\rm thr}(H_0,m_{\rm thr}(\Omega),z)}^{M_{\rm max}(H_0)} \de M {\rm Sch}(M;\alpha+\epsilon) \frac{\de V_c}{\de z \de \Omega} \\  
    &= \phi_*(H_0) \Gamma_{\rm inc}(\alpha+\epsilon+1,x_{\rm max},x_{\rm thr}) \frac{\de V_c}{\de z \de \Omega}, 
    \label{eq:emitters}
\end{split}
\end{equation}
where $x_{\rm thr}=10^{0.4(M_*(H_0)-M_{\rm thr}(H_0))}$. The bright end of the integration is given by the absolute magnitude threshold, which is computed from the sky-dependent apparent magnitude threshold as a function of redshift and $H_0$. 
Before explaining the physical sense of the above result, let us note that  Eq.~\ref{eq:emitters} is not actually dependent on $H_0$. On one hand, $\phi_*(H_0) \propto H_0^{3}$, while $\frac{\de V_c}{\de z \de \Omega} \propto H_0^{-3}$, therefore the $H_0$-dependence will cancel. On the other hand, the incomplete gamma function depends on the slope parameters $\alpha$ and $\epsilon$ (which are cosmology-independent) and on the parameter $x$. The parameter $x$ is function of $M_*(H_0)-M_{\rm max/thr}(H_0)$, and this difference is $H_0$-independent as the two absolute magnitudes will scale in the same way with $H_0$.

Eq.~\ref{eq:emitters} can be thought of as an ``effective'' density of galaxy per steradian per redshift, which are the CBC emitters that are missing from the catalog. As an example, if the catalog is 0\% complete  (no galaxies are present and $M_{\rm thr}=-\infty$) and no luminosity weight is applied  $(\epsilon=0)$, then Eq.~\ref{eq:emitters} will simply reduce to the integral of the Schechter function times the differential of the comoving volume, which is by definition the number density of galaxies per redshift per steradian. If the galaxy catalog is 100\% complete (all the galaxies are present and $M_{\rm thr}=\infty$), then the integral in Eq.~\ref{eq:emitters} will evaluate to 0 as by definition the Schechter function is null for $M$ fainter than $M_{\rm max}$. 

Using Eq.~\ref{eq:emitters} we can also give a more formal definition of ``incompleteness''. This can be defined as the fraction of galaxies that are CBC emitters that we think are not present in the catalog part. More formally, we can define the incompleteness $\mathcal{I}_{\rm CBC}(z,\Omega)$ as: 
\begin{eqnarray}
        \mathcal{I}_{\rm CBC}(z,\Omega) &=& \frac{\int_{M_{\rm thr}(H_0,m_{\rm thr}(\Omega),z)}^{M_{\rm max}(H_0)} \de M {\rm Sch}(M;\alpha+\epsilon) \frac{\de V_c}{\de z \de \Omega}}{\int_{M_{\rm min}}^{M_{\rm max}(H_0)} \de M {\rm Sch}(M;\alpha+\epsilon) \frac{\de V_c}{\de z \de \Omega}} \nn \\ 
        &=& \frac{ \Gamma_{\rm inc}(\alpha+\epsilon+1,x_{\rm max},x_{\rm thr}(\Omega,z)) }{ \Gamma_{\rm inc}(\alpha+\epsilon+1,x_{\rm max},x_{\rm min}) }.
        \label{eq:inco}
\end{eqnarray}
In analogy with the discussion in the previous paragraph, we note that the incompleteness is also a quantity that does not depend on $H_0$.

\subsubsection{The catalog part}

We now calculate the ``in-catalog'' part of Eq.~\ref{eq:total1}, which is built starting from the galaxies present in the catalog. The discussion in this section follows the same logic as presented in \cite{2022arXiv221208694G}. In analogy to what we discussed in the previous section, even for the catalog part, we can define an effective number of CBC emitters:
\begin{equation}\label{eq:cat_part}
\begin{split}
    &\frac{\de N^{\rm eff}_{\rm gal,cat}}{\de z \de \Omega} = \\ &= \int_{M_{\rm min}}^{M_{\rm thr}(z,m_{\rm thr}(\Omega),H_0)} \de M \, 10^{0.4 \epsilon(M_*(H_0)-M)}  \frac{\de N_{\rm gal,cat}}{\de z \de \Omega \de M}.
\end{split}
\end{equation}
The integral in Eq.~\ref{eq:cat_part} is limited up to a faint end identified by $M_{\rm thr}(z,m_{\rm thr}(\Omega),H_0)$, since by definition we can not detect galaxies that are fainter than a given threshold value (which are accounted for by the completeness correction). Since galaxy surveys usually provide lists of galaxy redshifts, sky locations, and apparent magnitudes, it is more convenient to express the above integral in terms of apparent magnitudes $m$ as: 
\begin{equation}
\begin{split}
    &\frac{\de N^{\rm eff}_{\rm gal,cat}}{\de z \de \Omega} = \\ 
    &= \int_{m_{\rm min}}^{m_{\rm thr}(\Omega)} \de m \, 10^{0.4 \epsilon(M_*(H_0)-M(H_0,m,z))} \frac{\de N_{\rm gal,cat}}{\de z \de \Omega \de m}.
    \label{eq:neffcat}
\end{split}
\end{equation}
The galaxy number density can be built from the observed list of galaxies as the product of the total number of galaxies observed in the catalog and the probability of finding a galaxy at redshift $z$, sky position $\Omega$, and apparent magnitude $m$, given our observation of the catalog (labeling by $C$ the observed quantities):
\begin{equation}
    \frac{\de N_{\rm gal,cat}}{\de z \de \Omega \de m}= \Ngal p_{\rm gal}(z,\Omega,m|C).
\end{equation}
If the galaxy catalog provides us with perfectly measured redshift, sky position, and apparent magnitudes, $p_{\rm gal}(z,\Omega,m|C)$ would simply be a \new{sum of delta functions centered at the galaxy redshift, sky position, and apparent magnitude}. However, we are not usually able to measure perfectly some of these quantities, in particular redshift, so we need to account for this.

Following \cite{2022arXiv221208694G}, the probability of finding a galaxy at redshift $z$, sky position $\Omega$, and apparent magnitude $m$ given the observed survey can be written as:
\begin{equation}\label{eq:pgal}    
\begin{split}
    & p_{\rm gal}(z,\Omega,m|C) = \int \de \{z\}_{\rm gal} \de \{\Omega\}_{\rm gal} \de \{m\}_{\rm gal} \, \times \\
    & \times p_{\rm loc}(z,\Omega,m|\{z\}_{\rm gal}, \{\Omega\}_{\rm gal},\{m\}_{\rm gal}, C) \times \\
    & \times p_{\rm cat}(\{z\}_{\rm gal}, \{\Omega\}_{\rm gal},\{m\}_{\rm gal}|C). 
\end{split}
\end{equation}
In Eq.~\ref{eq:pgal} $\{z_{\rm gal}\}$, $\{\Omega_{\rm gal}\}$, and $\{m_{\rm gal}\}$ indicate a set of \textit{true} redshifts, sky positions, and apparent magnitudes for all the $\Ngal$ galaxies reported in the survey. 
The term $p_{\rm loc}$ is the probability of having a galaxy at $z,\Omega,m$ once we know the true $z_{\rm gal},\Omega_{\rm gal}$, and $m_{\rm gal}$ of all the other galaxies. This probability does not actually depend on the galaxy catalog and is simply: 
\begin{equation}
\begin{split}
    &p_{\rm loc}(z,\Omega,m|\{z\}_{\rm gal}, \{\Omega\}_{\rm gal},\{m\}_{\rm gal})= \\ 
    &= \frac{1}{\Ngal} \sum_{j=1}^{\Ngal}\delta(z-z^j_{\rm gal}) \delta(\Omega-\Omega^j_{\rm gal}) \delta(m-m^j_{\rm gal}).
    \label{eq:ploc}
\end{split}
\end{equation}
The term $p_{\rm cat}$ represents the probability that the galaxies have their true values of redshift, magnitude, and sky location of the galaxies $\{z_{\rm gal}\}$, $\{\Omega_{\rm gal}\}$, and $\{m_{\rm gal}\}$, given the \textit{observed} galaxy survey.
Generally, the galaxy catalog is made of a collection of galaxies, each of them observed \textit{independently} and reported with \textit{observed} apparent magnitude, redshift, and sky location $\{m\}_{\rm obs},\{\Omega\}_{\rm obs},\{z\}_{\rm obs}$ and typical uncertainties $\{\sigma_m\}_{\rm obs},\{\sigma_\Omega\}_{\rm obs},\{\sigma_z\}_{\rm obs}$.
Given the \textit{observed} quantities $C$, the probability that a given galaxy $j$  has a \textit{true} redshift $z^j_{\rm gal}$, sky location $\Omega^j_{\rm gal}$, and apparent magnitude $m^j_{\rm gal}$ can be written as: 
\begin{equation}\label{eq:uncgal}
\begin{split}
   p(z^j_{\rm gal},\Omega^j_{\rm gal},m^j_{\rm gal}|C) &= p(z^j_{\rm gal}|z^j_{\rm obs},\sigma^j_{\rm z, obs})  \times  \\ 
   &\times p(\Omega^j_{\rm gal}|\Omega^j_{\rm obs},\sigma^j_{\rm \Omega, obs}) \times\\
   &\times p(m^j_{\rm gal}|m^j_{\rm obs},\sigma^j_{\rm m, obs}),
\end{split}
\end{equation}
and since galaxies are observed independently from each other, we can write: 
\begin{equation}\label{eq:pcat}
\begin{split}
    &p_{\rm cat} (\{z\}_{\rm gal}, \{\Omega\}_{\rm gal},\{m\}_{\rm gal}|C) =\prod_{j=0}^{\Ngal} p(z^j_{\rm gal}|z^j_{\rm obs},\sigma^j_{\rm z, obs})\times \\ 
    & \times p(\Omega^j_{\rm gal}|\Omega^j_{\rm obs},\sigma^j_{\rm \Omega, obs}) p(m^j_{\rm gal}|m^j_{\rm obs},\sigma^j_{\rm m, obs}).
\end{split}
\end{equation}
By substituting Eqs.~\ref{eq:ploc} and \ref{eq:pcat} into Eq.~\ref{eq:pgal}, we find that the reconstructed galaxy number density is:
\begin{eqnarray}
   \frac{dN_{\rm gal,cat}}{dz d\Omega dm}=  && \sum_{j}^{N_{\rm gal} (\Omega)} p(z|z^j_{\rm obs},\sigma^j_{\rm z,obs}) p(\Omega|\Omega^j_{\rm obs},\sigma^j_{\rm \Omega, obs})  \times \nn \\ && \times p(m|m^j_{\rm obs},\sigma^j_{\rm m,obs}).
    \label{eq:galrr}
\end{eqnarray}
We now use Eq.~\ref{eq:galrr} to calculate the effective CBC emitters hosts in Eq.~\ref{eq:neffcat}. If we now assume that the apparent magnitude and sky position of each galaxy are measured perfectly (this is a reasonable assumption for galaxy surveys), and that the sky is divided in equal-size pixels of area $\Delta \Omega$, we can write the effective CBC emitters hosts in a given sky direction as:
\begin{eqnarray}
    \frac{\de N^{\rm eff}_{\rm gal,cat}}{\de z \de \Omega} \approx \frac{1}{\Delta \Omega} \sum_{j}^{\Ngal(\Omega)} && 10^{0.4 \epsilon(M_*(H_0)-M(H_0,m_j,z))} \times \nn \\ && \times p(z|z^j_{\rm obs},\sigma^j_{\rm z,obs})
    \label{eq:dngaleffcat},
\end{eqnarray}
where the sum is only on galaxies in a given sky direction.
Eq.~\ref{eq:dngaleffcat} is the relation needed to calculate the effective number of CBC hosts from the galaxy survey. The quantity $p(z|z^j_{\rm obs}, \sigma^j_{\rm z,obs})$ is the distribution of what true redshift we expect the galaxy to have, namely
\begin{equation}
        p(z|z^j_{\rm obs}, \sigma^j_{\rm z,obs}) \propto \mathcal{N}(z^j_{\rm obs},\sigma^j_{\rm z,obs}) \frac{\de V_c}{\de z}.
        \label{eq:correctinter}
\end{equation}
where $\mathcal{N}$ is a Gaussian. In Eq.\ref{eq:correctinter}, we have assumed that the galaxy redshift is measured from the catalog with gaussian uncertainty and that we apply a prior proportional to $\frac{\de V_c}{\de z}$ \citep{2022arXiv221208694G}.

We now make the following remarks. Eq.~\ref{eq:dngaleffcat} does not depend on $H_0$ for the same motivations explained for the out-of-catalog part, see the discussion below Eq.~\ref{eq:emitters}. Therefore, Eq.~\ref{eq:dngaleffcat} can be computed only once for each observed galaxy catalog, thus decreasing significantly the computational cost of the analysis. 
Strictly speaking, as discussed in \cite{2022arXiv221208694G}, in principle, the \textit{true} redshift of each galaxy should be treated as a ``population'' parameter for the hierarchical inference. In other words, in the limit that the galaxy redshifts are not observed perfectly, the effective density of CBC hosts can not be reconstructed \new{as a sum of redshift probability density distributions centered around the redshift values provided by the galaxy catalog}. However, in the limit that the number density of observed CBC events is much lower than the number density of CBC hosts, Eq.~\ref{eq:dngaleffcat} should still be valid to perform the hierarchical inference. Finally, we adopt also the same color corrections used in \cite{gwtc3_H0} to calculate the absolute magnitude for the apparent magnitude. 

\section{Tests with well-localized dark sirens}
\label{sec:4}

In this section, we provide some working examples of the methodology discussed in the previous sections using the well-localized standard sirens GW170817 \cite{ligobns} and GW190814 \cite{2020ApJ...896L..44A}. We use the \textsc{glade+} galaxy catalog \cite{gladeplus} and the galaxy luminosities reported for the infrared K-band. \new{\textsc{glade+} already reports redshift values which are corrected for galaxies peculiar motion. Additionaly, we consider the peculiar motion uncertainties addition by summing it in quadrature to the reported spectroscopic or photometric redshift uncertainty.}
In this section we apply Eq.~\ref{eq:fund_scalefree} to infer $H_0$ only, while we fix all the other population parameters (see App.~\ref{app:moredetails} for more details).

To compute the completeness correction in Eq.~\ref{eq:emitters}, we adopt a Schechter function specified by the parameters $M_{\rm min}=-27.85$, $M_{\rm max}=-19.84$, $\alpha=-1.09$,  $\phi_*=0.03 \,{\rm Mpc^{-3}}$ (for $H_0= 67.7 \, \hu$). To calculate the effective hosts' number from the galaxy catalog in Eq.~\ref{eq:dngaleffcat}, we subdivide the sky into 49152 pixels\footnote{calculated with \texttt{healpy} and a choice of \texttt{nside=64}} and a choice of equal area of 0.8 deg$^2$. The choice of the pixel size is driven by the fact that it should be smaller than the localization areas of GW190814, which is 18.5 deg$^2$~\citep{2020ApJ...896L..44A}. While for GW170817, since we will fix the sky position, we should take a pixel size large enough to contain the galaxies' localization errors and angular size.

\new{To calculate the apparent magnitude threshold, we follow the same procedure as in \cite{gwtc3_H0}. The apparent magnitude threshold is computed all over the sky using equal-sized pixels of 3.35 deg$^2$. On one hand, the pixel size for the calculation of the apparent magnitude threshold should be small enough to appreciate the different characteristics of the surveys composing \textsc{glade+}. On the other hand, the pixel size should be high enough to contain a reasonable number of galaxies to calculate the apparent magnitude threshold. The apparent magnitude threshold is defined as the median of the apparent magnitude of the galaxies reported in each pixel. This is an over-conservative threshold that is applied to avoid any systematic that could be potentially introduced by galaxy catalogs selection biases. We find that in a pixel of 3.35 deg$^2$ the median number of galaxies present is 80, which is enough to calculate an apparent magnitude threshold. All the galaxies fainter than the apparent magnitude threshold are removed from the calculation of the CBC host redshift density.}

\subsection{GW170817 as quasi-dark siren: the pencil beam approach}
\label{sec:4a}

So far, GW170817 is the only GW source for which it was possible to detect an associated electromagnetic counterpart and identify its host galaxy: NGC4993 \cite{2017ApJ...848L..12A}. GW170817 is also the only standard siren observed with EM counterpart. Using the GW luminosity distance estimation and the redshift of NGC4993 it was possible to estimate $H_0=70^{+19}_{-8}\, \hu$ \cite{2017Natur.551...85A,2019PhRvX...9a1001A}. An analysis of GW170817 as ``dark siren'' was also performed in \cite{2019ApJ...871L..13F} with the \textsc{glade} \cite{2018MNRAS.479.2374D} catalog using B-band luminosities, obtaining $H_0=77^{+37}_{-18}\, \hu$ \cite{2019ApJ...871L..13F}.

Here, we employ GW170817 as a ``quasi-dark siren'', namely, we assume the sky position to be perfectly known but the host galaxy not identified. \new{This is type of methodology is also referred to as the ``pencil beam'' approach \cite{chen17}}.

We made this choice in order to show the connection between the framework presented in Sec.~\ref{sec:3} and the various physical quantities involved. For this study, we use the parameter estimation (PE) samples from the \textsc{IMRPhenomPv2NRT} waveform model with low spin priors \footnote{\href{https://dcc.ligo.org/LIGO-P1800370/public}{https://dcc.ligo.org/LIGO-P1800370/public}} released for GWTC-1 \cite{gwtc1}. The PE samples are generated by fixing GW170817 sky location to the one its EM counterpart. We take all the galaxies falling within the 0.8 deg$^2$ pixel of GW170817's counterpart, for which we obtain an apparent magnitude threshold of $m_{\rm thr}=13.68$ (defined as the median of the apparent magnitude threshold of all the galaxies reported in the pixel).
\begin{table}[t]
  \centering
  \caption{Galaxies with K-band luminosities from the \textsc{glade+} galaxy catalog reported in the 0.8 deg$^2$ pixel of GW170817. From left to right columns, the reported values are observed redshift, redshift uncertainty (including peculiar motion), apparent magnitude in the K-band, and absolute magnitude calculated for a Planck15 \citep{2016A&A...594A..13P} cosmology. The first entry is NGC4993, the host of  GW170817. The apparent magnitude threshold for this particular pixel is $m_{\rm thr}=13.68$.}
  \begin{tabular}{c c c c}
    \hline
    $z_{\rm obs}$ & $\sigma_{\rm z, obs}$ & $m_{\rm K}$ & $M_{\rm K}$ \\ 
    \hline\hline
    0.0092 & 0.0005 & 9.3 & -23.7 \\ 
    0.0301 & 0.0148 & 13.0 & -22.6 \\ 
    0.0423 & 0.0014 & 12.9 & -23.4 \\ 
    0.0613 & 0.0153 & 13.5 & -23.7 \\ 
    0.0709 & 0.0152 & 13.0 & -24.4 \\ 
    0.1207 & 0.0151 & 13.4 & -25.1 \\ 
    0.1505 & 0.0151 & 13.4 & -25.6 \\ 
    \hline
  \end{tabular}
  \label{tab:tabgw}
\end{table}
After removing all the galaxies fainter than $m_{\rm thr}$ from \textsc{glade+}, we obtain only 7 galaxies with redshift spanning from $0.0092$ to $0.15$ in the pixel of GW170817. The observed redshifts and luminosities are reported in  Tab.~\ref{tab:tabgw}, where the first row of the table is NGC4993, the host of GW170817. We can immediately notice that farther galaxies display a lower (brighter) absolute magnitude. This is due to the presence of a selection bias, for which at high redshift we can observe only the brightest galaxies, and which is accounted using the completeness correction.

\begin{figure}
    \centering
    \hspace{-.8cm}
    \includegraphics[scale=0.51]{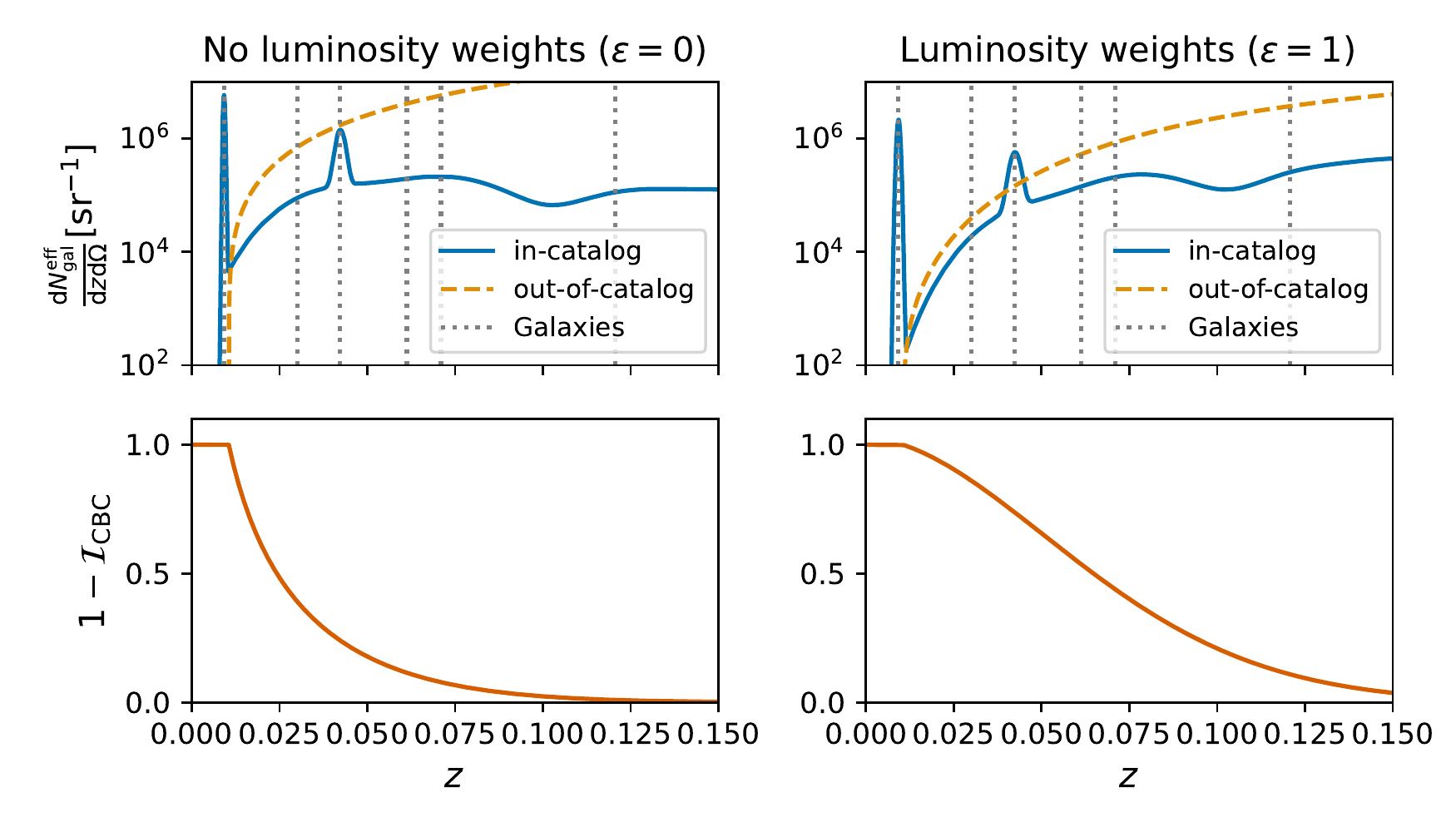}
    \caption{\textit{Top row}: Effective number of in-catalog (solid lines) and out-of-catalog (dashed line) CBC emitters per redshift per steradian for GW170817 as a quasi-dark siren. The blue solid line is the contribution from the galaxy catalog, while the orange dashed line is the completeness correction.
    The vertical dotted lines indicate the observed redshift values reported in Tab.~\ref{tab:tabgw}. \textit{Bottom row:} CBC completeness $1 - \mathcal{I}_{\rm CBC}$, with $\mathcal{I}_{\rm CBC}$ given by Eq.~\ref{eq:inco}. \new{The luminosity weights are assigned using a probability $p(L|\epsilon) \propto L^\epsilon$.}.
    The first column is generated using a CBC emission rate independent of galaxy luminosity ($\epsilon=0$), while the second is generated using a rate linearly proportional to the galaxy luminosity ($\epsilon=1$).}
    \label{fig:GW170817_catalog_info}
\end{figure} 
Using the galaxies reported in Tab.~\ref{tab:tabgw} and the apparent magnitude threshold previously computed, we calculate the effective CBC hosts density present in the catalog (Eq.~\ref{eq:dngaleffcat}) and their completeness correction (Eq.~\ref{eq:emitters}). The top panels of Fig.~\ref{fig:GW170817_catalog_info} show a comparison between the effective number density of CBC hosts ``in'' and ``out'' of the galaxy catalog. 
The figure displays two cases: one where we assume that the CBC merger rate does not depend on the intrinsic galaxy luminosity ($\epsilon=0$, left column), and one  where we assume the CBC merger rate is proportional to the galaxy luminosity ($\epsilon=1$, right column). 
Importantly, we note that in the $\epsilon=1$ case, the effective number of CBC hosts in the catalog is lower. This is because in this model galaxies fainter than $M_{*}$, which are the majority of galaxies according to the Schechter function, host fewer GW sources.
We also note that in the $\epsilon=1$ model, the CBC completeness of the galaxy catalog (second row in Fig.~\ref{fig:GW170817_catalog_info}) is higher since the brightest galaxies included in the catalog are the majority fraction of the CBC emitters. 
Finally, from Fig.~\ref{fig:GW170817_catalog_info} we notice that from the in-catalog part, only two galaxies can be well localized in redshift and provide a redshift scale (the first peak is NGC4993). The other galaxies have large redshift uncertainties (cf. Tab.~\ref{tab:tabgw}) and will not significantly contribute to the measure. 

\begin{figure}
    \centering
    \includegraphics[scale=0.7]{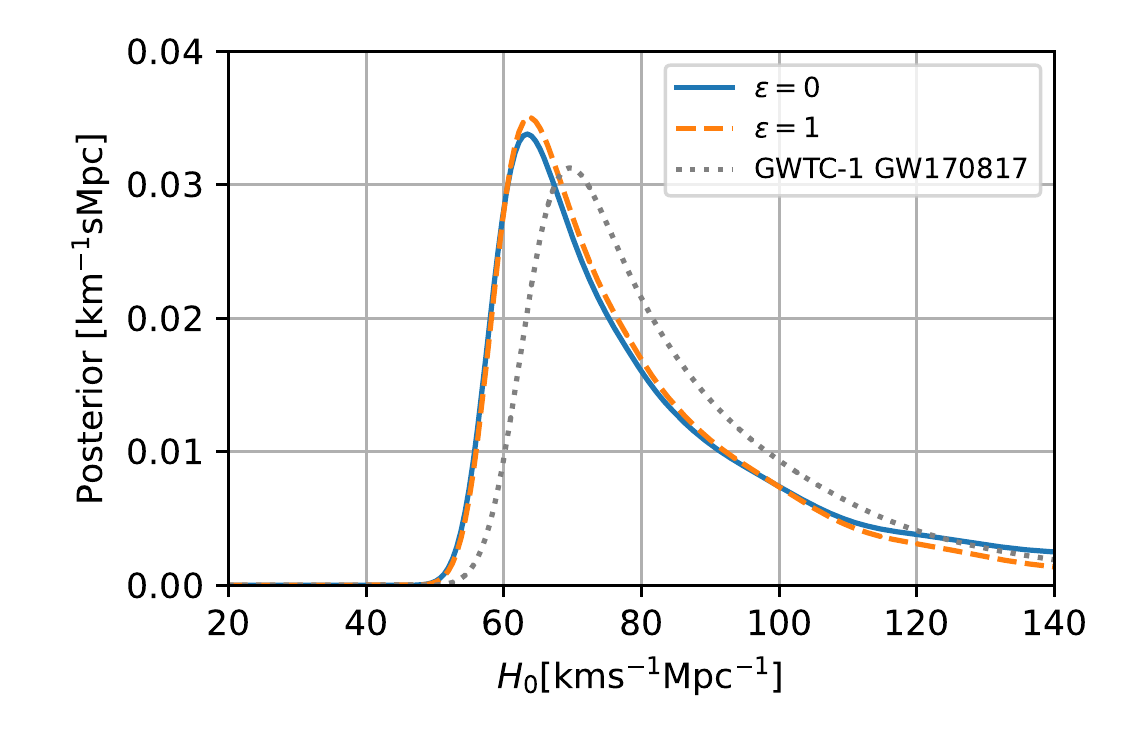}
    \caption{Hubble constant posterior distribution obtained for GW170817 as quasi-dark siren. The blue solid and orange dashed lines indicate the posterior obtained using a CBC merger rate independent of the galaxies' luminosity and proportional to it, respectively. \new{The luminosity weights are assigned using a probability $p(L|\epsilon) \propto L^\epsilon$.} The gray dotted line is the posterior of GW170817 obtained assuming the identification of the host galaxy with $z=0.0100 \pm 0.0005$ \cite{LVC_O2_StS}.}
    \label{fig:GW170817_comparison}
\end{figure}
We use the CBC host densities shown in Fig.~\ref{fig:GW170817_catalog_info} to 
calculate the CBC rate  in Eq.~\ref{eq:total1}, which is then used to evaluate the scale-free hierarchical likelihood in Eq.~\ref{eq:fund_scalefree} and
obtain a posterior distribution on the value of $H_0$. We consider a single population model for CBC rates in terms of masses and redshift. In App.~\ref{app:moredetails} we provide more details about the CBC rate models for masses and redshift used, as well as the corrections of selection biases. 
Fig.~\ref{fig:GW170817_comparison} shows the $H_0$ posterior obtained for the different cases for which we computed the in-catalog and out-of-catalog CBC hosts densities. We observe that, for GW170817, the $H_0$ measure is nearly independent of the choice of using or not luminosity weights. This is a consequence of the fact that the only structure in the CBC host density contributing to the $H_0$ estimation is the one associated with NGC4993. 
Compared to the posterior obtained by the LIGO-Virgo Collaboration using GW posterior samples from GWTC-1 \cite{LVC_O2_StS}, we obtain an $H_0$ posterior slightly narrower with a maximum a posteriori shifted to smaller values. This is due to the fact that in \textsc{glade+}, NGC4993 redshift is slightly lower \new{($z=0.0092 \pm 0.0005$)} than the value of $z=0.0100 \pm 0.0005$ used for the analysis in \cite{LVC_O2_StS}. \new{Moreover, in the pencil-beam approach, the $H_0$ inferece gets some contribution from galaxies not well localized at high redshifts.}

Moreover, in our analysis, we also have the contribution from galaxies at higher redshifts which are not well localized (see Fig.~\ref{fig:GW170817_catalog_info}): this introduces a slight preference for lower values of $H_0$.

\subsection{GW190814 as a typical dark siren}
\label{sec:4b}

GW190814 \cite{2020ApJ...896L..44A} is one of the best localized GW events so far, second only to GW170817. The 90\% credible interval sky area of GW190814 covers about 20 deg$^2$ and its localization volume (assuming the Planck15 cosmology \citep{2016A&A...594A..13P}) is about 900 Mpc$^3$\cite{gwtc3_H0}. Previous $H_0$ studies for this event using the galaxy catalog method with the \textsc{glade} (B-band) and the Dark Energy Survey \cite{palmese20_sts} found a posterior with a maximum a posteriori in the Hubble constant tension region. However, it was not possible to constrain the $H_0$ value in the prior range of $[20,140]\, \hu$. One of the peculiar aspects of this event is that the mass of the lighter object falls between the expected gap between neutron stars and black hole masses. 

In our test case, we classify GW190814 as a massive neutron star black hole merger. \new{Note that, when considering cosmology unknown we are intrinsically changing the source frame mass of the secondary object, so one should use a population model that includes neutron stars and black holes altogether. This is due to the fact that the secondary object might end up in the neutron star or black hole mass region depending on the cosmological model.
However, we do not expect results for this event to strongly depend on the population model of the neutron star masses. For the prior extremes of $H_0$ assumed, this event would change its source frame mass from a median value of $2.6 \, \Msol$ (for Planck cosmology), \footnote{value taken from \url{https://gwosc.org/eventapi/html/allevents/}} to a range of $2.46 \Msol - 2.67 \Msol$ (depending on which $H_0$ extreme we are using). This range is well between then ``massive'' neutron star region and it is also smaller than the typical statistical uncertainties.}

The population model we fix for neutron star black hole mergers is described in App.~\ref{app:moredetails}. For the GW190814 PE samples, we use the ``Publication Samples''\footnote{\href{https://dcc.ligo.org/LIGO-P2000223/public}{https://dcc.ligo.org/LIGO-P2000223/public}} that are a combination of the samples obtained with the two \textsc{IMRPhenom} and \textsc{EOB} waveform families  released with GWTC-2 \cite{gwtc2}.

Consistently with \cite{gwtc3_H0}, we find that GW190814's 90\% credible sky localization contains 90 galaxies with K-band luminosities reported in \textsc{glade+}, while we find a median apparent magnitude threshold equal to $m_{\rm thr}=13.55$.
\begin{figure}
    \centering
    \hspace{-.8cm}
    \includegraphics[scale=0.51]{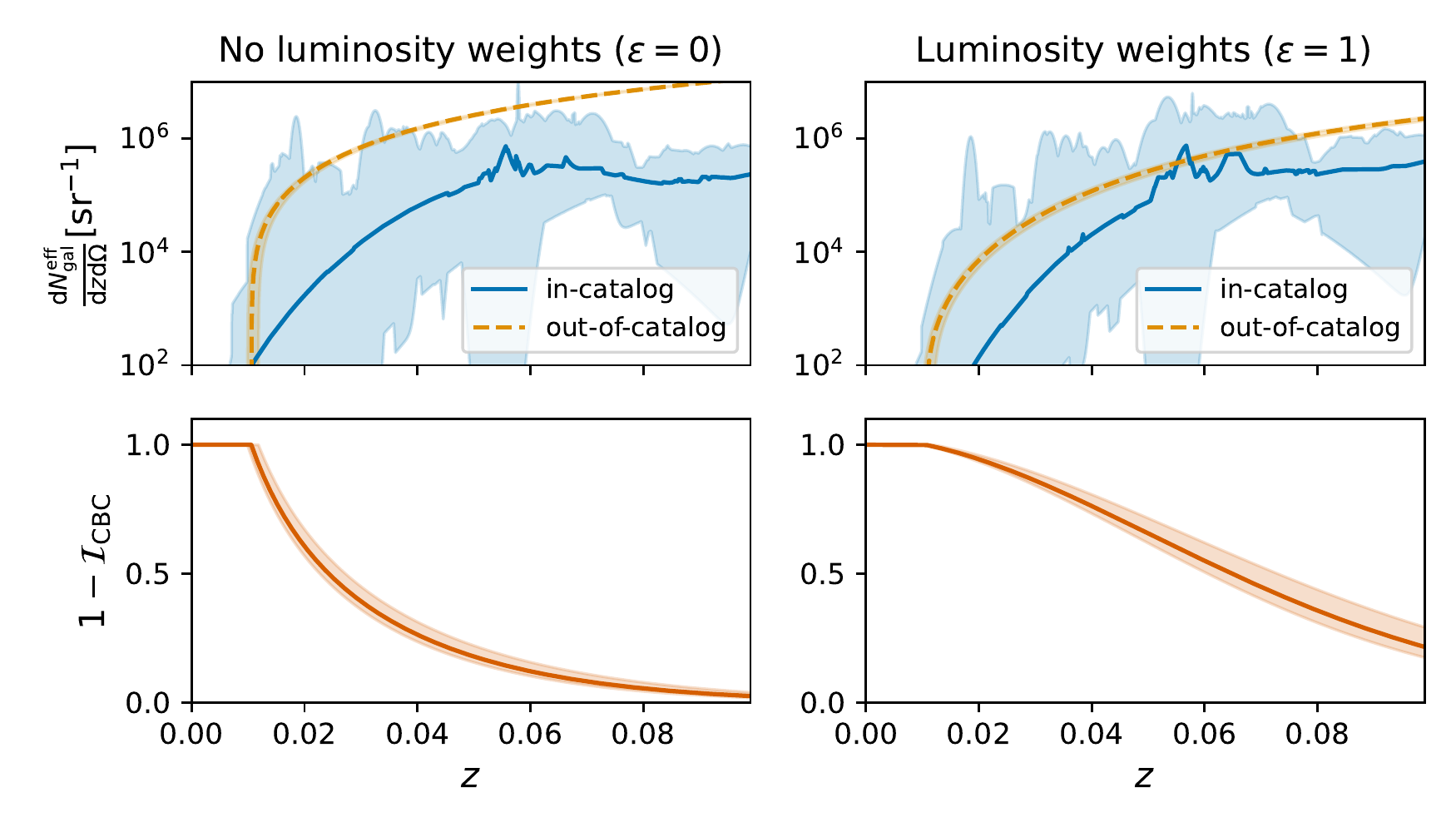}
    \caption{\textit{Top row}: Effective number of in-catalog (solid lines) and out-of-catalog (dashed line) CBC emitters per redshift per steradian in the 90\% credible interval sky area of GW190814. \textit{Bottom row:} CBC completeness  $1- \mathcal{I}_{\rm CBC}$, with $\mathcal{I}_{\rm CBC}$ given by Eq.~\ref{eq:inco}. The shaded areas indicate the contours identified by the 90\% credible sky area of GW190814, while the lines correspond to the median values. \new{The luminosity weights are assigned using a probability $p(L|\epsilon) \propto L^\epsilon$.} The first column is generated using a CBC emission rate independent of galaxy luminosity ($\epsilon=0$), while the second is generated using a rate linearly proportional to the galaxy luminosity ($\epsilon=1$).}
    \label{fig:GW190814_catalog_info}
\end{figure}
Fig.~\ref{fig:GW190814_catalog_info} displays the effective number density of in-catalog and out-of-catalog CBC hosts calculated for different models. Given the larger sky area covered by GW190814, in Fig.~\ref{fig:GW190814_catalog_info} we show the 90\% credible intervals of the different quantities. We observe that there is an overdensity of galaxies in the redshift region from 0.05 to 0.06, which is the redshift expected for GW190814 if one assumes a cosmological model with $H_0$ priors in the Hubble tension region.

\begin{figure}
    \centering
    \includegraphics[scale=0.7]{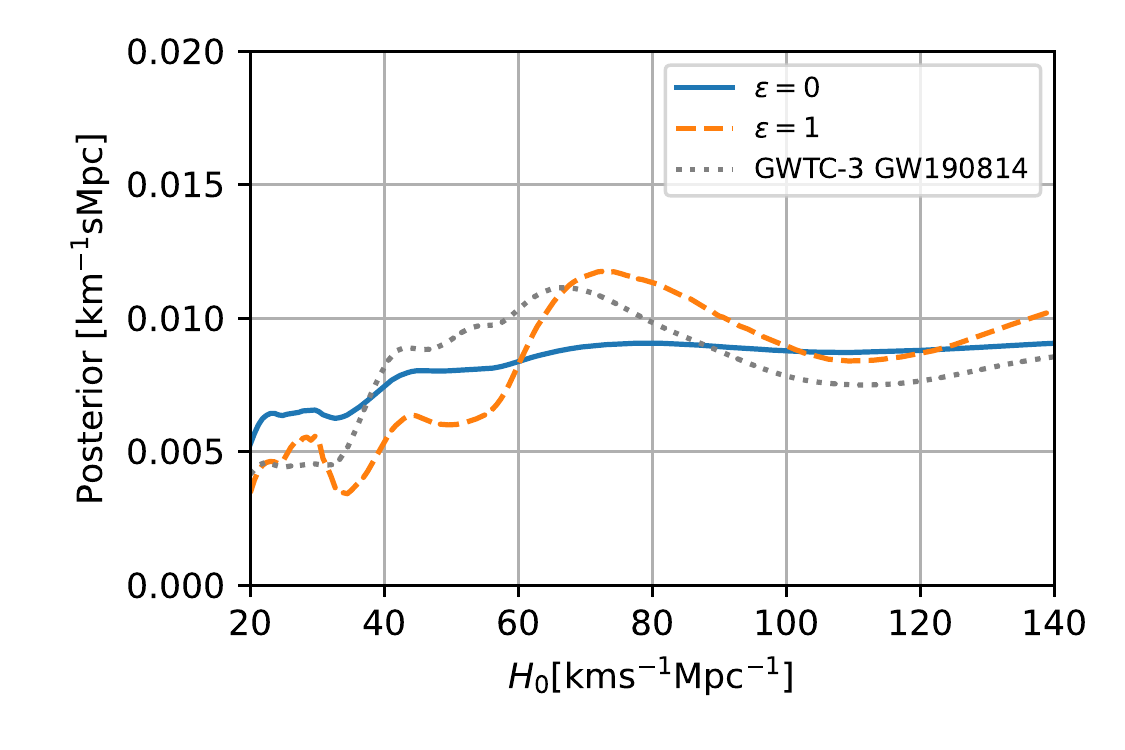}
    \caption{Hubble constant posterior distribution obtained for the dark siren GW190814. The blue solid and orange dashed lines indicate the posterior obtained using a CBC merger rate independent of the galaxies' luminosity and proportional to it. \new{The luminosity weights are assigned using a probability $p(L|\epsilon) \propto L^\epsilon$.} The gray dotted line is the posterior of GW190814 obtained by the LVK collaboration in \cite{gwtc3_H0}.}
    \label{fig:GW190814_comparison}
\end{figure}

In Fig.~\ref{fig:GW190814_comparison} we show the posteriors obtained for $H_0$ using different prescriptions for the galaxy catalog. 
We note that when we assume a CBC rate proportional to the galaxies' intrinsic luminosity ($\varepsilon=1$), the Hubble constant posterior displays a local maximum in the $H_0$-tension region. The local maximum is still present, but less pronounced when a constant CBC rate with respect to the galaxies' luminosity is assumed ($\varepsilon=0$), the reason being that the CBC completeness is lower for the uniform rate, so that the out-of-catalog term washes out the impact of redshift features introduced by the in-catalog term. 

Regarding the comparison with the GWTC-2 estimate, the analysis in \citep{gwtc3_H0} also uses a sky pixel of 0.8  deg$^2$, but it makes the choice of considering pixels with less than 10 galaxies as ``empty''. This choice translates to a higher incompleteness of the galaxy catalog for GW190814, resulting in a less informative $H_0$ posterior. 

\section{Application to binary black holes in GWTC-3}
\label{sec:5}

We now apply our framework to probe the current expansion rate of the Universe using BBHs reported in GWTC-3.  In \cite{gwtc3_H0,2022PhRvD.105f4030M}, it has been shown that BBHs can be used to probe cosmology either by using the standard galaxy catalog approach, or knowledge from the source-frame mass distribution. Although in \cite{gwtc3_H0,2022PhRvD.105f4030M} these two aspects are considered separately, it is important to consider them conjointly \cite{mastrogiovanni_2021}, in order to obtain an unbiased estimate of the cosmological parameters. For the analyses presented in this section, and consistently with \cite{gwtc3_H0}, we use galaxies with K-band luminosities reported in the \textsc{glade+} catalog. To calculate the apparent magnitude threshold, we divide the sky into pixels of equal area\footnote{calculated with \texttt{healpy} and \texttt{nside=16}.} of 13 deg$^2$. This resolution is suitable for the best localized BBHs detections, which are usually constrained in a sky region larger than $50-100$ deg$^2$ \cite{gwtc3_H0}. Smaller pixel areas will not affect the results also because, as we will see later, most of the $H_0$ information will actually come from the completeness correction.
In this section, we apply Eq.~\ref{eq:form2} to infer $H_0$ jointly with the population parameters.

\subsection{Catalog-only benchmark analysis}
\label{sec:5a}

We first perform a benchmark analysis. We fix the BBH population model (source mass spectrum) to the one used by the galaxy catalog analysis in \cite{gwtc3_H0} (see App.~\ref{app:moredetails} for the population models employed).
We select 42 BBHs from GWTC-3 with a signal-to-noise ratio (SNR) above 11 and inverse false alarm rate (IFAR) higher than $4$ yr. 
\begin{figure}
    \centering
    \includegraphics[scale=0.6]{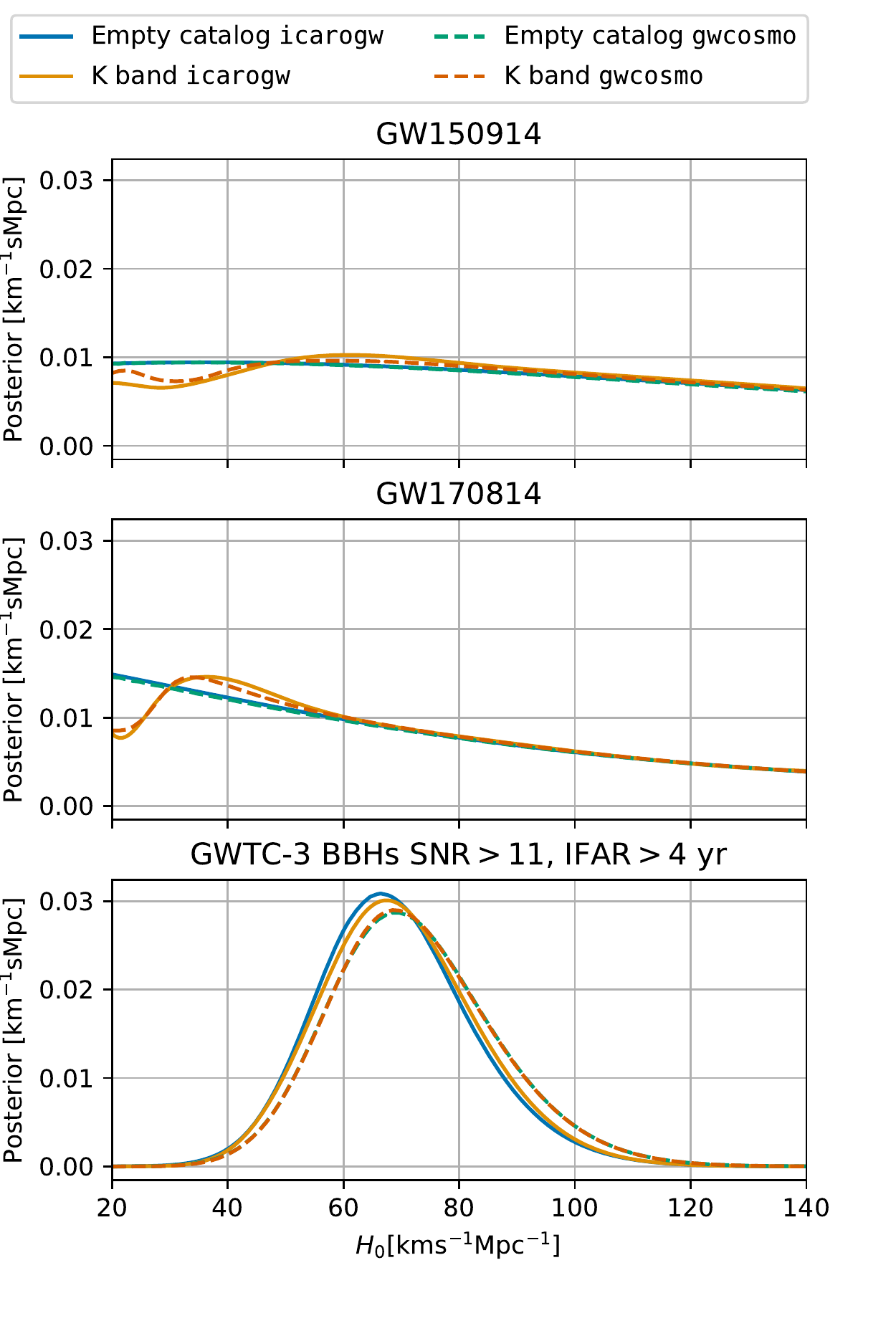}
    \caption{Hubble constant posteriors\ distributions for an analysis of the 42 BBHs from GWTC-3 with SNR above 11 and IFAR higher than $4$, fixing the BBH population model in the inference. The dashed lines are the $H_0$ posterior obtained by \textsc{gwcosmo} in \cite{gwtc3_H0}, while the solid lines are the ones obtained in this work. The ``empty catalog'' assumes a 0\% complete galaxy catalog, where the inference is driven mostly by BBH population assumptions. \textit{Top panel}: $H_0$ posterior for GW150914. \textit{Middle panel}: $H_0$ posterior for GW170814. \textit{Bottom panel}: $H_0$ posterior for the combined 42 BBHs considered in the analysis.}
    \label{fig:BBHs_differences}
\end{figure}
In Fig.~\ref{fig:BBHs_differences} we show the $H_0$ posterior obtained in this analysis for two events (GW150914 and GW170814) and the combination of all 42 dark sirens. In general, we find a good agreement between the results in \cite{gwtc3_H0} generated by \textsc{gwcosmo} and our approach. The plots show two test cases: the ``empty catalog'' case and the ``full catalog'' case. The former is generated by artificially assuming that the galaxy catalog contains no galaxies, while the latter uses the entire galaxy catalog. The empty catalog case is used to quantify the impact of population assumptions as the redshift information is dominated by the completeness correction. 
Comparing the empty and full catalog curves in Fig.~\ref{fig:BBHs_differences}, we see that, for the combined posterior, the $H_0$ inference is completely dominated by the BBH population prescription, while on a single-event level, adding redshift information from the galaxy catalog we are able to disfavor lower values of $H_0$. This is explained by the fact that for low $H_0$ values, the redshift of the GW events is lower and is located in a region of the catalog that is more complete.

\subsection{Full catalog and BBH population analysis}
\label{sec:5b}

Next, we perform a joint fit of $H_0$ and BBH population properties, this time using information also from the galaxy catalog. Following \cite{gwtc3_H0}, for the BBH population model, we adopt a \textsc{power law + peak} phenomenological mass model and a Madau-Dickinson \cite{2014ARA&A..52..415M} model $\psi(z;\Lambda_p) =  \psi_{\rm MD}(z;\gamma,k,z_p)$ for the CBC merger rate as a function of redshift. 
Also for this analysis, we selected 42 BBHs from GWTC--3 with SNR above 11 and IFAR higher than 4 yr. More details about the redshift-related parameters as well as the BBH population models with their prior ranges can be found 
 in App.~\ref{app:moredetails}.
In this analysis, we model the CBC merger rate per galaxy as: 
\begin{equation}
    \frac{\de \Ncbc}{\de \Ngal \de t_s}= \Rgal\left( \frac{L}{L_*}\right)^\epsilon \psi_{\rm MD}(z;\gamma,k,z_p).
    \label{eq:cbcrate_1}
\end{equation}
We consider three scenarios characterized by different assumptions on the luminosity weight: $\epsilon=1$, $\epsilon=0$, and $\epsilon=1$ but with the further assumption that the galaxy catalog is 0\% complete (empty catalog case). note that for the ``empty catalog'' case, changing $\epsilon$ would only result in a different \Rgal, the GW population parameters are only determined by the completeness correction.

\begin{table*}
    \centering
    \caption{Maximum a posteriori and symmetric 68.3\% credible intervals of the marginal distributions for the mass-related population parameters using 3 galaxy catalog prescriptions used in the analysis of the 42 BBHs from GWTC-3 having SNR above 12 and IFAR higher than 4 yr. The reported population parameters are used to build the BBH mass spectrum, see App.~\ref{app:moredetails} for more details. Parameters with no entry are not constrained within their prior ranges.}
    \label{tab:mass}
    \begin{tabular}{ccccccccc}
        \hline
		Model & $\alpha$ & $\beta$ & $m_{\rm min}$ & $m_{\rm max}$ & $\delta_{\rm m}$ [$M_\odot$] & $\mu_{\rm g}$ [$M_\odot$] & $\sigma_{\rm g}$[$M_\odot$] & $\lambda$ \\ 
		\hline
		$\epsilon=1$ & $3.85^{+0.50}_{-0.52}$ & $0.7^{+1.3}_{-1.1}$ & $5.17^{+0.69}_{-0.95}$ & - & $4.8\pm 2.4$ & $32.3^{+2.8}_{-4.3}$ & $1.9^{+3.4}_{-1.5}$ & $0.019^{+0.030}_{-0.017}$ \\ 
		$\epsilon=0$ & $3.81^{+0.49}_{-0.48}$ & $0.7\pm 1.2$ & $5.16^{+0.66}_{-0.96}$ & - & $4.7^{+2.9}_{-1.8}$ & $33.2^{+2.8}_{-4.3}$ & $2.8^{+3.0}_{-2.3}$ & $0.018^{+0.029}_{-0.016}$ \\ 
		Empty ($\epsilon=1$) & $3.82^{+0.46}_{-0.51}$ & $0.56^{+1.47}_{-0.97}$ & $4.99^{+0.95}_{-0.68}$ & - & $5.0^{+2.4}_{-2.2}$ & $33.0^{+3.0}_{-4.0}$ & $2.8^{+3.0}_{-2.4}$ & $0.019^{+0.030}_{-0.017}$ \\ 
		\hline
    \end{tabular}
\end{table*}
In all the cases, we find that the population parameters related to the mass distribution do not strongly depend on the prescription used for the galaxy catalog (see Tab.~\ref{tab:mass}). 
This is a consequence of the fact that our CBC mass rate model in Eq.~\ref{eq:cbcrate} does not explicitly depend on galaxies' redshift and absolute magnitude. 
\begin{table*}
    \centering
    \caption{Maximum a posteriori and symmetric 68.3\% credible intervals of the marginal distributions of the cosmology and redshift-related parameters for the 3 galaxy catalog prescriptions used in the analysis of the 42 BBHs from GWTC-3 having SNR above 12 and IFAR higher than 4 yr. The reported population parameters are used to build the BBH rate as function of redshift, see App.~\ref{app:moredetails} for more details. Parameters with no entry are not constrained within their prior ranges.}
    \label{tab:rate_params}
    \begin{tabular}{cccccc}
        \hline
		Model & $H_0 \, {\rm[km \,s^{-1}\, Mpc{-1}]}$ & $\gamma$ & $k$ & $z_p$ & $\log_{10} \left[\frac{R^*_{\rm gal,0}}{{\rm yr^{-1}}}\right]$ \\ 
		\hline
		$\epsilon=1$ & $71^{+35}_{-30}$ & $4.5^{+2.0}_{-1.7}$ & -  & -  & $-5.67^{+0.63}_{-0.50}$ \\ 
		$\epsilon=0$ & $43^{+48}_{-18}$ & $4.9^{+2.2}_{-2.1}$ &  - & -  & $-6.10^{+0.81}_{-0.51}$ \\ 
		Empty ($\epsilon=1$) & $52^{+36}_{-25}$ & $4.9^{+2.3}_{-2.1}$ & -  & - & $-5.48^{+0.80}_{-0.50}$ \\ 
		\hline
    \end{tabular}
\end{table*}

We  also note that the parameters related to the shape of CBC rate as a function of redshift $\psi_{\rm MD}$ do not strongly depend on the catalog prescription (see Tab.~\ref{tab:rate_params}).  Redshift-related population parameters are independent of the catalog prescription since \textit{(i)} GW events are in general not precisely localised in the redshift space, thus they are associated to thousands of galaxies and \textit{(ii)} the galaxy catalog is highly incomplete at $z >0.1$, and the CBC redshift rate contribution is mostly calculated with the completeness correction (dependent only on the comoving volume).

\begin{figure}
    \centering
    \includegraphics[scale=0.7]{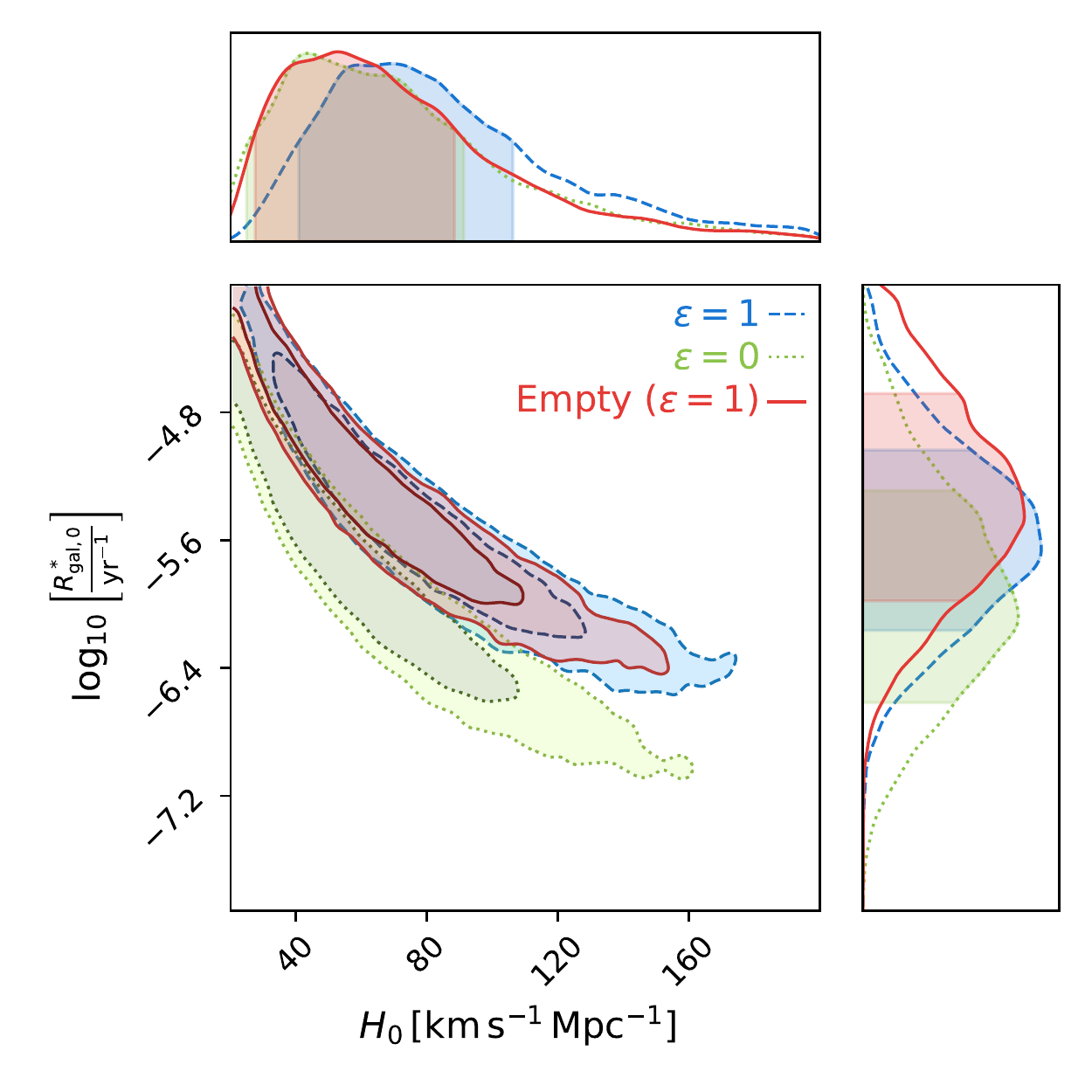}
    \caption{Two dimensional and one-dimensional marginal distributions on the logarithm of the CBC merger rate per galaxy today \Rgal and $H_0$. The inner contour of the two-dimensional posterior indicates the 68.3\% credible intervals while the outer contour is the $90\%$. \new{The luminosity weights are assigned using a probability $p(L|\epsilon) \propto L^\epsilon$.} The marginal 1-dimensional distribution reports the 68.3\% credible intervals. Different colors correspond to the three galaxy catalog prescriptions.}
    \label{fig:posteriors_Kband_BBHs_small}
\end{figure}

\new{Let us now focus on the estimation of \Rgal, the CBC merger rate per galaxy with a reference luminosity $L_*$, jointly with $H_0$}. In Fig.~\ref{fig:posteriors_Kband_BBHs_small} we report the joint posterior distribution for the 42 BBHs selected in the analysis and the various galaxy catalog and rate prescriptions.
Firstly, we find that the value of \Rgal is strongly dependent on the prescription used for the CBC merger rate as a function of the galaxy luminosity ($\epsilon$). In particular, the CBC merger rate per galaxy should be higher when we assume that brighter galaxies are more likely to host GW events. This is expected as brighter galaxies are significantly lower in number than fainter galaxies, therefore to keep constant the expected number of CBC detections per year \Rgal must significantly increase. 
Secondly, we find that there is a strong anti-correlation between \Rgal and $H_0$. Correlations are introduced when the variation of a certain combination of population-level parameters keeps the hierarchical likelihood in Eq.~\ref{eq:fund} (or Eq.~\ref{eq:form2}) constant. For the particular case of \Rgal and $H_0$, the two parameters are degenerate for the calculation of the expected number of CBC detections \Nexp. The expected number of CBC detections \Nexp is linearly dependent on \Rgal and is also dependent on $H_0$, as for lower values of $H_0$ the GW detection range can include galaxies at higher redshift.  In other words:
\begin{equation}
    \Nexp \approx \Rgal \Tobs \Ngal(z<z(d_L^H,H_0)),
\end{equation}
where $d_L^H$ is a typical luminosity distance horizon for GW detection and $\Ngal(z<z(d_L^H,H_0))$ the number of galaxies enclosed in a redshift sphere with radius $z(d_L^H,H_0)$. The number of galaxies enclosed in the GW detection horizon will roughly scale as $\Ngal \propto [d_L^H H_0]^3$. This implies that, to keep constant the value of \Nexp we can either increase \Rgal and decrease $H_0$ or viceversa. 
Note that this type of correlation is not usually observed between $H_0$ and the CBC merger rate per comoving volume $R_0$, which is the parametrization typically used in population studies. This is due to the fact that the $R_0$-parameterization depends on the differential of the comoving volume (which scales as $H_0^{-3}$):
\begin{equation}
    \Nexp \approx R_0 \Tobs V_c(z<z(d_L^H,H_0)),
\end{equation}
where $V_c(z<z(d_L^H,H_0))$ is the comoving volume within a redshift shell with radius $z(d_L^H,H_0)$. As the comoving volume scales as $V_c \propto [z(d_L^H,H_0)]^3/H_0^3$ and $z(d_L^H,H_0) \propto d_L^H H_0$, then $V_c$ will not strongly scale with $H_0$.  As a consequence $R_0$ is not correlated with $H_0$ during the inference. In App.~\ref{app:emptycat} we discuss in more detail how $R_0$ and $\Rgal$ are related to each other.

Let us finally comment about $H_0$. On one hand, the marginal posterior for $H_0$ is almost identical in the empty catalog and $\epsilon=0$ case. In fact, as we have shown in Sec.~\ref{sec:4}, assuming $\epsilon=0$ corresponds to a galaxy catalog that is significantly less complete from the CBC hosts' point of view (see Fig.~\ref{fig:GW190814_catalog_info} as an example). On the other hand, assuming $\epsilon=1$ with the galaxy catalog we obtain a marginal $H_0$ posterior which disfavors low values of $H_0$. This is due to the fact that for small $H_0$ values, GW events are located at lower redshifts, where the galaxy catalog is more complete (especially for $\epsilon=1$). Galaxy catalogs, if complete, can provide additional constraints to lower values of $H_0$, therefore it is crucial to use them to improve current constraints.

\subsection{GWTC--3 and modified gravity}
\label{sec:5c}

Proposed modifications to general relativity (GR) that include a friction term for the GW propagation can result in different expressions for the GW luminosity distance $d^{\rm GW}_L$ and the classical (EM) luminosity distance $d^{\rm EM}_L$.
Different parametrizations of $d^{\rm GW}_L$ have been proposed. Here we focus on three models: the $\Xi_0$-model \cite{PhysRevD.98.023510}, the extra-dimensions model \cite{Corman:2021avn}, and the running Planck mass model \cite{Lagos:2019kds}. The $\Xi_0$-model adds the two phenomenological parameters $\Xi_0$ and $n_{\Xi_0}$ to describe the distance relation. The extra-dimensions model adds three parameters: the number of spacetime dimensions $D$, the screening distance $R_c$, and a scaling factor $n_D$. The running Planck mass adds only one extra parameter $c_M$.
These models parametrize the relation between $d^{\rm GW}_L$ and $d^{\rm EM}_L$ as: 
\begin{eqnarray}
    \label{eq: def Xi parametrization dl}
    d_{L}^{\mathrm{GW}} &=& d_{L}^{\mathrm{EM}}\left[\Xi_0 + \frac{1-\Xi_0}{(1+z)^{n_{\scriptscriptstyle \Xi_0}}}\right], \label{eq:xi0} \\ 
    d_L^{\rm GW} &=& d_L^{\rm EM}\left[1+\left(\frac{d_L^{\rm EM}}{(1+z)R_c}\right)^n \right] ^{\frac{D-4}{2n_{\scriptscriptstyle D}}},
    \label{eq:dpg} \\ 
    d_L^{\rm{GW}} &=& d_L^{\rm{EM}} {\rm{exp}} \left[ \frac{c_M}{2} \int_0^{z} \frac{1}{(1+z')E^{2}(z')} dz' \right].
\end{eqnarray}
Bright sirens are often used to put constraints on these models (see~\cite{2022JCAP...09..012L} and references therein). However, even the galaxy catalog method \citep{finke2021cosmology}  and the spectral siren method \cite{PhysRevD.107.084033,2021PhLB..82236665E,2022PhRvD.105f4030M,2022JCAP...09..012L} can be used. 

Here we repeat the full spectral siren and spectral+catalog analyses for these three models. To do so, we use the same selection of events used in Sec.~\ref{sec:5a} and Sec.~\ref{sec:5b}. This time, we fix the value of $H_0$ to 67.7\, \hu because the determination of $H_0$ and the modified gravity parameters is strongly degenerate and with the current number of events we are unable to jointly constrain both. For the  $\Xi_0$-model, we use a uniform prior in $[0.1,50]$ for $\Xi_0$ and  a uniform prior in $[1,10]$ for $n_{\Xi_0}$. For the extra dimension model, we use a uniform prior in $[3.6,8]$ for $D$, a log uniform prior in $[0.1,100]$ for $n_D$, and a log uniform prior in $[10,10^5]$ Mpc for $R_c$. For the running Planck mass model, we use a uniform prior in  $[-10,50]$. The rest of the priors assumed for the population model are equal to the ones used in Sec.~\ref{sec:5b}. 

\begin{figure}
    \centering
    \includegraphics[scale=0.7]{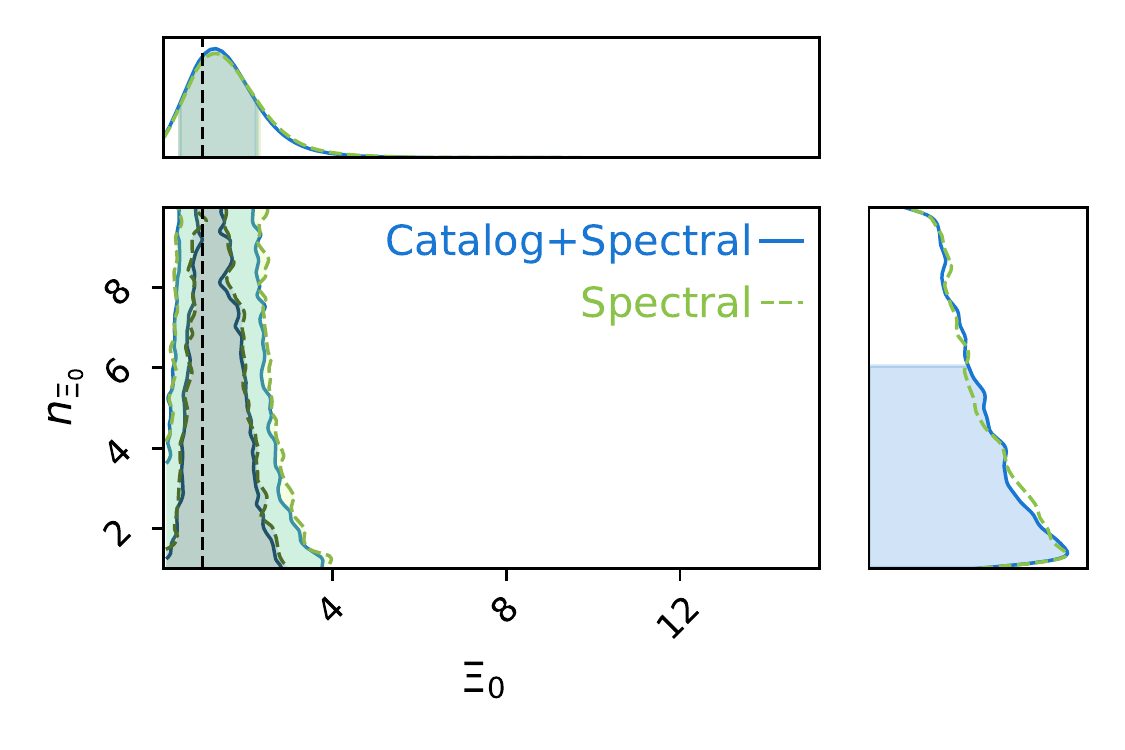}
    \caption{ Marginal posterior distribution for the parameters of the $\Xi_0$ model for the catalog+spectral (blue solid line) and spectral (green dashed line) siren analyses. The black dashed line is the GR value ($\Xi_0 = 0$). The plots report the $68.3\%$ and $90\%$ credible intervals. The marginal 1-dimensional distribution reports the 68.3\% credible intervals.}
    \label{fig:Xi0_cat}
\end{figure}

\begin{figure}
    \centering
    \includegraphics[scale=0.7]{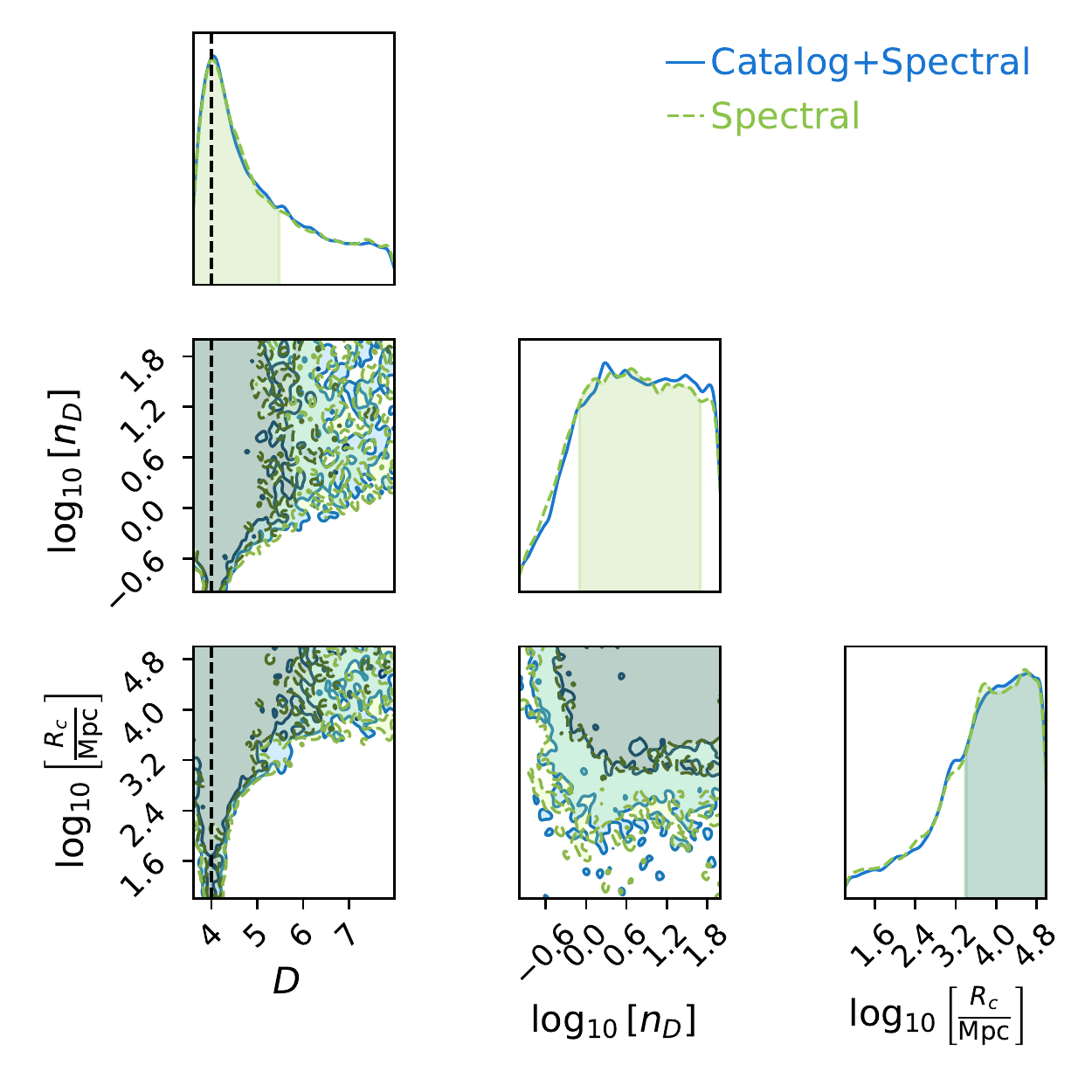}
    \caption{Marginal posterior distribution for the parameters of the extra-dimensions model for the catalog+spectral (blue solid line) and spectral (green dashed line) siren analyses. The black dashed line is the GR value ($D = 4$). The plots report the $68.3\%$ and $90\%$ credible intervals. The marginal 1-dimensional distribution reports the 68.3\% credible intervals.}
    \label{fig:extraD_pos}
\end{figure}

\begin{figure}
    \centering
    \includegraphics[scale=0.7]{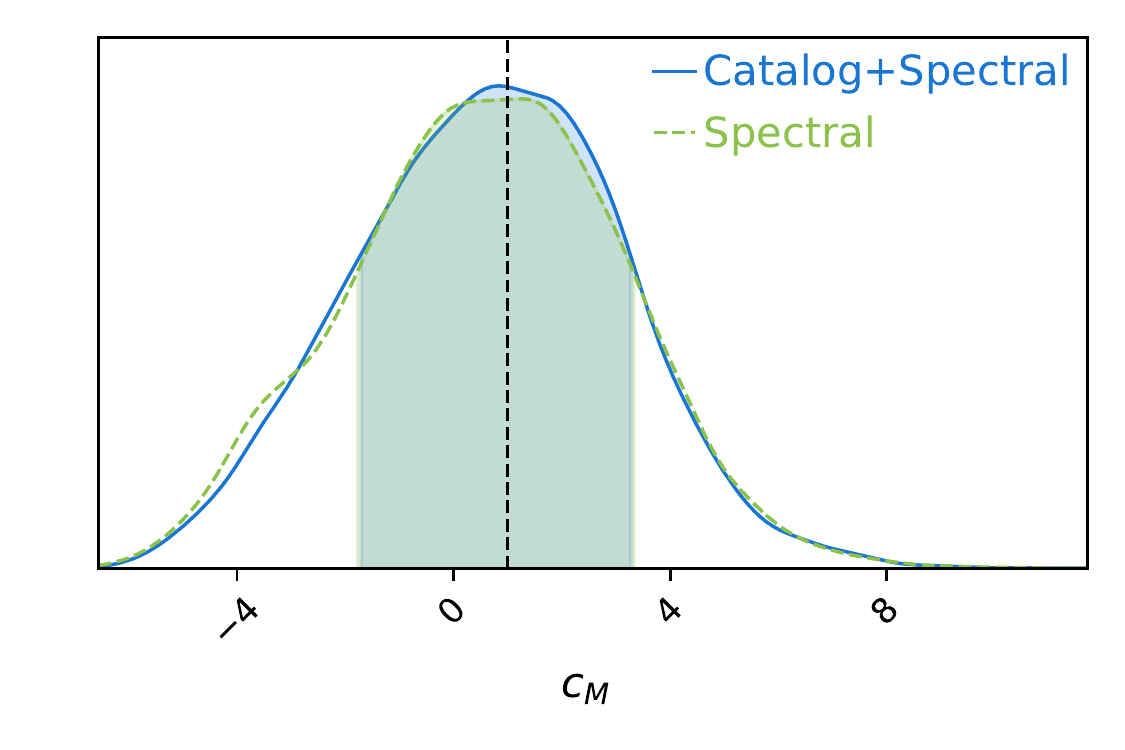}
    \caption{Marginal posterior distribution for the parameter of the running Planck mass model for the catalog+spectral (blue solid line) and spectral (green dashed line) siren analyses. The black dashed line is the GR value ($c_M = 1$).  The plots report the $68.3\%$ and $90\%$ credible intervals. The marginal 1-dimensional distribution reports the 68.3\% credible intervals.}
    \label{fig:cM_cat}
\end{figure}

The marginal posterior distributions that we obtain on the modified gravity parameters are displayed in Figs.~\ref{fig:Xi0_cat}-\ref{fig:extraD_pos}-\ref{fig:cM_cat}. As we can see from the marginal posterior distributions, the constraints on the modified gravity parameters are equal when only using source mass information (spectral) and the galaxy catalog information. This is due to the fact that the galaxy catalog is strongly incomplete and all the information on the redshift distribution is obtained from the source frame mass distribution. In all the cases, we obtain marginal posterior distributions consistent with no deviations from GR.

For the running Planck mass model, we obtain a constraint of $1.0^{+2.6}_{-3.4}$ and $1.6^{+2.2}_{-4.0}$ on $c_M$ for the catalog+spectral and spectral siren analyses respectively (credible intervals reported at $68.3\%$ credible interval unless stated otherwise). For the $\Xi_0$ model we obtain a bound of 
$1.44^{+1.17}_{-0.93}$ and $1.37^{+1.36}_{-0.90}$ on $\Xi_0$ for the catalog+spectral and spectral siren cases, but we are unable to constrain $n_{\Xi_0}$. For the extra dimensions model, we are not able to place strong bounds on any of the parameters as the number of space time dimensions and screening distance are strongly degenerate. If the screening distance is smaller than $\sim 100$ Mpc, then the number of spacetime dimensions is strongly constrained to 4. The constraint on the rest of the population parameters is reported in Tab.~\ref{tab:MG}

\begin{table*}
    \centering
    \caption{Constraints on the population parameters for the spectral and catalog+spectral analyses performed with 42 BBHs and various modified gravity models. The values are reported as the median with symmetric 68.3\% credible intervals.}
    \label{tab:MG}
    \begin{tabular}{cccccccccccc}
       \hline
       \hline
		Model & $\alpha$ & $\beta$ & $m_{\rm min}$ & $m_{\rm max}$ & $\delta_{\rm m}$ [$M_\odot$] & $\mu_{\rm g}$ [$M_\odot$] & $\sigma_{\rm g}$[$M_\odot$] & $\lambda$ & $\gamma$ & $k$ & $z_p$ \\ 
		\hline
  \hline
		$\Xi_0$ (Catalog+Spectral) & $3.74^{+0.49}_{-0.41}$ & $0.8^{+1.2}_{-1.1}$ & $4.98^{+0.95}_{-0.64}$ & - & $5.0^{+2.6}_{-2.2}$ & $32.9^{+3.2}_{-2.2}$ & -  & $0.019^{+0.032}_{-0.018}$ & - & - & - \\ 
		Extra dimensions (Catalog+Spectral) & $3.80^{+0.45}_{-0.48}$ & $0.6^{+1.2}_{-1.1}$ & $5.37^{+0.55}_{-1.02}$ & -  & $4.6^{+2.7}_{-2.0}$ & $33.1^{+2.1}_{-2.4}$ & $2.3^{+3.0}_{-1.9}$ & $0.017^{+0.031}_{-0.015}$ & - & - & - \\ 
		$c_M$ (Catalog+Spectral) & $3.72^{+0.50}_{-0.45}$ & $0.41^{+1.49}_{-0.82}$ & $5.08^{+0.78}_{-0.81}$ & - & - & $32.9^{+2.7}_{-2.8}$ & - & $0.018^{+0.031}_{-0.017}$ & - & - & -  \\ 
  \hline
  
		$\Xi_0$ (Spectral) & $3.71^{+0.49}_{-0.40}$ & $0.7^{+1.4}_{-1.1}$ & $5.14^{+0.77}_{-0.79}$ & - & $5.2\pm 2.4$ & $33.3^{+2.7}_{-3.1}$ & $3.7^{+3.2}_{-2.5}$ & $0.018^{+0.035}_{-0.015}$ & - & - & - \\ 
		Extra dimensions (Spectral) & $3.86^{+0.42}_{-0.50}$ & $0.7^{+1.2}_{-1.1}$ & $5.13^{+0.84}_{-0.70}$ & - & $4.9^{+2.5}_{-2.1}$ & $33.0\pm 2.3$ & $2.4^{+3.0}_{-2.0}$ & $0.016^{+0.032}_{-0.014}$ & - & - & - \\ 
		$c_M$ (Spectral) & $3.79^{+0.46}_{-0.50}$ & $0.7^{+1.2}_{-1.1}$ & $5.22^{+0.64}_{-0.90}$ & - & $5.3^{+2.3}_{-2.4}$ & $32.7^{+2.6}_{-2.7}$ & $3.0^{+2.4}_{-2.6}$ & $0.017^{+0.032}_{-0.015}$ & -  & - & -  \\ 
		\hline
  \hline
    \end{tabular}
\end{table*}

\section{Conclusion}
\label{sec:6}

In this paper, we have presented a new methodology to conjointly estimate the CBC merger rate and cosmological parameters using GW data and galaxies surveys.

In Sec.~\ref{sec:3}, we presented a new parametrization for the CBC merger rate as a function of galaxies' number density. We have discussed how the galaxies' number density can be constructed from a real survey also accounting for a completeness correction and errors in the redshift evaluation of each galaxy. The proposed CBC merger rate parameterization introduces new astrophysical quantities such as \Rgal, the CBC merger rate per galaxy. The parametrization allows us to fit the number density of galaxies once for the entire analysis, thus making the study computationally feasible, while at the same time linking CBC physics to galaxy physics.

In Sec.~\ref{sec:4} and Sec.~\ref{sec:5}, we have shown how this method could be applied to well-localized dark sirens and to a population of GW events, inferring both cosmology and population properties. We have discussed extensively the relation between $H_0$ and the CBC merger rate per galaxy \Rgal. We constrained the value of  $\log_{10} \Rgal$ to  
 $-5.67^{+0.63}_{-0.50}$ and $-6.10^{+0.81}_{-0.51}$ at 90\% CI in the two scenarios where more luminous galaxies are more likely to emit GW events and where there is no such dependency, respectively. We have also shown that, if more luminous galaxies are more likely to host GW sources, then we can exclude lower values of $H_0$ just by using the 42 BBHs used in the LVK analysis in \cite{gwtc3_H0}.

\begin{acknowledgments}

The authors are grateful for computational resources provided by the LIGO Laboratory and supported by National Science Foundation Grants PHY-0757058 and PHY-0823459.
This research has made use of data or software obtained from the Gravitational Wave Open Science Center (gwosc.org), a service of LIGO Laboratory, the LIGO Scientific Collaboration, the Virgo Collaboration, and KAGRA. LIGO Laboratory and Advanced LIGO are funded by the United States National Science Foundation (NSF) as well as the Science and Technology Facilities Council (STFC) of the United Kingdom, the Max-Planck-Society (MPS), and the State of Niedersachsen/Germany for support of the construction of Advanced LIGO and construction and operation of the GEO600 detector. Additional support for Advanced LIGO was provided by the Australian Research Council. Virgo is funded, through the European Gravitational Observatory (EGO), by the French Centre National de Recherche Scientifique (CNRS), the Italian Istituto Nazionale di Fisica Nucleare (INFN) and the Dutch Nikhef, with contributions by institutions from Belgium, Germany, Greece, Hungary, Ireland, Japan, Monaco, Poland, Portugal, Spain. KAGRA is supported by Ministry of Education, Culture, Sports, Science and Technology (MEXT), Japan Society for the Promotion of Science (JSPS) in Japan; National Research Foundation (NRF) and Ministry of Science and ICT (MSIT) in Korea; Academia Sinica (AS) and National Science and Technology Council (NSTC) in Taiwan.

This material is based upon work supported by NSF's LIGO Laboratory which is a major facility fully funded by the National Science Foundation.

The research of AG is supported by the Ghent University BOF project BOF/STA/202009/040 and the Fonds Wetenschappelijk Onderzoek (FWO) iBOF project BOF20/IBF/124. RG was supported by ERC starting grant SHADE 949572 and STFC grant ST/V005634/1.

\end{acknowledgments}

\appendix

\newpage
\clearpage

\onecolumngrid

\section{CBC mass and redshift models}
\label{app:moredetails}

\subsection{Redshift model}
In this paper, we employ a CBC rate model as a function of redshift following the Madau-Dickinson star formation rate \cite{2014ARA&A..52..415M}. We parametrize the CBC merger rate evolution $\psi(z;\Lambda_p)$ as: 
\begin{equation}
    \psi_{\rm MD}(z;\gamma,k,z_p)=[1+(1+z_p)^{-\gamma-k}] \frac{(1+z)^\gamma}{1+\left(\frac{1+z}{1+z_p}\right)^{\gamma+k}}\,.
\end{equation}
For the analysis with GW170817 in Sec.~\ref{sec:4a} we fix $\gamma=0, k=0, z_p=2.47$ a uniform in comoving volume rate. For the analysis with G190814 in Sec.~\ref{sec:4b} and the catalog analysis in Sec.~\ref{sec:5a} we fix $\gamma=4.59, k=2.86, z_p=2.47$ \citep{gwtc3_H0}, while for the full catalog and population analysis presented in Sec.~\ref{sec:5b} we assume priors equal to the ones used in \cite{gwtc3_H0}, see Tab.~\ref{tab:prior_r}.

\begin{table}[h!]
    \centering
    \begin{tabular}{ c p{11cm} p{2mm} p{3cm} }
        \hline
        {\bf Parameter} & \textbf{Description} &  & \textbf{Prior} \\\hline\hline
        $R_0$& BBH merger rate today in Gpc$^{-3}$ yr$^{-1}$&  & $\mathcal{U}$($\Rnotmin$, $\Rnotmax$) \\
        $\gamma$ & Slope of the power law regime for the rate evolution before the point $z_p$ &  & $\mathcal{U}$($\gammamin$, $\gammamax$) \\
        $k$ & Slope of the power law regime for the rate evolution after the point $z_{\rm p}$ &  & $\mathcal{U}$($\kappamin$, $\kappamax$)\\
        $z_p$ & Redshift turning point between the power law regimes with $\gamma$ and $k$ &  & $\mathcal{U}$($\zpmin$, $\zpmax$) \\
        \hline
        \hline 
    \end{tabular}
    \caption{
    Summary of the prior hyper-parameters used for the merger rate evolution models adopted in this paper.}
  \label{tab:prior_r}
\end{table}

\subsection{Mass models}

The source-frame mass models we use are composed of two statistical distributions: a truncated power law distribution:
\begin{equation}
\mathcal{P}(x|x_{\rm min},x_{\rm max},\alpha) \propto 
\begin{cases}
    x^\alpha, & \left(x_{\rm min}\leqslant x \leqslant x_{\rm max}\right), \\
    0, & \mathrm{otherwise},
\end{cases}
\end{equation}
and a Gaussian distribution with  mean $\mu$ and standard deviation $\sigma$:
\begin{equation}
\mathcal{G}(x|\mu,\sigma)=\frac{1}{\sigma\sqrt{2\pi}} \exp{\left[ -\frac{(x-\mu)^2}{2\sigma^2}
\right]}\,.
\end{equation}
The distribution of the source frame masses $m_{1},m_{2}$ is factorized as:
\begin{equation}
\pi(m_{1},m_{2}|\Phi_m)=\pi(m_{1}|\Lambda_p)\pi(m_{2}|m_{1},\Lambda_p),
\end{equation}
where $\pi(m_{1}|\Phi_m)$ is the primary mass distribution (different for the various cases of the paper, as specified below). The secondary mass distribution $\pi(m_{2}|m_{1},\Phi_m)$ is a truncated power law distribution. The secondary mass is conditioned to the primary mass as $m_{2}<m_{1}$:
\begin{equation}
    \pi(m_{2}|m_{1},m_{\rm min},\alpha)=\mathcal{P}(m_{2}|m_{\rm min},m_{1},\beta)\,.
\end{equation}

Below we list the primary mass models and priors used in this paper:

\begin{itemize}
    \item Sec.~\ref{sec:4a} (GW170817 as a quasi-dark siren): The primary mass model is a truncated power law with $\alpha=0, m_{\rm max}=3 M_{\odot}, m_{\rm min}=1 M_{\rm \odot}$; the secondary mass model is a truncated power law with $\beta=0.81$, $m_{\rm max}=m_1, m_{\rm min}=1 M_{\rm \odot}$ \citep{gwtc3_H0}.
    \item Sec.~\ref{sec:4b} (GW190814 as dark siren): For the primary mass we use the \textsc{Power Law+peak} model from \cite{gwtc3_H0} to describe the source frame distribution. The \textsc{Power Law+Peak} distribution is:
\begin{align}
    \pi(m_{1}|m_{\rm min},m_{\rm max},\alpha,\lambda_{\rm g},\mu_{\rm g},\sigma_{\rm g})=  (1-\lambda_{\rm g})\mathcal{P}(m_{1}|m_{\rm min},m_{\rm max},-\alpha) +  \lambda_{\rm g} \mathcal{G}(m_{1}|\mu_{\rm g},\sigma_{\rm g}),
    \label{eq:PLG}
\end{align}
where the power law part has slope $-\alpha$ between $m_{\rm min}$ and $m_{\rm max}$, while the Gaussian component has  mean $\mu_{\rm g}$ and standard deviation $\sigma_{\rm g}$ and 
accounts $\lambda_g$ of the total fraction of the distribution. We also apply an additional smoothing at the lower edge of the distribution:
\begin{eqnarray}
\pi(m_{1},m_{2}|\Phi_m)=[\pi(m_{1}|\Phi_m)S(m_{1}|\delta_m,m_{\rm min})] [\pi(m_{2}|m_{1},\Phi_m)S(m_{2}|\delta_m,m_{\rm min})], 
\end{eqnarray}
where $S$ is a sigmoid-like window function as described in \cite{LIGOScientific:2020kqk}.  The population parameters fixed for the analysis are $\alpha=3.78, \beta=0.81, m_{\rm max}=112.5 M_{\odot}, m_{\rm min}=4.98 M_{\rm \odot}, \delta_m=4.8 M_{\odot}, \sigma_g=3.88 M_{\odot}, \mu_g=342.27 M_{\odot}, \lambda_g=0.03$. For the secondary mass (the neutron star) we use a truncated power law with $m_{\rm min, NS}=1 M_{\rm \odot}, m_{\rm max, NS}=3 M_{\rm \odot}$ \cite{gwtc3_H0}.

    \item Sec.~\ref{sec:5a} (catalog-only BBH analysis): For the primary mass we use the same \textsc{Power law + peak} population model used in Sec.~\ref{sec:4b}. For the secondary mass, we use a truncated power law with $\beta=0.81$ and $m_{\rm min}=4.88 M_{\rm \odot}, m_{\rm max}=m_1$.

    \item Sec.~\ref{sec:5b} (full catalog and population BBH analysis): the baseline model of the primary mass is always the \textsc{Power Law + peak}, but this time the population parameters of the model are also sampled using priors equal to the ones used in \cite{gwtc3_H0}.

\begin{table}[h!]
    \centering
    \begin{tabular}{ c p{10cm} p{2mm} p{4cm} }
    \hline
    \hline
        {\bf Parameter} & \textbf{Description} &  & \textbf{Prior} \\\hline\hline
        $\alpha$ & Spectral index for the PL of the primary mass distribution. &  & $\mathcal{U}$($\PLGalphamin$, $\PLGalphamax$)\\
        $\beta$ & Spectral index for the PL of the mass ratio distribution. &  & $\mathcal{U}$($\PLGbetamin$, $\PLGbetamax$)\\
        $\mmin$ & Minimum mass of the PL component of the primary mass distribution. &  & $\mathcal{U}$($\PLGmminmin\, \Msol$, $\PLGmminmax\, \Msol$)\\
        $\mmax$ &  Maximum mass of the PL component of the primary mass distribution. &  & $\mathcal{U}$($\PLGmmaxmin\, \Msol$, $\PLGmmaxmax\, \Msol$)\\
        $\lambda_{\rm g}$ & Fraction of the model in the Gaussian component. &  & $\mathcal{U}$($\PLGlambdapeakmin$, $\PLGlambdapeakmax$) \\
        $\mu_{\rm g}$ & Mean of the Gaussian component in the primary mass distribution.  &  & $\mathcal{U}$($\PLGmugmin\, \Msol$, $\PLGmugmax\, \Msol$) \\
        $\sigma_{\rm g}$ & Width of the Gaussian component in the primary mass distribution.  &  & $\mathcal{U}$($\PLGsigmagmin\, \Msol$, $\PLGsigmagmax\, \Msol$)\\
        $\delta_{m}$ & Range of mass tapering at the lower end of the mass distribution.  &  & $\mathcal{U}$($\PLGdeltammin\, \Msol$, $\PLGdeltammax\, \Msol$)\\
        \hline 
        \hline 
    \end{tabular}
    \caption{
    Summary of the priors used for the population hyper-parameters for the three phenomenological mass models.}
  \label{tab:priors}
\end{table}

\end{itemize}

\subsection{Correction of selection biases}
To correct the selection biases, we need to calculate the expected number of detections
\begin{equation}
    \Nexp (\Lambda)= \Tobs \int \de \theta \de z \; \Pdet(z,\theta,\Lambda_c) \frac{\de \Ncbc}{\de z  \de \theta \de t_s}(\Lambda) \frac{1}{1+z}.
\end{equation}
The integral above can be approximated as a Monte Carlo sum over a set of simulated injections
\begin{equation}
    N_{\rm exp}  \approx \frac{T_{\rm obs}}{N_{\rm gen}} \sum_{j=1}^{N_{\rm det}} \frac{1}{\pi_{\rm inj}(z_j, \theta_j)}  \frac{1}{1+z_j} \frac{\de \Ncbc}{\de z  \de \theta \de t_s} \bigg|_j \equiv \frac{T_{\rm obs}}{N_{\rm gen}} \sum_{j=1}^{N_{\rm det}} s_j,
\end{equation}
where $T_{\rm obs}$ is the observing time, $N_{\rm gen}$ the number of simulated signals (also the non-detected ones), $N_{\rm det}$ the number of detected signals, and $\pi_{\rm inj}(z_j, \theta_j)$ the prior used to generate the injections. 
Following \citet{2019RNAAS...3...66F}, we also define a numerical stability estimator for the selection bias. We require that the effective number of injections used to evaluate the selection bias $N_{\rm eff,inj}$ is 4 times higher than the observed number of GW signals. The effective number of injections is given by
\begin{equation}
    N_{\rm eff,inj} =  \frac{\left[\sum_j^{N_{\rm det}} s_j \right]^2}{\left[\sum_j^{N_{\rm det}} s_j^2 - 
 N_{\rm gen}^{-1} (\sum_j^{N_{\rm det}} s_j)^2 \right]}.
 \label{eq:neffinj}
\end{equation}

In this work, we used different injection sets. For the GW170817 analysis, we used the BNSs and NSBHs injections released after O3 \footnote{https://zenodo.org/record/5546676}. For the BBH analysis we use the injection set released with \cite{gwtc3_H0}.

\section{The relation between $R_0$ and \Rgal for the empty catalog case}
\label{app:emptycat}

In this Appendix we show how the CBC merger rate based on galaxy number density in Sec.~\ref{sec:3b} relates to the spectral siren parameterization in Sec.~\ref{sec:3a}. The CBC merger rate written in terms of galaxy number density can be written as 
\begin{eqnarray}
    \frac{\de N_{\rm CBC}}{\de \vec{m}_s \de \theta d\Omega \de z \de t_s} = \Rgal \ppop(\vec{m}_s,\theta|\Lambda_p) \psi(z;\Lambda_p)  \int \de M \, 10^{0.4 \epsilon(M_*-M)} \left[\frac{\de N_{\rm gal,cat}}{\de z \de \Omega \de M}+\frac{\de N_{\rm gal,out}}{\de z \de \Omega \de M} \right].
    \end{eqnarray}
Let us assume that the catalog is completely empty and our rate is determined by the completeness correction; then we have:
\begin{equation}
    \frac{\de N_{\rm CBC}}{\de \vec{m}_s \de \theta d\Omega \de z \de t_s} = \Rgal \ppop(\vec{m}_s,\theta|\Lambda_p) \psi(z;\Lambda_p) \phi_*(H_0) \Gamma_{\rm inc}(\alpha+\epsilon+1,x_{\rm max},x_{\rm min}) \frac{\de V_c}{\de z \de \Omega},
    \label{eq:B2}
\end{equation}
where $\phi_*(H_0) \Gamma_{\rm inc}(\alpha+\epsilon+1,x_{\rm max},x_{\rm min})$ is simply the overall integral of the entire Schechter function (with the addition of the luminosity weight). The spectral siren rate parameterization is: 
\begin{equation}
    \frac{\de N_{\rm CBC}}{\de \vec{m}_s \de \theta d\Omega \de z \de t_s} = R_0 \ppop(\vec{m}_s,\theta|\Lambda_p) \psi(z;\Lambda_p) \frac{\de V_c}{\de z \de \Omega}.
    \label{eq:B3}
\end{equation}
By equating Eq.~\ref{eq:B2} and Eq.~\ref{eq:B3}, we obtain the relation between the spectral siren and galaxy rate, namely:
\begin{equation}
    R_0 = \Rgal \phi_*(H_0) \Gamma_{\rm inc}(\alpha+\epsilon+1,x_{\rm max},x_{\rm min}) \propto \Rgal H_0^3.  
\end{equation}
The above relation also shows why in the full population parameter estimation of Sec.~\ref{sec:5b} \Rgal and $H_0$ are correlated. 

\newpage
\clearpage
\twocolumngrid


\bibliography{apssamp}

\begin{thebibliography}{68}%
\makeatletter
\providecommand \@ifxundefined [1]{%
 \@ifx{#1\undefined}
}%
\providecommand \@ifnum [1]{%
 \ifnum #1\expandafter \@firstoftwo
 \else \expandafter \@secondoftwo
 \fi
}%
\providecommand \@ifx [1]{%
 \ifx #1\expandafter \@firstoftwo
 \else \expandafter \@secondoftwo
 \fi
}%
\providecommand \natexlab [1]{#1}%
\providecommand \enquote  [1]{``#1''}%
\providecommand \bibnamefont  [1]{#1}%
\providecommand \bibfnamefont [1]{#1}%
\providecommand \citenamefont [1]{#1}%
\providecommand \href@noop [0]{\@secondoftwo}%
\providecommand \href [0]{\begingroup \@sanitize@url \@href}%
\providecommand \@href[1]{\@@startlink{#1}\@@href}%
\providecommand \@@href[1]{\endgroup#1\@@endlink}%
\providecommand \@sanitize@url [0]{\catcode `\\12\catcode `\$12\catcode
  `\&12\catcode `\#12\catcode `\^12\catcode `\_12\catcode `\%12\relax}%
\providecommand \@@startlink[1]{}%
\providecommand \@@endlink[0]{}%
\providecommand \url  [0]{\begingroup\@sanitize@url \@url }%
\providecommand \@url [1]{\endgroup\@href {#1}{\urlprefix }}%
\providecommand \urlprefix  [0]{URL }%
\providecommand \Eprint [0]{\href }%
\providecommand \doibase [0]{https://doi.org/}%
\providecommand \selectlanguage [0]{\@gobble}%
\providecommand \bibinfo  [0]{\@secondoftwo}%
\providecommand \bibfield  [0]{\@secondoftwo}%
\providecommand \translation [1]{[#1]}%
\providecommand \BibitemOpen [0]{}%
\providecommand \bibitemStop [0]{}%
\providecommand \bibitemNoStop [0]{.\EOS\space}%
\providecommand \EOS [0]{\spacefactor3000\relax}%
\providecommand \BibitemShut  [1]{\csname bibitem#1\endcsname}%
\let\auto@bib@innerbib\@empty
\bibitem [{\citenamefont {Abbott}\ \emph
  {et~al.}(2017{\natexlab{a}})\citenamefont {Abbott}, \citenamefont {Abbott},
  \citenamefont {Abbott}, \citenamefont {Acernese}, \citenamefont {Ackley}
  \emph {et~al.}}]{ligobns}%
  \BibitemOpen
  \bibfield  {author} {\bibinfo {author} {\bibfnamefont {B.~P.}\ \bibnamefont
  {Abbott}}, \bibinfo {author} {\bibfnamefont {R.}~\bibnamefont {Abbott}},
  \bibinfo {author} {\bibfnamefont {T.~D.}\ \bibnamefont {Abbott}}, \bibinfo
  {author} {\bibfnamefont {F.}~\bibnamefont {Acernese}}, \bibinfo {author}
  {\bibfnamefont {K.}~\bibnamefont {Ackley}}, \emph {et~al.} (\bibinfo
  {collaboration} {LIGO Scientific Collaboration and Virgo Collaboration}),\
  }\bibfield  {title} {\bibinfo {title} {Gw170817: Observation of gravitational
  waves from a binary neutron star inspiral},\ }\href
  {https://doi.org/10.1103/PhysRevLett.119.161101} {\bibfield  {journal}
  {\bibinfo  {journal} {Phys. Rev. Lett.}\ }\textbf {\bibinfo {volume} {119}},\
  \bibinfo {pages} {161101} (\bibinfo {year} {2017}{\natexlab{a}})}\BibitemShut
  {NoStop}%
\bibitem [{\citenamefont {Abbott}\ \emph
  {et~al.}(2017{\natexlab{b}})\citenamefont {Abbott}, \citenamefont {Abbott},
  \citenamefont {Abbott}, \citenamefont {Acernese}, \citenamefont {Ackley}
  \emph {et~al.}}]{MMApaper}%
  \BibitemOpen
  \bibfield  {author} {\bibinfo {author} {\bibfnamefont {B.~P.}\ \bibnamefont
  {Abbott}}, \bibinfo {author} {\bibfnamefont {R.}~\bibnamefont {Abbott}},
  \bibinfo {author} {\bibfnamefont {T.~D.}\ \bibnamefont {Abbott}}, \bibinfo
  {author} {\bibfnamefont {F.}~\bibnamefont {Acernese}}, \bibinfo {author}
  {\bibfnamefont {K.}~\bibnamefont {Ackley}}, \emph {et~al.},\ }\bibfield
  {title} {\bibinfo {title} {{Multi-messenger Observations of a Binary Neutron
  Star Merger}},\ }\href {https://doi.org/10.3847/2041-8213/aa91c9} {\bibfield
  {journal} {\bibinfo  {journal} {\apjl}\ }\textbf {\bibinfo {volume} {848}},\
  \bibinfo {eid} {L12} (\bibinfo {year} {2017}{\natexlab{b}})},\ \Eprint
  {https://arxiv.org/abs/1710.05833} {arXiv:1710.05833 [astro-ph.HE]}
  \BibitemShut {NoStop}%
\bibitem [{\citenamefont {Aasi}\ \emph {et~al.}(2015)\citenamefont {Aasi} \emph
  {et~al.}}]{LIGOScientific:2014pky}%
  \BibitemOpen
  \bibfield  {author} {\bibinfo {author} {\bibfnamefont {J.}~\bibnamefont
  {Aasi}} \emph {et~al.} (\bibinfo {collaboration} {LIGO Scientific}),\
  }\bibfield  {title} {\bibinfo {title} {{Advanced LIGO}},\ }\href
  {https://doi.org/10.1088/0264-9381/32/7/074001} {\bibfield  {journal}
  {\bibinfo  {journal} {Class. Quant. Grav.}\ }\textbf {\bibinfo {volume}
  {32}},\ \bibinfo {pages} {074001} (\bibinfo {year} {2015})},\ \Eprint
  {https://arxiv.org/abs/1411.4547} {arXiv:1411.4547 [gr-qc]} \BibitemShut
  {NoStop}%
\bibitem [{\citenamefont {Acernese}\ \emph {et~al.}(2015)\citenamefont
  {Acernese} \emph {et~al.}}]{VIRGO:2014yos}%
  \BibitemOpen
  \bibfield  {author} {\bibinfo {author} {\bibfnamefont {F.}~\bibnamefont
  {Acernese}} \emph {et~al.} (\bibinfo {collaboration} {VIRGO}),\ }\bibfield
  {title} {\bibinfo {title} {{Advanced Virgo: a second-generation
  interferometric gravitational wave detector}},\ }\href
  {https://doi.org/10.1088/0264-9381/32/2/024001} {\bibfield  {journal}
  {\bibinfo  {journal} {Class. Quant. Grav.}\ }\textbf {\bibinfo {volume}
  {32}},\ \bibinfo {pages} {024001} (\bibinfo {year} {2015})},\ \Eprint
  {https://arxiv.org/abs/1408.3978} {arXiv:1408.3978 [gr-qc]} \BibitemShut
  {NoStop}%
\bibitem [{\citenamefont {{LIGO Scientific Collaboration}}\ \emph
  {et~al.}(2021)\citenamefont {{LIGO Scientific Collaboration}}, \citenamefont
  {{Virgo Collaboration}},\ and\ \citenamefont {{KAGRA
  Collaboration}}}]{gwtc3}%
  \BibitemOpen
  \bibfield  {author} {\bibinfo {author} {\bibnamefont {{LIGO Scientific
  Collaboration}}}, \bibinfo {author} {\bibnamefont {{Virgo Collaboration}}},\
  and\ \bibinfo {author} {\bibnamefont {{KAGRA Collaboration}}},\ }\href@noop
  {} {\bibinfo {title} {Gwtc-3: Compact binary coalescences observed by ligo
  and virgo during the second part of the third observing run}} (\bibinfo
  {year} {2021}),\ \Eprint {https://arxiv.org/abs/2111.03606} {arXiv:2111.03606
  [gr-qc]} \BibitemShut {NoStop}%
\bibitem [{\citenamefont {Oguri}(2016)}]{PhysRevD.93.083511}%
  \BibitemOpen
  \bibfield  {author} {\bibinfo {author} {\bibfnamefont {M.}~\bibnamefont
  {Oguri}},\ }\bibfield  {title} {\bibinfo {title} {Measuring the
  distance-redshift relation with the cross-correlation of gravitational wave
  standard sirens and galaxies},\ }\href
  {https://doi.org/10.1103/PhysRevD.93.083511} {\bibfield  {journal} {\bibinfo
  {journal} {Phys. Rev. D}\ }\textbf {\bibinfo {volume} {93}},\ \bibinfo
  {pages} {083511} (\bibinfo {year} {2016})}\BibitemShut {NoStop}%
\bibitem [{\citenamefont {Mukherjee}\ \emph {et~al.}(2020)\citenamefont
  {Mukherjee}, \citenamefont {Wandelt},\ and\ \citenamefont
  {Silk}}]{Mukherjee:2019wcg}%
  \BibitemOpen
  \bibfield  {author} {\bibinfo {author} {\bibfnamefont {S.}~\bibnamefont
  {Mukherjee}}, \bibinfo {author} {\bibfnamefont {B.~D.}\ \bibnamefont
  {Wandelt}},\ and\ \bibinfo {author} {\bibfnamefont {J.}~\bibnamefont
  {Silk}},\ }\bibfield  {title} {\bibinfo {title} {{Probing the theory of
  gravity with gravitational lensing of gravitational waves and galaxy
  surveys}},\ }\href {https://doi.org/10.1093/mnras/staa827} {\bibfield
  {journal} {\bibinfo  {journal} {Mon. Not. Roy. Astron. Soc.}\ }\textbf
  {\bibinfo {volume} {494}},\ \bibinfo {pages} {1956} (\bibinfo {year}
  {2020})},\ \Eprint {https://arxiv.org/abs/1908.08951} {arXiv:1908.08951
  [astro-ph.CO]} \BibitemShut {NoStop}%
\bibitem [{\citenamefont {Mukherjee}\ \emph {et~al.}(2021)\citenamefont
  {Mukherjee}, \citenamefont {Wandelt}, \citenamefont {Nissanke},\ and\
  \citenamefont {Silvestri}}]{Mukherjee:2020hyn}%
  \BibitemOpen
  \bibfield  {author} {\bibinfo {author} {\bibfnamefont {S.}~\bibnamefont
  {Mukherjee}}, \bibinfo {author} {\bibfnamefont {B.~D.}\ \bibnamefont
  {Wandelt}}, \bibinfo {author} {\bibfnamefont {S.~M.}\ \bibnamefont
  {Nissanke}},\ and\ \bibinfo {author} {\bibfnamefont {A.}~\bibnamefont
  {Silvestri}},\ }\bibfield  {title} {\bibinfo {title} {{Accurate precision
  Cosmology with redshift unknown gravitational wave sources}},\ }\href
  {https://doi.org/10.1103/PhysRevD.103.043520} {\bibfield  {journal} {\bibinfo
   {journal} {Phys. Rev. D}\ }\textbf {\bibinfo {volume} {103}},\ \bibinfo
  {pages} {043520} (\bibinfo {year} {2021})},\ \Eprint
  {https://arxiv.org/abs/2007.02943} {arXiv:2007.02943 [astro-ph.CO]}
  \BibitemShut {NoStop}%
\bibitem [{\citenamefont {Bera}\ \emph {et~al.}(2020)\citenamefont {Bera},
  \citenamefont {Rana}, \citenamefont {More},\ and\ \citenamefont
  {Bose}}]{Bera:2020jhx}%
  \BibitemOpen
  \bibfield  {author} {\bibinfo {author} {\bibfnamefont {S.}~\bibnamefont
  {Bera}}, \bibinfo {author} {\bibfnamefont {D.}~\bibnamefont {Rana}}, \bibinfo
  {author} {\bibfnamefont {S.}~\bibnamefont {More}},\ and\ \bibinfo {author}
  {\bibfnamefont {S.}~\bibnamefont {Bose}},\ }\bibfield  {title} {\bibinfo
  {title} {{Incompleteness Matters Not: Inference of $H_0$ from Binary Black
  Hole\textendash{}Galaxy Cross-correlations}},\ }\href
  {https://doi.org/10.3847/1538-4357/abb4e0} {\bibfield  {journal} {\bibinfo
  {journal} {Astrophys. J.}\ }\textbf {\bibinfo {volume} {902}},\ \bibinfo
  {pages} {79} (\bibinfo {year} {2020})},\ \Eprint
  {https://arxiv.org/abs/2007.04271} {arXiv:2007.04271 [astro-ph.CO]}
  \BibitemShut {NoStop}%
\bibitem [{\citenamefont {Diaz}\ and\ \citenamefont
  {Mukherjee}(2022)}]{Diaz:2021pem}%
  \BibitemOpen
  \bibfield  {author} {\bibinfo {author} {\bibfnamefont {C.~C.}\ \bibnamefont
  {Diaz}}\ and\ \bibinfo {author} {\bibfnamefont {S.}~\bibnamefont
  {Mukherjee}},\ }\bibfield  {title} {\bibinfo {title} {{Mapping the cosmic
  expansion history from LIGO-Virgo-KAGRA in synergy with DESI and SPHEREx}},\
  }\href {https://doi.org/10.1093/mnras/stac208} {\bibfield  {journal}
  {\bibinfo  {journal} {Mon. Not. Roy. Astron. Soc.}\ }\textbf {\bibinfo
  {volume} {511}},\ \bibinfo {pages} {2782} (\bibinfo {year} {2022})},\ \Eprint
  {https://arxiv.org/abs/2107.12787} {arXiv:2107.12787 [astro-ph.CO]}
  \BibitemShut {NoStop}%
\bibitem [{\citenamefont {{Chatterjee}}\ \emph {et~al.}(2021)\citenamefont
  {{Chatterjee}}, \citenamefont {{Hegade K.~R.}}, \citenamefont {{Holder}},
  \citenamefont {{Holz}}, \citenamefont {{Perkins}}, \citenamefont {{Yagi}},\
  and\ \citenamefont {{Yunes}}}]{2021PhRvD.104h3528C}%
  \BibitemOpen
  \bibfield  {author} {\bibinfo {author} {\bibfnamefont {D.}~\bibnamefont
  {{Chatterjee}}}, \bibinfo {author} {\bibfnamefont {A.}~\bibnamefont {{Hegade
  K.~R.}}}, \bibinfo {author} {\bibfnamefont {G.}~\bibnamefont {{Holder}}},
  \bibinfo {author} {\bibfnamefont {D.~E.}\ \bibnamefont {{Holz}}}, \bibinfo
  {author} {\bibfnamefont {S.}~\bibnamefont {{Perkins}}}, \bibinfo {author}
  {\bibfnamefont {K.}~\bibnamefont {{Yagi}}},\ and\ \bibinfo {author}
  {\bibfnamefont {N.}~\bibnamefont {{Yunes}}},\ }\bibfield  {title} {\bibinfo
  {title} {{Cosmology with Love: Measuring the Hubble constant using neutron
  star universal relations}},\ }\href
  {https://doi.org/10.1103/PhysRevD.104.083528} {\bibfield  {journal} {\bibinfo
   {journal} {PhRvD}\ }\textbf {\bibinfo {volume} {104}},\ \bibinfo {eid}
  {083528} (\bibinfo {year} {2021})},\ \Eprint
  {https://arxiv.org/abs/2106.06589} {arXiv:2106.06589 [gr-qc]} \BibitemShut
  {NoStop}%
\bibitem [{\citenamefont {Ghosh}\ \emph {et~al.}(2022)\citenamefont {Ghosh},
  \citenamefont {Biswas},\ and\ \citenamefont {Bose}}]{Ghosh:2022muc}%
  \BibitemOpen
  \bibfield  {author} {\bibinfo {author} {\bibfnamefont {T.}~\bibnamefont
  {Ghosh}}, \bibinfo {author} {\bibfnamefont {B.}~\bibnamefont {Biswas}},\ and\
  \bibinfo {author} {\bibfnamefont {S.}~\bibnamefont {Bose}},\ }\bibfield
  {title} {\bibinfo {title} {{Simultaneous inference of neutron star equation
  of state and the Hubble constant with a population of merging neutron
  stars}},\ }\href {https://doi.org/10.1103/PhysRevD.106.123529} {\bibfield
  {journal} {\bibinfo  {journal} {Phys. Rev. D}\ }\textbf {\bibinfo {volume}
  {106}},\ \bibinfo {pages} {123529} (\bibinfo {year} {2022})},\ \Eprint
  {https://arxiv.org/abs/2203.11756} {arXiv:2203.11756 [astro-ph.CO]}
  \BibitemShut {NoStop}%
\bibitem [{\citenamefont {Bambi}\ \emph {et~al.}(2021)\citenamefont {Bambi},
  \citenamefont {Katsanevas},\ and\ \citenamefont {Kokkotas}}]{Bambi:2020tsh}%
  \BibitemOpen
  \bibinfo {editor} {\bibfnamefont {C.}~\bibnamefont {Bambi}}, \bibinfo
  {editor} {\bibfnamefont {S.}~\bibnamefont {Katsanevas}},\ and\ \bibinfo
  {editor} {\bibfnamefont {K.~D.}\ \bibnamefont {Kokkotas}},\ eds.,\ \href
  {https://doi.org/10.1007/978-981-15-4702-7} {\emph {\bibinfo {title}
  {{Handbook of Gravitational Wave Astronomy}}}}\ (\bibinfo  {publisher}
  {Springer Singapore},\ \bibinfo {year} {2021})\BibitemShut {NoStop}%
\bibitem [{\citenamefont {{Moresco}}\ \emph {et~al.}(2022)\citenamefont
  {{Moresco}}, \citenamefont {{Amati}}, \citenamefont {{Amendola}},
  \citenamefont {{Birrer}}, \citenamefont {{Blakeslee}}, \citenamefont
  {{Cantiello}}, \citenamefont {{Cimatti}}, \citenamefont {{Darling}},
  \citenamefont {{Della Valle}}, \citenamefont {{Fishbach}},\ and\
  \citenamefont {et~al.}}]{2022LRR....25....6M}%
  \BibitemOpen
  \bibfield  {author} {\bibinfo {author} {\bibfnamefont {M.}~\bibnamefont
  {{Moresco}}}, \bibinfo {author} {\bibfnamefont {L.}~\bibnamefont {{Amati}}},
  \bibinfo {author} {\bibfnamefont {L.}~\bibnamefont {{Amendola}}}, \bibinfo
  {author} {\bibfnamefont {S.}~\bibnamefont {{Birrer}}}, \bibinfo {author}
  {\bibfnamefont {J.~P.}\ \bibnamefont {{Blakeslee}}}, \bibinfo {author}
  {\bibfnamefont {M.}~\bibnamefont {{Cantiello}}}, \bibinfo {author}
  {\bibfnamefont {A.}~\bibnamefont {{Cimatti}}}, \bibinfo {author}
  {\bibfnamefont {J.}~\bibnamefont {{Darling}}}, \bibinfo {author}
  {\bibfnamefont {M.}~\bibnamefont {{Della Valle}}}, \bibinfo {author}
  {\bibfnamefont {M.}~\bibnamefont {{Fishbach}}},\ and\ \bibinfo {author}
  {\bibnamefont {et~al.}},\ }\bibfield  {title} {\bibinfo {title} {{Unveiling
  the Universe with emerging cosmological probes}},\ }\href
  {https://doi.org/10.1007/s41114-022-00040-z} {\bibfield  {journal} {\bibinfo
  {journal} {LRR}\ }\textbf {\bibinfo {volume} {25}},\ \bibinfo {eid} {6}
  (\bibinfo {year} {2022})},\ \Eprint {https://arxiv.org/abs/2201.07241}
  {arXiv:2201.07241 [astro-ph.CO]} \BibitemShut {NoStop}%
\bibitem [{\citenamefont {{Schutz}}(1986)}]{schutz}%
  \BibitemOpen
  \bibfield  {author} {\bibinfo {author} {\bibfnamefont {B.~F.}\ \bibnamefont
  {{Schutz}}},\ }\bibfield  {title} {\bibinfo {title} {{Determining the Hubble
  constant from gravitational wave observations}},\ }\href
  {https://doi.org/10.1038/323310a0} {\bibfield  {journal} {\bibinfo  {journal}
  {\nat}\ }\textbf {\bibinfo {volume} {323}},\ \bibinfo {pages} {310} (\bibinfo
  {year} {1986})}\BibitemShut {NoStop}%
\bibitem [{\citenamefont {{Holz}}\ and\ \citenamefont
  {{Hughes}}(2005)}]{2005ApJ...629...15H}%
  \BibitemOpen
  \bibfield  {author} {\bibinfo {author} {\bibfnamefont {D.~E.}\ \bibnamefont
  {{Holz}}}\ and\ \bibinfo {author} {\bibfnamefont {S.~A.}\ \bibnamefont
  {{Hughes}}},\ }\bibfield  {title} {\bibinfo {title} {{Using
  Gravitational-Wave Standard Sirens}},\ }\href
  {https://doi.org/10.1086/431341} {\bibfield  {journal} {\bibinfo  {journal}
  {\apj}\ }\textbf {\bibinfo {volume} {629}},\ \bibinfo {pages} {15} (\bibinfo
  {year} {2005})},\ \Eprint {https://arxiv.org/abs/astro-ph/0504616}
  {astro-ph/0504616} \BibitemShut {NoStop}%
\bibitem [{\citenamefont {Dalal}\ \emph {et~al.}(2006)\citenamefont {Dalal},
  \citenamefont {Holz}, \citenamefont {Hughes},\ and\ \citenamefont
  {Jain}}]{Dalal:2006qt}%
  \BibitemOpen
  \bibfield  {author} {\bibinfo {author} {\bibfnamefont {N.}~\bibnamefont
  {Dalal}}, \bibinfo {author} {\bibfnamefont {D.~E.}\ \bibnamefont {Holz}},
  \bibinfo {author} {\bibfnamefont {S.~A.}\ \bibnamefont {Hughes}},\ and\
  \bibinfo {author} {\bibfnamefont {B.}~\bibnamefont {Jain}},\ }\bibfield
  {title} {\bibinfo {title} {{Short grb and binary black hole standard sirens
  as a probe of dark energy}},\ }\href
  {https://doi.org/10.1103/PhysRevD.74.063006} {\bibfield  {journal} {\bibinfo
  {journal} {Phys. Rev. D}\ }\textbf {\bibinfo {volume} {74}},\ \bibinfo
  {pages} {063006} (\bibinfo {year} {2006})},\ \Eprint
  {https://arxiv.org/abs/astro-ph/0601275} {arXiv:astro-ph/0601275}
  \BibitemShut {NoStop}%
\bibitem [{\citenamefont {Del~Pozzo}(2012)}]{PhysRevD.86.043011}%
  \BibitemOpen
  \bibfield  {author} {\bibinfo {author} {\bibfnamefont {W.}~\bibnamefont
  {Del~Pozzo}},\ }\bibfield  {title} {\bibinfo {title} {Inference of
  cosmological parameters from gravitational waves: Applications to second
  generation interferometers},\ }\href
  {https://doi.org/10.1103/PhysRevD.86.043011} {\bibfield  {journal} {\bibinfo
  {journal} {Phys. Rev. D}\ }\textbf {\bibinfo {volume} {86}},\ \bibinfo
  {pages} {043011} (\bibinfo {year} {2012})}\BibitemShut {NoStop}%
\bibitem [{\citenamefont {{Chen}}\ \emph {et~al.}(2018)\citenamefont {{Chen}},
  \citenamefont {{Fishbach}},\ and\ \citenamefont {{Holz}}}]{chen17}%
  \BibitemOpen
  \bibfield  {author} {\bibinfo {author} {\bibfnamefont {H.-Y.}\ \bibnamefont
  {{Chen}}}, \bibinfo {author} {\bibfnamefont {M.}~\bibnamefont {{Fishbach}}},\
  and\ \bibinfo {author} {\bibfnamefont {D.~E.}\ \bibnamefont {{Holz}}},\
  }\bibfield  {title} {\bibinfo {title} {{A two per cent Hubble constant
  measurement from standard sirens within five years}},\ }\href
  {https://doi.org/10.1038/s41586-018-0606-0} {\bibfield  {journal} {\bibinfo
  {journal} {\nat}\ }\textbf {\bibinfo {volume} {562}},\ \bibinfo {pages} {545}
  (\bibinfo {year} {2018})},\ \Eprint {https://arxiv.org/abs/1712.06531}
  {arXiv:1712.06531 [astro-ph.CO]} \BibitemShut {NoStop}%
\bibitem [{\citenamefont {Fishbach}\ \emph
  {et~al.}(2019{\natexlab{a}})\citenamefont {Fishbach} \emph
  {et~al.}}]{fishbach}%
  \BibitemOpen
  \bibfield  {author} {\bibinfo {author} {\bibfnamefont {M.}~\bibnamefont
  {Fishbach}} \emph {et~al.} (\bibinfo {collaboration} {LIGO Scientific,
  Virgo}),\ }\bibfield  {title} {\bibinfo {title} {{A Standard Siren
  Measurement of the Hubble Constant from GW170817 without the Electromagnetic
  Counterpart}},\ }\href {https://doi.org/10.3847/2041-8213/aaf96e} {\bibfield
  {journal} {\bibinfo  {journal} {Astrophys. J. Lett.}\ }\textbf {\bibinfo
  {volume} {871}},\ \bibinfo {pages} {L13} (\bibinfo {year}
  {2019}{\natexlab{a}})},\ \Eprint {https://arxiv.org/abs/1807.05667}
  {arXiv:1807.05667 [astro-ph.CO]} \BibitemShut {NoStop}%
\bibitem [{\citenamefont {Gray}\ \emph {et~al.}(2020)\citenamefont {Gray} \emph
  {et~al.}}]{2020PhRvD.101l2001G}%
  \BibitemOpen
  \bibfield  {author} {\bibinfo {author} {\bibfnamefont {R.}~\bibnamefont
  {Gray}} \emph {et~al.},\ }\bibfield  {title} {\bibinfo {title} {{Cosmological
  inference using gravitational wave standard sirens: A mock data analysis}},\
  }\href {https://doi.org/10.1103/PhysRevD.101.122001} {\bibfield  {journal}
  {\bibinfo  {journal} {Phys. Rev. D}\ }\textbf {\bibinfo {volume} {101}},\
  \bibinfo {pages} {122001} (\bibinfo {year} {2020})},\ \Eprint
  {https://arxiv.org/abs/1908.06050} {arXiv:1908.06050 [gr-qc]} \BibitemShut
  {NoStop}%
\bibitem [{\citenamefont {Gray}\ \emph {et~al.}(2022)\citenamefont {Gray},
  \citenamefont {Messenger},\ and\ \citenamefont {Veitch}}]{Gray2022}%
  \BibitemOpen
  \bibfield  {author} {\bibinfo {author} {\bibfnamefont {R.}~\bibnamefont
  {Gray}}, \bibinfo {author} {\bibfnamefont {C.}~\bibnamefont {Messenger}},\
  and\ \bibinfo {author} {\bibfnamefont {J.}~\bibnamefont {Veitch}},\
  }\bibfield  {title} {\bibinfo {title} {{A pixelated approach to galaxy
  catalogue incompleteness: improving the dark siren measurement of the Hubble
  constant}},\ }\href {https://doi.org/10.1093/mnras/stac366} {\bibfield
  {journal} {\bibinfo  {journal} {Monthly Notices of the Royal Astronomical
  Society}\ }\textbf {\bibinfo {volume} {512}},\ \bibinfo {pages} {1127}
  (\bibinfo {year} {2022})},\ \Eprint
  {https://arxiv.org/abs/https://academic.oup.com/mnras/article-pdf/512/1/1127/45303118/stac366.pdf}
  {https://academic.oup.com/mnras/article-pdf/512/1/1127/45303118/stac366.pdf}
  \BibitemShut {NoStop}%
\bibitem [{\citenamefont {Leandro}\ \emph {et~al.}(2022)\citenamefont
  {Leandro}, \citenamefont {Marra},\ and\ \citenamefont
  {Sturani}}]{PhysRevD.105.023523}%
  \BibitemOpen
  \bibfield  {author} {\bibinfo {author} {\bibfnamefont {H.}~\bibnamefont
  {Leandro}}, \bibinfo {author} {\bibfnamefont {V.}~\bibnamefont {Marra}},\
  and\ \bibinfo {author} {\bibfnamefont {R.}~\bibnamefont {Sturani}},\
  }\bibfield  {title} {\bibinfo {title} {Measuring the hubble constant with
  black sirens},\ }\href {https://doi.org/10.1103/PhysRevD.105.023523}
  {\bibfield  {journal} {\bibinfo  {journal} {Phys. Rev. D}\ }\textbf {\bibinfo
  {volume} {105}},\ \bibinfo {pages} {023523} (\bibinfo {year}
  {2022})}\BibitemShut {NoStop}%
\bibitem [{\citenamefont {{Chernoff}}\ and\ \citenamefont
  {{Finn}}(1993)}]{1993ApJ...411L...5C}%
  \BibitemOpen
  \bibfield  {author} {\bibinfo {author} {\bibfnamefont {D.~F.}\ \bibnamefont
  {{Chernoff}}}\ and\ \bibinfo {author} {\bibfnamefont {L.~S.}\ \bibnamefont
  {{Finn}}},\ }\bibfield  {title} {\bibinfo {title} {{Gravitational Radiation,
  Inspiraling Binaries, and Cosmology}},\ }\href
  {https://doi.org/10.1086/186898} {\bibfield  {journal} {\bibinfo  {journal}
  {\apjl}\ }\textbf {\bibinfo {volume} {411}},\ \bibinfo {pages} {L5} (\bibinfo
  {year} {1993})},\ \Eprint {https://arxiv.org/abs/gr-qc/9304020}
  {arXiv:gr-qc/9304020 [gr-qc]} \BibitemShut {NoStop}%
\bibitem [{\citenamefont {Taylor}\ \emph {et~al.}(2012)\citenamefont {Taylor},
  \citenamefont {Gair},\ and\ \citenamefont {Mandel}}]{Taylor_2012}%
  \BibitemOpen
  \bibfield  {author} {\bibinfo {author} {\bibfnamefont {S.~R.}\ \bibnamefont
  {Taylor}}, \bibinfo {author} {\bibfnamefont {J.~R.}\ \bibnamefont {Gair}},\
  and\ \bibinfo {author} {\bibfnamefont {I.}~\bibnamefont {Mandel}},\
  }\bibfield  {title} {\bibinfo {title} {Cosmology using advanced
  gravitational-wave detectors alone},\ }\bibfield  {journal} {\bibinfo
  {journal} {Physical Review D}\ }\textbf {\bibinfo {volume} {85}},\ \href
  {https://doi.org/10.1103/physrevd.85.023535} {10.1103/physrevd.85.023535}
  (\bibinfo {year} {2012})\BibitemShut {NoStop}%
\bibitem [{\citenamefont {Farr}\ \emph {et~al.}(2019)\citenamefont {Farr},
  \citenamefont {Fishbach}, \citenamefont {Ye},\ and\ \citenamefont
  {Holz}}]{Farr_2019}%
  \BibitemOpen
  \bibfield  {author} {\bibinfo {author} {\bibfnamefont {W.~M.}\ \bibnamefont
  {Farr}}, \bibinfo {author} {\bibfnamefont {M.}~\bibnamefont {Fishbach}},
  \bibinfo {author} {\bibfnamefont {J.}~\bibnamefont {Ye}},\ and\ \bibinfo
  {author} {\bibfnamefont {D.~E.}\ \bibnamefont {Holz}},\ }\bibfield  {title}
  {\bibinfo {title} {A future percent-level measurement of the hubble expansion
  at redshift 0.8 with advanced {LIGO}},\ }\href
  {https://doi.org/10.3847/2041-8213/ab4284} {\bibfield  {journal} {\bibinfo
  {journal} {The Astrophysical Journal}\ }\textbf {\bibinfo {volume} {883}},\
  \bibinfo {pages} {L42} (\bibinfo {year} {2019})}\BibitemShut {NoStop}%
\bibitem [{\citenamefont {{Mar{\'\i}a Ezquiaga}}\ and\ \citenamefont
  {{Holz}}(2020)}]{2020arXiv200602211M}%
  \BibitemOpen
  \bibfield  {author} {\bibinfo {author} {\bibfnamefont {J.}~\bibnamefont
  {{Mar{\'\i}a Ezquiaga}}}\ and\ \bibinfo {author} {\bibfnamefont {D.~E.}\
  \bibnamefont {{Holz}}},\ }\bibfield  {title} {\bibinfo {title} {{Jumping the
  gap: searching for LIGO's biggest black holes}},\ }\href@noop {} {\bibfield
  {journal} {\bibinfo  {journal} {arXiv}\ ,\ \bibinfo {eid} {arXiv:2006.02211}}
  (\bibinfo {year} {2020})},\ \Eprint {https://arxiv.org/abs/2006.02211}
  {arXiv:2006.02211 [astro-ph.HE]} \BibitemShut {NoStop}%
\bibitem [{\citenamefont {{Mastrogiovanni}}\ \emph {et~al.}(2021)\citenamefont
  {{Mastrogiovanni}}, \citenamefont {{Leyde}}, \citenamefont {{Karathanasis}},
  \citenamefont {{Chassande-Mottin}}, \citenamefont {{Steer}}, \citenamefont
  {{Gair}}, \citenamefont {{Ghosh}}, \citenamefont {{Gray}}, \citenamefont
  {{Mukherjee}},\ and\ \citenamefont {{Rinaldi}}}]{mastrogiovanni_2021}%
  \BibitemOpen
  \bibfield  {author} {\bibinfo {author} {\bibfnamefont {S.}~\bibnamefont
  {{Mastrogiovanni}}}, \bibinfo {author} {\bibfnamefont {K.}~\bibnamefont
  {{Leyde}}}, \bibinfo {author} {\bibfnamefont {C.}~\bibnamefont
  {{Karathanasis}}}, \bibinfo {author} {\bibfnamefont {E.}~\bibnamefont
  {{Chassande-Mottin}}}, \bibinfo {author} {\bibfnamefont {D.~A.}\ \bibnamefont
  {{Steer}}}, \bibinfo {author} {\bibfnamefont {J.}~\bibnamefont {{Gair}}},
  \bibinfo {author} {\bibfnamefont {A.}~\bibnamefont {{Ghosh}}}, \bibinfo
  {author} {\bibfnamefont {R.}~\bibnamefont {{Gray}}}, \bibinfo {author}
  {\bibfnamefont {S.}~\bibnamefont {{Mukherjee}}},\ and\ \bibinfo {author}
  {\bibfnamefont {S.}~\bibnamefont {{Rinaldi}}},\ }\bibfield  {title} {\bibinfo
  {title} {{On the importance of source population models for
  gravitational-wave cosmology}},\ }\href
  {https://doi.org/10.1103/PhysRevD.104.062009} {\bibfield  {journal} {\bibinfo
   {journal} {\prd}\ }\textbf {\bibinfo {volume} {104}},\ \bibinfo {eid}
  {062009} (\bibinfo {year} {2021})},\ \Eprint
  {https://arxiv.org/abs/2103.14663} {arXiv:2103.14663 [gr-qc]} \BibitemShut
  {NoStop}%
\bibitem [{\citenamefont {Mukherjee}(2022)}]{Mukherjee:2021rtw}%
  \BibitemOpen
  \bibfield  {author} {\bibinfo {author} {\bibfnamefont {S.}~\bibnamefont
  {Mukherjee}},\ }\bibfield  {title} {\bibinfo {title} {{The redshift
  dependence of black hole mass distribution: is it reliable for standard
  sirens cosmology?}},\ }\href {https://doi.org/10.1093/mnras/stac2152}
  {\bibfield  {journal} {\bibinfo  {journal} {Mon. Not. Roy. Astron. Soc.}\
  }\textbf {\bibinfo {volume} {515}},\ \bibinfo {pages} {5495} (\bibinfo {year}
  {2022})},\ \Eprint {https://arxiv.org/abs/2112.10256} {arXiv:2112.10256
  [astro-ph.CO]} \BibitemShut {NoStop}%
\bibitem [{\citenamefont {{Ezquiaga}}(2021)}]{2021PhLB..82236665E}%
  \BibitemOpen
  \bibfield  {author} {\bibinfo {author} {\bibfnamefont {J.~M.}\ \bibnamefont
  {{Ezquiaga}}},\ }\bibfield  {title} {\bibinfo {title} {{Hearing gravity from
  the cosmos: GWTC-2 probes general relativity at cosmological scales}},\
  }\href {https://doi.org/10.1016/j.physletb.2021.136665} {\bibfield  {journal}
  {\bibinfo  {journal} {Physics Letters B}\ }\textbf {\bibinfo {volume}
  {822}},\ \bibinfo {eid} {136665} (\bibinfo {year} {2021})},\ \Eprint
  {https://arxiv.org/abs/2104.05139} {arXiv:2104.05139 [astro-ph.CO]}
  \BibitemShut {NoStop}%
\bibitem [{\citenamefont {{Leyde}}\ \emph {et~al.}(2022)\citenamefont
  {{Leyde}}, \citenamefont {{Mastrogiovanni}}, \citenamefont {{Steer}},
  \citenamefont {{Chassande-Mottin}},\ and\ \citenamefont
  {{Karathanasis}}}]{2022JCAP...09..012L}%
  \BibitemOpen
  \bibfield  {author} {\bibinfo {author} {\bibfnamefont {K.}~\bibnamefont
  {{Leyde}}}, \bibinfo {author} {\bibfnamefont {S.}~\bibnamefont
  {{Mastrogiovanni}}}, \bibinfo {author} {\bibfnamefont {D.~A.}\ \bibnamefont
  {{Steer}}}, \bibinfo {author} {\bibfnamefont {E.}~\bibnamefont
  {{Chassande-Mottin}}},\ and\ \bibinfo {author} {\bibfnamefont
  {C.}~\bibnamefont {{Karathanasis}}},\ }\bibfield  {title} {\bibinfo {title}
  {{Current and future constraints on cosmology and modified gravitational wave
  friction from binary black holes}},\ }\href
  {https://doi.org/10.1088/1475-7516/2022/09/012} {\bibfield  {journal}
  {\bibinfo  {journal} {\jcap}\ }\textbf {\bibinfo {volume} {2022}},\ \bibinfo
  {eid} {012} (\bibinfo {year} {2022})},\ \Eprint
  {https://arxiv.org/abs/2203.11680} {arXiv:2203.11680 [gr-qc]} \BibitemShut
  {NoStop}%
\bibitem [{\citenamefont {Ezquiaga}\ and\ \citenamefont
  {Holz}(2022)}]{Ezquiaga_2022}%
  \BibitemOpen
  \bibfield  {author} {\bibinfo {author} {\bibfnamefont {J.~M.}\ \bibnamefont
  {Ezquiaga}}\ and\ \bibinfo {author} {\bibfnamefont {D.~E.}\ \bibnamefont
  {Holz}},\ }\bibfield  {title} {\bibinfo {title} {{Spectral Sirens: Cosmology
  from the Full Mass Distribution of Compact Binaries}},\ }\href
  {https://doi.org/10.1103/PhysRevLett.129.061102} {\bibfield  {journal}
  {\bibinfo  {journal} {Phys. Rev. Lett.}\ }\textbf {\bibinfo {volume} {129}},\
  \bibinfo {pages} {061102} (\bibinfo {year} {2022})},\ \Eprint
  {https://arxiv.org/abs/2202.08240} {arXiv:2202.08240 [astro-ph.CO]}
  \BibitemShut {NoStop}%
\bibitem [{\citenamefont {{Karathanasis}}\ \emph {et~al.}(2022)\citenamefont
  {{Karathanasis}}, \citenamefont {{Mukherjee}},\ and\ \citenamefont
  {{Mastrogiovanni}}}]{Karathanasis:2022rtr}%
  \BibitemOpen
  \bibfield  {author} {\bibinfo {author} {\bibfnamefont {C.}~\bibnamefont
  {{Karathanasis}}}, \bibinfo {author} {\bibfnamefont {S.}~\bibnamefont
  {{Mukherjee}}},\ and\ \bibinfo {author} {\bibfnamefont {S.}~\bibnamefont
  {{Mastrogiovanni}}},\ }\bibfield  {title} {\bibinfo {title} {{Binary black
  holes population and cosmology in new lights: Signature of PISN mass and
  formation channel in GWTC-3}},\ }\href
  {https://doi.org/10.48550/arXiv.2204.13495} {\bibfield  {journal} {\bibinfo
  {journal} {arXiv}\ ,\ \bibinfo {eid} {arXiv:2204.13495}} (\bibinfo {year}
  {2022})},\ \Eprint {https://arxiv.org/abs/2204.13495} {arXiv:2204.13495
  [astro-ph.CO]} \BibitemShut {NoStop}%
\bibitem [{\citenamefont {{Ding}}\ \emph {et~al.}(2019)\citenamefont {{Ding}},
  \citenamefont {{Biesiada}}, \citenamefont {{Zheng}}, \citenamefont {{Liao}},
  \citenamefont {{Li}},\ and\ \citenamefont {{Zhu}}}]{2019JCAP...04..033D}%
  \BibitemOpen
  \bibfield  {author} {\bibinfo {author} {\bibfnamefont {X.}~\bibnamefont
  {{Ding}}}, \bibinfo {author} {\bibfnamefont {M.}~\bibnamefont {{Biesiada}}},
  \bibinfo {author} {\bibfnamefont {X.}~\bibnamefont {{Zheng}}}, \bibinfo
  {author} {\bibfnamefont {K.}~\bibnamefont {{Liao}}}, \bibinfo {author}
  {\bibfnamefont {Z.}~\bibnamefont {{Li}}},\ and\ \bibinfo {author}
  {\bibfnamefont {Z.-H.}\ \bibnamefont {{Zhu}}},\ }\bibfield  {title} {\bibinfo
  {title} {{Cosmological inference from standard sirens without redshift
  measurements}},\ }\href {https://doi.org/10.1088/1475-7516/2019/04/033}
  {\bibfield  {journal} {\bibinfo  {journal} {JCAP}\ }\textbf {\bibinfo
  {volume} {2019}}\bibfield  {number} {\bibinfo  {number} { (4)},\ \bibinfo
  {eid} {033}},\ }\Eprint {https://arxiv.org/abs/1801.05073} {arXiv:1801.05073
  [astro-ph.CO]} \BibitemShut {NoStop}%
\bibitem [{\citenamefont {{Ye}}\ and\ \citenamefont
  {{Fishbach}}(2021)}]{2021PhRvD.104d3507Y}%
  \BibitemOpen
  \bibfield  {author} {\bibinfo {author} {\bibfnamefont {C.}~\bibnamefont
  {{Ye}}}\ and\ \bibinfo {author} {\bibfnamefont {M.}~\bibnamefont
  {{Fishbach}}},\ }\bibfield  {title} {\bibinfo {title} {{Cosmology with
  standard sirens at cosmic noon}},\ }\href
  {https://doi.org/10.1103/PhysRevD.104.043507} {\bibfield  {journal} {\bibinfo
   {journal} {PhRvD}\ }\textbf {\bibinfo {volume} {104}},\ \bibinfo {eid}
  {043507} (\bibinfo {year} {2021})},\ \Eprint
  {https://arxiv.org/abs/2103.14038} {arXiv:2103.14038 [astro-ph.CO]}
  \BibitemShut {NoStop}%
\bibitem [{\citenamefont {Abbott}\ \emph
  {et~al.}(2021{\natexlab{a}})\citenamefont {Abbott}, \citenamefont {Abbott},
  \citenamefont {Abbott}, \citenamefont {Acernese}, \citenamefont {Ackley}
  \emph {et~al.}}]{gwtc3_H0}%
  \BibitemOpen
  \bibfield  {author} {\bibinfo {author} {\bibfnamefont {B.~P.}\ \bibnamefont
  {Abbott}}, \bibinfo {author} {\bibfnamefont {R.}~\bibnamefont {Abbott}},
  \bibinfo {author} {\bibfnamefont {T.~D.}\ \bibnamefont {Abbott}}, \bibinfo
  {author} {\bibfnamefont {F.}~\bibnamefont {Acernese}}, \bibinfo {author}
  {\bibfnamefont {K.}~\bibnamefont {Ackley}}, \emph {et~al.},\ }\href@noop {}
  {\bibinfo {title} {Constraints on the cosmic expansion history from gwtc-3}}
  (\bibinfo {year} {2021}{\natexlab{a}}),\ \Eprint
  {https://arxiv.org/abs/2111.03604} {arXiv:2111.03604 [astro-ph.CO]}
  \BibitemShut {NoStop}%
\bibitem [{\citenamefont {{Mandel}}\ \emph {et~al.}(2019)\citenamefont
  {{Mandel}}, \citenamefont {{Farr}},\ and\ \citenamefont {{Gair}}}]{mandel}%
  \BibitemOpen
  \bibfield  {author} {\bibinfo {author} {\bibfnamefont {I.}~\bibnamefont
  {{Mandel}}}, \bibinfo {author} {\bibfnamefont {W.~M.}\ \bibnamefont
  {{Farr}}},\ and\ \bibinfo {author} {\bibfnamefont {J.~R.}\ \bibnamefont
  {{Gair}}},\ }\bibfield  {title} {\bibinfo {title} {{Extracting distribution
  parameters from multiple uncertain observations with selection biases}},\
  }\href {https://doi.org/10.1093/mnras/stz896} {\bibfield  {journal} {\bibinfo
   {journal} {\mnras}\ }\textbf {\bibinfo {volume} {486}},\ \bibinfo {pages}
  {1086} (\bibinfo {year} {2019})},\ \Eprint {https://arxiv.org/abs/1809.02063}
  {arXiv:1809.02063 [physics.data-an]} \BibitemShut {NoStop}%
\bibitem [{\citenamefont {{Vitale}}\ \emph {et~al.}(2022)\citenamefont
  {{Vitale}}, \citenamefont {{Gerosa}}, \citenamefont {{Farr}},\ and\
  \citenamefont {{Taylor}}}]{2022hgwa.bookE..45V}%
  \BibitemOpen
  \bibfield  {author} {\bibinfo {author} {\bibfnamefont {S.}~\bibnamefont
  {{Vitale}}}, \bibinfo {author} {\bibfnamefont {D.}~\bibnamefont {{Gerosa}}},
  \bibinfo {author} {\bibfnamefont {W.~M.}\ \bibnamefont {{Farr}}},\ and\
  \bibinfo {author} {\bibfnamefont {S.~R.}\ \bibnamefont {{Taylor}}},\
  }\bibfield  {title} {\bibinfo {title} {{Inferring the Properties of a
  Population of Compact Binaries in Presence of Selection Effects}},\ }in\
  \href {https://doi.org/10.1007/978-981-15-4702-7_45-1} {\emph {\bibinfo
  {booktitle} {Handbook of Gravitational Wave Astronomy. Edited by C. Bambi}}}\
  (\bibinfo {year} {2022})\ p.~\bibinfo {pages} {45}\BibitemShut {NoStop}%
\bibitem [{\citenamefont {Abbott}\ \emph
  {et~al.}(2021{\natexlab{b}})\citenamefont {Abbott} \emph
  {et~al.}}]{LIGOScientific:2020kqk}%
  \BibitemOpen
  \bibfield  {author} {\bibinfo {author} {\bibfnamefont {R.}~\bibnamefont
  {Abbott}} \emph {et~al.} (\bibinfo {collaboration} {LIGO Scientific,
  Virgo}),\ }\bibfield  {title} {\bibinfo {title} {{Population Properties of
  Compact Objects from the Second LIGO-Virgo Gravitational-Wave Transient
  Catalog}},\ }\href {https://doi.org/10.3847/2041-8213/abe949} {\bibfield
  {journal} {\bibinfo  {journal} {Astrophys. J. Lett.}\ }\textbf {\bibinfo
  {volume} {913}},\ \bibinfo {pages} {L7} (\bibinfo {year}
  {2021}{\natexlab{b}})},\ \Eprint {https://arxiv.org/abs/2010.14533}
  {arXiv:2010.14533 [astro-ph.HE]} \BibitemShut {NoStop}%
\bibitem [{\citenamefont {{Abbott}}\ \emph {et~al.}(2023)\citenamefont
  {{Abbott}}, \citenamefont {{Abbott}}, \citenamefont {{Acernese}},
  \citenamefont {{Ackley}}, \citenamefont {{Adams}}, \citenamefont
  {{Adhikari}}, \citenamefont {{Adhikari}}, \citenamefont {{Adya}},
  \citenamefont {{Affeldt}}, \citenamefont {{Agarwal}},\ and\ \citenamefont
  {et~al.}}]{LIGOScientific:2021psn}%
  \BibitemOpen
  \bibfield  {author} {\bibinfo {author} {\bibfnamefont {R.}~\bibnamefont
  {{Abbott}}}, \bibinfo {author} {\bibfnamefont {T.~D.}\ \bibnamefont
  {{Abbott}}}, \bibinfo {author} {\bibfnamefont {F.}~\bibnamefont
  {{Acernese}}}, \bibinfo {author} {\bibfnamefont {K.}~\bibnamefont
  {{Ackley}}}, \bibinfo {author} {\bibfnamefont {C.}~\bibnamefont {{Adams}}},
  \bibinfo {author} {\bibfnamefont {N.}~\bibnamefont {{Adhikari}}}, \bibinfo
  {author} {\bibfnamefont {R.~X.}\ \bibnamefont {{Adhikari}}}, \bibinfo
  {author} {\bibfnamefont {V.~B.}\ \bibnamefont {{Adya}}}, \bibinfo {author}
  {\bibfnamefont {C.}~\bibnamefont {{Affeldt}}}, \bibinfo {author}
  {\bibfnamefont {D.}~\bibnamefont {{Agarwal}}},\ and\ \bibinfo {author}
  {\bibnamefont {et~al.}},\ }\bibfield  {title} {\bibinfo {title} {{Population
  of Merging Compact Binaries Inferred Using Gravitational Waves through
  GWTC-3}},\ }\href {https://doi.org/10.1103/PhysRevX.13.011048} {\bibfield
  {journal} {\bibinfo  {journal} {PhRvX}\ }\textbf {\bibinfo {volume} {13}},\
  \bibinfo {eid} {011048} (\bibinfo {year} {2023})},\ \Eprint
  {https://arxiv.org/abs/2111.03634} {arXiv:2111.03634 [astro-ph.HE]}
  \BibitemShut {NoStop}%
\bibitem [{\citenamefont {{Mancarella}}\ \emph {et~al.}(2022)\citenamefont
  {{Mancarella}}, \citenamefont {{Genoud-Prachex}},\ and\ \citenamefont
  {{Maggiore}}}]{2022PhRvD.105f4030M}%
  \BibitemOpen
  \bibfield  {author} {\bibinfo {author} {\bibfnamefont {M.}~\bibnamefont
  {{Mancarella}}}, \bibinfo {author} {\bibfnamefont {E.}~\bibnamefont
  {{Genoud-Prachex}}},\ and\ \bibinfo {author} {\bibfnamefont {M.}~\bibnamefont
  {{Maggiore}}},\ }\bibfield  {title} {\bibinfo {title} {{Cosmology and
  modified gravitational wave propagation from binary black hole population
  models}},\ }\href {https://doi.org/10.1103/PhysRevD.105.064030} {\bibfield
  {journal} {\bibinfo  {journal} {PhRvD}\ }\textbf {\bibinfo {volume} {105}},\
  \bibinfo {eid} {064030} (\bibinfo {year} {2022})},\ \Eprint
  {https://arxiv.org/abs/2112.05728} {arXiv:2112.05728 [gr-qc]} \BibitemShut
  {NoStop}%
\bibitem [{\citenamefont {{Fishbach}}\ and\ \citenamefont
  {{Holz}}(2017)}]{2017ApJ...851L..25F}%
  \BibitemOpen
  \bibfield  {author} {\bibinfo {author} {\bibfnamefont {M.}~\bibnamefont
  {{Fishbach}}}\ and\ \bibinfo {author} {\bibfnamefont {D.~E.}\ \bibnamefont
  {{Holz}}},\ }\bibfield  {title} {\bibinfo {title} {{Where Are
  LIGO{\textquoteright}s Big Black Holes?}},\ }\href
  {https://doi.org/10.3847/2041-8213/aa9bf6} {\bibfield  {journal} {\bibinfo
  {journal} {\apjl}\ }\textbf {\bibinfo {volume} {851}},\ \bibinfo {eid} {L25}
  (\bibinfo {year} {2017})},\ \Eprint {https://arxiv.org/abs/1709.08584}
  {arXiv:1709.08584 [astro-ph.HE]} \BibitemShut {NoStop}%
\bibitem [{\citenamefont {{Fishbach}}\ \emph {et~al.}(2021)\citenamefont
  {{Fishbach}}, \citenamefont {{Doctor}}, \citenamefont {{Callister}},
  \citenamefont {{Edelman}}, \citenamefont {{Ye}}, \citenamefont {{Essick}},
  \citenamefont {{Farr}}, \citenamefont {{Farr}},\ and\ \citenamefont
  {{Holz}}}]{2021ApJ...912...98F}%
  \BibitemOpen
  \bibfield  {author} {\bibinfo {author} {\bibfnamefont {M.}~\bibnamefont
  {{Fishbach}}}, \bibinfo {author} {\bibfnamefont {Z.}~\bibnamefont
  {{Doctor}}}, \bibinfo {author} {\bibfnamefont {T.}~\bibnamefont
  {{Callister}}}, \bibinfo {author} {\bibfnamefont {B.}~\bibnamefont
  {{Edelman}}}, \bibinfo {author} {\bibfnamefont {J.}~\bibnamefont {{Ye}}},
  \bibinfo {author} {\bibfnamefont {R.}~\bibnamefont {{Essick}}}, \bibinfo
  {author} {\bibfnamefont {W.~M.}\ \bibnamefont {{Farr}}}, \bibinfo {author}
  {\bibfnamefont {B.}~\bibnamefont {{Farr}}},\ and\ \bibinfo {author}
  {\bibfnamefont {D.~E.}\ \bibnamefont {{Holz}}},\ }\bibfield  {title}
  {\bibinfo {title} {{When Are LIGO/Virgo's Big Black Hole Mergers?}},\ }\href
  {https://doi.org/10.3847/1538-4357/abee11} {\bibfield  {journal} {\bibinfo
  {journal} {ApJ}\ }\textbf {\bibinfo {volume} {912}},\ \bibinfo {eid} {98}
  (\bibinfo {year} {2021})},\ \Eprint {https://arxiv.org/abs/2101.07699}
  {arXiv:2101.07699 [astro-ph.HE]} \BibitemShut {NoStop}%
\bibitem [{\citenamefont {{van Son}}\ \emph {et~al.}(2022)\citenamefont {{van
  Son}}, \citenamefont {{de Mink}}, \citenamefont {{Callister}}, \citenamefont
  {{Justham}}, \citenamefont {{Renzo}}, \citenamefont {{Wagg}}, \citenamefont
  {{Broekgaarden}}, \citenamefont {{Kummer}}, \citenamefont {{Pakmor}},\ and\
  \citenamefont {{Mandel}}}]{2022ApJ...931...17V}%
  \BibitemOpen
  \bibfield  {author} {\bibinfo {author} {\bibfnamefont {L.~A.~C.}\
  \bibnamefont {{van Son}}}, \bibinfo {author} {\bibfnamefont {S.~E.}\
  \bibnamefont {{de Mink}}}, \bibinfo {author} {\bibfnamefont {T.}~\bibnamefont
  {{Callister}}}, \bibinfo {author} {\bibfnamefont {S.}~\bibnamefont
  {{Justham}}}, \bibinfo {author} {\bibfnamefont {M.}~\bibnamefont {{Renzo}}},
  \bibinfo {author} {\bibfnamefont {T.}~\bibnamefont {{Wagg}}}, \bibinfo
  {author} {\bibfnamefont {F.~S.}\ \bibnamefont {{Broekgaarden}}}, \bibinfo
  {author} {\bibfnamefont {F.}~\bibnamefont {{Kummer}}}, \bibinfo {author}
  {\bibfnamefont {R.}~\bibnamefont {{Pakmor}}},\ and\ \bibinfo {author}
  {\bibfnamefont {I.}~\bibnamefont {{Mandel}}},\ }\bibfield  {title} {\bibinfo
  {title} {{The Redshift Evolution of the Binary Black Hole Merger Rate: A
  Weighty Matter}},\ }\href {https://doi.org/10.3847/1538-4357/ac64a3}
  {\bibfield  {journal} {\bibinfo  {journal} {ApJ}\ }\textbf {\bibinfo {volume}
  {931}},\ \bibinfo {eid} {17} (\bibinfo {year} {2022})},\ \Eprint
  {https://arxiv.org/abs/2110.01634} {arXiv:2110.01634 [astro-ph.HE]}
  \BibitemShut {NoStop}%
\bibitem [{\citenamefont {{Biscoveanu}}\ \emph {et~al.}(2022)\citenamefont
  {{Biscoveanu}}, \citenamefont {{Callister}}, \citenamefont {{Haster}},
  \citenamefont {{Ng}}, \citenamefont {{Vitale}},\ and\ \citenamefont
  {{Farr}}}]{2022ApJ...932L..19B}%
  \BibitemOpen
  \bibfield  {author} {\bibinfo {author} {\bibfnamefont {S.}~\bibnamefont
  {{Biscoveanu}}}, \bibinfo {author} {\bibfnamefont {T.~A.}\ \bibnamefont
  {{Callister}}}, \bibinfo {author} {\bibfnamefont {C.-J.}\ \bibnamefont
  {{Haster}}}, \bibinfo {author} {\bibfnamefont {K.~K.~Y.}\ \bibnamefont
  {{Ng}}}, \bibinfo {author} {\bibfnamefont {S.}~\bibnamefont {{Vitale}}},\
  and\ \bibinfo {author} {\bibfnamefont {W.~M.}\ \bibnamefont {{Farr}}},\
  }\bibfield  {title} {\bibinfo {title} {{The Binary Black Hole Spin
  Distribution Likely Broadens with Redshift}},\ }\href
  {https://doi.org/10.3847/2041-8213/ac71a8} {\bibfield  {journal} {\bibinfo
  {journal} {ApJL}\ }\textbf {\bibinfo {volume} {932}},\ \bibinfo {eid} {L19}
  (\bibinfo {year} {2022})},\ \Eprint {https://arxiv.org/abs/2204.01578}
  {arXiv:2204.01578 [astro-ph.HE]} \BibitemShut {NoStop}%
\bibitem [{\citenamefont {{Bavera}}\ \emph {et~al.}(2022)\citenamefont
  {{Bavera}}, \citenamefont {{Fishbach}}, \citenamefont {{Zevin}},
  \citenamefont {{Zapartas}},\ and\ \citenamefont
  {{Fragos}}}]{2022A&A...665A..59B}%
  \BibitemOpen
  \bibfield  {author} {\bibinfo {author} {\bibfnamefont {S.~S.}\ \bibnamefont
  {{Bavera}}}, \bibinfo {author} {\bibfnamefont {M.}~\bibnamefont
  {{Fishbach}}}, \bibinfo {author} {\bibfnamefont {M.}~\bibnamefont {{Zevin}}},
  \bibinfo {author} {\bibfnamefont {E.}~\bibnamefont {{Zapartas}}},\ and\
  \bibinfo {author} {\bibfnamefont {T.}~\bibnamefont {{Fragos}}},\ }\bibfield
  {title} {\bibinfo {title} {{The {\ensuremath{\chi}}$_{eff}$ {\ensuremath{-}}
  z correlation of field binary black hole mergers and how 3G
  gravitational-wave detectors can constrain it}},\ }\href
  {https://doi.org/10.1051/0004-6361/202243724} {\bibfield  {journal} {\bibinfo
   {journal} {A\&A}\ }\textbf {\bibinfo {volume} {665}},\ \bibinfo {eid} {A59}
  (\bibinfo {year} {2022})},\ \Eprint {https://arxiv.org/abs/2204.02619}
  {arXiv:2204.02619 [astro-ph.HE]} \BibitemShut {NoStop}%
\bibitem [{\citenamefont {Singer}\ \emph {et~al.}(2016)\citenamefont {Singer}
  \emph {et~al.}}]{Singer:2016eax}%
  \BibitemOpen
  \bibfield  {author} {\bibinfo {author} {\bibfnamefont {L.~P.}\ \bibnamefont
  {Singer}} \emph {et~al.},\ }\bibfield  {title} {\bibinfo {title} {{Going the
  Distance: Mapping Host Galaxies of LIGO and Virgo Sources in Three Dimensions
  Using Local Cosmography and Targeted Follow-up}},\ }\href
  {https://doi.org/10.3847/2041-8205/829/1/L15} {\bibfield  {journal} {\bibinfo
   {journal} {Astrophys. J. Lett.}\ }\textbf {\bibinfo {volume} {829}},\
  \bibinfo {pages} {L15} (\bibinfo {year} {2016})},\ \Eprint
  {https://arxiv.org/abs/1603.07333} {arXiv:1603.07333 [astro-ph.HE]}
  \BibitemShut {NoStop}%
\bibitem [{\citenamefont {{Blanton}}\ \emph {et~al.}(2001)\citenamefont
  {{Blanton}}, \citenamefont {{Dalcanton}}, \citenamefont {{Eisenstein}},
  \citenamefont {{Loveday}}, \citenamefont {{Strauss}}, \citenamefont
  {{SubbaRao}}, \citenamefont {{Weinberg}}, \citenamefont {{Anderson}},
  \citenamefont {{Annis}}, \citenamefont {{Bahcall}},\ and\ \citenamefont
  {et~al.}}]{2001AJ....121.2358B}%
  \BibitemOpen
  \bibfield  {author} {\bibinfo {author} {\bibfnamefont {M.~R.}\ \bibnamefont
  {{Blanton}}}, \bibinfo {author} {\bibfnamefont {J.}~\bibnamefont
  {{Dalcanton}}}, \bibinfo {author} {\bibfnamefont {D.}~\bibnamefont
  {{Eisenstein}}}, \bibinfo {author} {\bibfnamefont {J.}~\bibnamefont
  {{Loveday}}}, \bibinfo {author} {\bibfnamefont {M.~A.}\ \bibnamefont
  {{Strauss}}}, \bibinfo {author} {\bibfnamefont {M.}~\bibnamefont
  {{SubbaRao}}}, \bibinfo {author} {\bibfnamefont {D.~H.}\ \bibnamefont
  {{Weinberg}}}, \bibinfo {author} {\bibfnamefont {J.}~\bibnamefont
  {{Anderson}}, \bibfnamefont {John~E.}}, \bibinfo {author} {\bibfnamefont
  {J.}~\bibnamefont {{Annis}}}, \bibinfo {author} {\bibfnamefont {N.~A.}\
  \bibnamefont {{Bahcall}}},\ and\ \bibinfo {author} {\bibnamefont {et~al.}},\
  }\bibfield  {title} {\bibinfo {title} {{The Luminosity Function of Galaxies
  in SDSS Commissioning Data}},\ }\href {https://doi.org/10.1086/320405}
  {\bibfield  {journal} {\bibinfo  {journal} {AJ}\ }\textbf {\bibinfo {volume}
  {121}},\ \bibinfo {pages} {2358} (\bibinfo {year} {2001})},\ \Eprint
  {https://arxiv.org/abs/astro-ph/0012085} {arXiv:astro-ph/0012085 [astro-ph]}
  \BibitemShut {NoStop}%
\bibitem [{\citenamefont {{Gair}}\ \emph {et~al.}(2022)\citenamefont {{Gair}},
  \citenamefont {{Ghosh}}, \citenamefont {{Gray}}, \citenamefont {{Holz}},
  \citenamefont {{Mastrogiovanni}}, \citenamefont {{Mukherjee}}, \citenamefont
  {{Palmese}}, \citenamefont {{Tamanini}}, \citenamefont {{Baker}},
  \citenamefont {{Beirnaert}},\ and\ \citenamefont
  {et~al.}}]{2022arXiv221208694G}%
  \BibitemOpen
  \bibfield  {author} {\bibinfo {author} {\bibfnamefont {J.~R.}\ \bibnamefont
  {{Gair}}}, \bibinfo {author} {\bibfnamefont {A.}~\bibnamefont {{Ghosh}}},
  \bibinfo {author} {\bibfnamefont {R.}~\bibnamefont {{Gray}}}, \bibinfo
  {author} {\bibfnamefont {D.~E.}\ \bibnamefont {{Holz}}}, \bibinfo {author}
  {\bibfnamefont {S.}~\bibnamefont {{Mastrogiovanni}}}, \bibinfo {author}
  {\bibfnamefont {S.}~\bibnamefont {{Mukherjee}}}, \bibinfo {author}
  {\bibfnamefont {A.}~\bibnamefont {{Palmese}}}, \bibinfo {author}
  {\bibfnamefont {N.}~\bibnamefont {{Tamanini}}}, \bibinfo {author}
  {\bibfnamefont {T.}~\bibnamefont {{Baker}}}, \bibinfo {author} {\bibfnamefont
  {F.}~\bibnamefont {{Beirnaert}}},\ and\ \bibinfo {author} {\bibnamefont
  {et~al.}},\ }\bibfield  {title} {\bibinfo {title} {{The Hitchhiker's guide to
  the galaxy catalog approach for gravitational wave cosmology}},\ }\href@noop
  {} {\bibfield  {journal} {\bibinfo  {journal} {arXiv}\ ,\ \bibinfo {eid}
  {arXiv:2212.08694}} (\bibinfo {year} {2022})},\ \Eprint
  {https://arxiv.org/abs/2212.08694} {arXiv:2212.08694 [gr-qc]} \BibitemShut
  {NoStop}%
\bibitem [{\citenamefont {{Abbott}}\ \emph {et~al.}(2020)\citenamefont
  {{Abbott}}, \citenamefont {{Abbott}}, \citenamefont {{Abraham}},
  \citenamefont {{Acernese}}, \citenamefont {{Ackley}}, \citenamefont
  {{Adams}}, \citenamefont {{Adhikari}}, \citenamefont {{Adya}}, \citenamefont
  {{Affeldt}}, \citenamefont {{Agathos}},\ and\ \citenamefont
  {et~al.}}]{2020ApJ...896L..44A}%
  \BibitemOpen
  \bibfield  {author} {\bibinfo {author} {\bibfnamefont {R.}~\bibnamefont
  {{Abbott}}}, \bibinfo {author} {\bibfnamefont {T.~D.}\ \bibnamefont
  {{Abbott}}}, \bibinfo {author} {\bibfnamefont {S.}~\bibnamefont {{Abraham}}},
  \bibinfo {author} {\bibfnamefont {F.}~\bibnamefont {{Acernese}}}, \bibinfo
  {author} {\bibfnamefont {K.}~\bibnamefont {{Ackley}}}, \bibinfo {author}
  {\bibfnamefont {C.}~\bibnamefont {{Adams}}}, \bibinfo {author} {\bibfnamefont
  {R.~X.}\ \bibnamefont {{Adhikari}}}, \bibinfo {author} {\bibfnamefont
  {V.~B.}\ \bibnamefont {{Adya}}}, \bibinfo {author} {\bibfnamefont
  {C.}~\bibnamefont {{Affeldt}}}, \bibinfo {author} {\bibfnamefont
  {M.}~\bibnamefont {{Agathos}}},\ and\ \bibinfo {author} {\bibnamefont
  {et~al.}},\ }\bibfield  {title} {\bibinfo {title} {{GW190814: Gravitational
  Waves from the Coalescence of a 23 Solar Mass Black Hole with a 2.6 Solar
  Mass Compact Object}},\ }\href {https://doi.org/10.3847/2041-8213/ab960f}
  {\bibfield  {journal} {\bibinfo  {journal} {ApJL}\ }\textbf {\bibinfo
  {volume} {896}},\ \bibinfo {eid} {L44} (\bibinfo {year} {2020})},\ \Eprint
  {https://arxiv.org/abs/2006.12611} {arXiv:2006.12611 [astro-ph.HE]}
  \BibitemShut {NoStop}%
\bibitem [{\citenamefont {Dálya}\ \emph {et~al.}(2021)\citenamefont {Dálya},
  \citenamefont {Díaz}, \citenamefont {Bouchet}, \citenamefont {Frei},
  \citenamefont {Jasche}, \citenamefont {Lavaux}, \citenamefont {Macas},
  \citenamefont {Mukherjee}, \citenamefont {Pálfi}, \citenamefont {de~Souza},
  \citenamefont {Wandelt}, \citenamefont {Bilicki},\ and\ \citenamefont
  {Raffai}}]{gladeplus}%
  \BibitemOpen
  \bibfield  {author} {\bibinfo {author} {\bibfnamefont {G.}~\bibnamefont
  {Dálya}}, \bibinfo {author} {\bibfnamefont {R.}~\bibnamefont {Díaz}},
  \bibinfo {author} {\bibfnamefont {F.~R.}\ \bibnamefont {Bouchet}}, \bibinfo
  {author} {\bibfnamefont {Z.}~\bibnamefont {Frei}}, \bibinfo {author}
  {\bibfnamefont {J.}~\bibnamefont {Jasche}}, \bibinfo {author} {\bibfnamefont
  {G.}~\bibnamefont {Lavaux}}, \bibinfo {author} {\bibfnamefont
  {R.}~\bibnamefont {Macas}}, \bibinfo {author} {\bibfnamefont
  {S.}~\bibnamefont {Mukherjee}}, \bibinfo {author} {\bibfnamefont
  {M.}~\bibnamefont {Pálfi}}, \bibinfo {author} {\bibfnamefont {R.~S.}\
  \bibnamefont {de~Souza}}, \bibinfo {author} {\bibfnamefont {B.~D.}\
  \bibnamefont {Wandelt}}, \bibinfo {author} {\bibfnamefont {M.}~\bibnamefont
  {Bilicki}},\ and\ \bibinfo {author} {\bibfnamefont {P.}~\bibnamefont
  {Raffai}},\ }\href@noop {} {\bibinfo {title} {Glade+: An extended galaxy
  catalogue for multimessenger searches with advanced gravitational-wave
  detectors}} (\bibinfo {year} {2021}),\ \Eprint
  {https://arxiv.org/abs/2110.06184} {arXiv:2110.06184 [astro-ph.CO]}
  \BibitemShut {NoStop}%
\bibitem [{\citenamefont {{Abbott}}\ \emph
  {et~al.}(2017{\natexlab{a}})\citenamefont {{Abbott}}, \citenamefont
  {{Abbott}}, \citenamefont {{Abbott}}, \citenamefont {{Acernese}},
  \citenamefont {{Ackley}}, \citenamefont {{Adams}}, \citenamefont {{Adams}},
  \citenamefont {{Addesso}}, \citenamefont {{Adhikari}}, \citenamefont
  {{Adya}},\ and\ \citenamefont {et~al.}}]{2017ApJ...848L..12A}%
  \BibitemOpen
  \bibfield  {author} {\bibinfo {author} {\bibfnamefont {B.~P.}\ \bibnamefont
  {{Abbott}}}, \bibinfo {author} {\bibfnamefont {R.}~\bibnamefont {{Abbott}}},
  \bibinfo {author} {\bibfnamefont {T.~D.}\ \bibnamefont {{Abbott}}}, \bibinfo
  {author} {\bibfnamefont {F.}~\bibnamefont {{Acernese}}}, \bibinfo {author}
  {\bibfnamefont {K.}~\bibnamefont {{Ackley}}}, \bibinfo {author}
  {\bibfnamefont {C.}~\bibnamefont {{Adams}}}, \bibinfo {author} {\bibfnamefont
  {T.}~\bibnamefont {{Adams}}}, \bibinfo {author} {\bibfnamefont
  {P.}~\bibnamefont {{Addesso}}}, \bibinfo {author} {\bibfnamefont {R.~X.}\
  \bibnamefont {{Adhikari}}}, \bibinfo {author} {\bibfnamefont {V.~B.}\
  \bibnamefont {{Adya}}},\ and\ \bibinfo {author} {\bibnamefont {et~al.}},\
  }\bibfield  {title} {\bibinfo {title} {{Multi-messenger Observations of a
  Binary Neutron Star Merger}},\ }\href
  {https://doi.org/10.3847/2041-8213/aa91c9} {\bibfield  {journal} {\bibinfo
  {journal} {ApJL}\ }\textbf {\bibinfo {volume} {848}},\ \bibinfo {eid} {L12}
  (\bibinfo {year} {2017}{\natexlab{a}})},\ \Eprint
  {https://arxiv.org/abs/1710.05833} {arXiv:1710.05833 [astro-ph.HE]}
  \BibitemShut {NoStop}%
\bibitem [{\citenamefont {{Abbott}}\ \emph
  {et~al.}(2017{\natexlab{b}})\citenamefont {{Abbott}}, \citenamefont
  {{Abbott}}, \citenamefont {{Abbott}}, \citenamefont {{Acernese}},
  \citenamefont {{Ackley}}, \citenamefont {{Adams}}, \citenamefont {{Adams}},
  \citenamefont {{Addesso}}, \citenamefont {{Adhikari}}, \citenamefont
  {{Adya}},\ and\ \citenamefont {et~al.}}]{2017Natur.551...85A}%
  \BibitemOpen
  \bibfield  {author} {\bibinfo {author} {\bibfnamefont {B.~P.}\ \bibnamefont
  {{Abbott}}}, \bibinfo {author} {\bibfnamefont {R.}~\bibnamefont {{Abbott}}},
  \bibinfo {author} {\bibfnamefont {T.~D.}\ \bibnamefont {{Abbott}}}, \bibinfo
  {author} {\bibfnamefont {F.}~\bibnamefont {{Acernese}}}, \bibinfo {author}
  {\bibfnamefont {K.}~\bibnamefont {{Ackley}}}, \bibinfo {author}
  {\bibfnamefont {C.}~\bibnamefont {{Adams}}}, \bibinfo {author} {\bibfnamefont
  {T.}~\bibnamefont {{Adams}}}, \bibinfo {author} {\bibfnamefont
  {P.}~\bibnamefont {{Addesso}}}, \bibinfo {author} {\bibfnamefont {R.~X.}\
  \bibnamefont {{Adhikari}}}, \bibinfo {author} {\bibfnamefont {V.~B.}\
  \bibnamefont {{Adya}}},\ and\ \bibinfo {author} {\bibnamefont {et~al.}},\
  }\bibfield  {title} {\bibinfo {title} {{A gravitational-wave standard siren
  measurement of the Hubble constant}},\ }\href
  {https://doi.org/10.1038/nature24471} {\bibfield  {journal} {\bibinfo
  {journal} {Natur}\ }\textbf {\bibinfo {volume} {551}},\ \bibinfo {pages} {85}
  (\bibinfo {year} {2017}{\natexlab{b}})},\ \Eprint
  {https://arxiv.org/abs/1710.05835} {arXiv:1710.05835 [astro-ph.CO]}
  \BibitemShut {NoStop}%
\bibitem [{\citenamefont {{Abbott}}\ \emph {et~al.}(2019)\citenamefont
  {{Abbott}}, \citenamefont {{Abbott}}, \citenamefont {{Abbott}}, \citenamefont
  {{Acernese}}, \citenamefont {{Ackley}}, \citenamefont {{Adams}},
  \citenamefont {{Adams}}, \citenamefont {{Addesso}}, \citenamefont
  {{Adhikari}}, \citenamefont {{Adya}},\ and\ \citenamefont
  {et~al.}}]{2019PhRvX...9a1001A}%
  \BibitemOpen
  \bibfield  {author} {\bibinfo {author} {\bibfnamefont {B.~P.}\ \bibnamefont
  {{Abbott}}}, \bibinfo {author} {\bibfnamefont {R.}~\bibnamefont {{Abbott}}},
  \bibinfo {author} {\bibfnamefont {T.~D.}\ \bibnamefont {{Abbott}}}, \bibinfo
  {author} {\bibfnamefont {F.}~\bibnamefont {{Acernese}}}, \bibinfo {author}
  {\bibfnamefont {K.}~\bibnamefont {{Ackley}}}, \bibinfo {author}
  {\bibfnamefont {C.}~\bibnamefont {{Adams}}}, \bibinfo {author} {\bibfnamefont
  {T.}~\bibnamefont {{Adams}}}, \bibinfo {author} {\bibfnamefont
  {P.}~\bibnamefont {{Addesso}}}, \bibinfo {author} {\bibfnamefont {R.~X.}\
  \bibnamefont {{Adhikari}}}, \bibinfo {author} {\bibfnamefont {V.~B.}\
  \bibnamefont {{Adya}}},\ and\ \bibinfo {author} {\bibnamefont {et~al.}},\
  }\bibfield  {title} {\bibinfo {title} {{Properties of the Binary Neutron Star
  Merger GW170817}},\ }\href {https://doi.org/10.1103/PhysRevX.9.011001}
  {\bibfield  {journal} {\bibinfo  {journal} {PhRvX}\ }\textbf {\bibinfo
  {volume} {9}},\ \bibinfo {eid} {011001} (\bibinfo {year} {2019})},\ \Eprint
  {https://arxiv.org/abs/1805.11579} {arXiv:1805.11579 [gr-qc]} \BibitemShut
  {NoStop}%
\bibitem [{\citenamefont {Fishbach}\ \emph
  {et~al.}(2019{\natexlab{b}})\citenamefont {Fishbach} \emph
  {et~al.}}]{2019ApJ...871L..13F}%
  \BibitemOpen
  \bibfield  {author} {\bibinfo {author} {\bibfnamefont {M.}~\bibnamefont
  {Fishbach}} \emph {et~al.} (\bibinfo {collaboration} {LIGO Scientific,
  Virgo}),\ }\bibfield  {title} {\bibinfo {title} {{A Standard Siren
  Measurement of the Hubble Constant from GW170817 without the Electromagnetic
  Counterpart}},\ }\href {https://doi.org/10.3847/2041-8213/aaf96e} {\bibfield
  {journal} {\bibinfo  {journal} {Astrophys. J. Lett.}\ }\textbf {\bibinfo
  {volume} {871}},\ \bibinfo {pages} {L13} (\bibinfo {year}
  {2019}{\natexlab{b}})},\ \Eprint {https://arxiv.org/abs/1807.05667}
  {arXiv:1807.05667 [astro-ph.CO]} \BibitemShut {NoStop}%
\bibitem [{\citenamefont {{D{\'a}lya}}\ \emph {et~al.}(2018)\citenamefont
  {{D{\'a}lya}}, \citenamefont {{Galg{\'o}czi}}, \citenamefont {{Dobos}},
  \citenamefont {{Frei}}, \citenamefont {{Heng}}, \citenamefont {{Macas}},
  \citenamefont {{Messenger}}, \citenamefont {{Raffai}},\ and\ \citenamefont
  {{de Souza}}}]{2018MNRAS.479.2374D}%
  \BibitemOpen
  \bibfield  {author} {\bibinfo {author} {\bibfnamefont {G.}~\bibnamefont
  {{D{\'a}lya}}}, \bibinfo {author} {\bibfnamefont {G.}~\bibnamefont
  {{Galg{\'o}czi}}}, \bibinfo {author} {\bibfnamefont {L.}~\bibnamefont
  {{Dobos}}}, \bibinfo {author} {\bibfnamefont {Z.}~\bibnamefont {{Frei}}},
  \bibinfo {author} {\bibfnamefont {I.~S.}\ \bibnamefont {{Heng}}}, \bibinfo
  {author} {\bibfnamefont {R.}~\bibnamefont {{Macas}}}, \bibinfo {author}
  {\bibfnamefont {C.}~\bibnamefont {{Messenger}}}, \bibinfo {author}
  {\bibfnamefont {P.}~\bibnamefont {{Raffai}}},\ and\ \bibinfo {author}
  {\bibfnamefont {R.~S.}\ \bibnamefont {{de Souza}}},\ }\bibfield  {title}
  {\bibinfo {title} {{GLADE: A galaxy catalogue for multimessenger searches in
  the advanced gravitational-wave detector era}},\ }\href
  {https://doi.org/10.1093/mnras/sty1703} {\bibfield  {journal} {\bibinfo
  {journal} {MNRAS}\ }\textbf {\bibinfo {volume} {479}},\ \bibinfo {pages}
  {2374} (\bibinfo {year} {2018})},\ \Eprint {https://arxiv.org/abs/1804.05709}
  {arXiv:1804.05709 [astro-ph.HE]} \BibitemShut {NoStop}%
\bibitem [{\citenamefont {Abbott}\ \emph {et~al.}(2019)\citenamefont {Abbott},
  \citenamefont {Abbott}, \citenamefont {Abbott}, \citenamefont {Abraham},
  \citenamefont {Acernese}, \citenamefont {Ackley}, \citenamefont {Adams},
  \citenamefont {Adhikari}, \citenamefont {Adya}, \citenamefont {Affeldt},\
  and\ \citenamefont {et~al.}}]{gwtc1}%
  \BibitemOpen
  \bibfield  {author} {\bibinfo {author} {\bibfnamefont {B.}~\bibnamefont
  {Abbott}}, \bibinfo {author} {\bibfnamefont {R.}~\bibnamefont {Abbott}},
  \bibinfo {author} {\bibfnamefont {T.}~\bibnamefont {Abbott}}, \bibinfo
  {author} {\bibfnamefont {S.}~\bibnamefont {Abraham}}, \bibinfo {author}
  {\bibfnamefont {F.}~\bibnamefont {Acernese}}, \bibinfo {author}
  {\bibfnamefont {K.}~\bibnamefont {Ackley}}, \bibinfo {author} {\bibfnamefont
  {C.}~\bibnamefont {Adams}}, \bibinfo {author} {\bibfnamefont
  {R.}~\bibnamefont {Adhikari}}, \bibinfo {author} {\bibfnamefont
  {V.}~\bibnamefont {Adya}}, \bibinfo {author} {\bibfnamefont {C.}~\bibnamefont
  {Affeldt}},\ and\ \bibinfo {author} {\bibnamefont {et~al.}},\ }\bibfield
  {title} {\bibinfo {title} {Gwtc-1: A gravitational-wave transient catalog of
  compact binary mergers observed by ligo and virgo during the first and second
  observing runs},\ }\bibfield  {journal} {\bibinfo  {journal} {Physical Review
  X}\ }\textbf {\bibinfo {volume} {9}},\ \href
  {https://doi.org/10.1103/physrevx.9.031040} {10.1103/physrevx.9.031040}
  (\bibinfo {year} {2019})\BibitemShut {NoStop}%
\bibitem [{\citenamefont {{Planck Collaboration}}\ \emph
  {et~al.}(2016)\citenamefont {{Planck Collaboration}}, \citenamefont {{Ade}},
  \citenamefont {{Aghanim}}, \citenamefont {{Arnaud}}, \citenamefont
  {{Ashdown}}, \citenamefont {{Aumont}}, \citenamefont {{Baccigalupi}},
  \citenamefont {{Banday}}, \citenamefont {{Barreiro}}, \citenamefont
  {{Bartlett}},\ and\ \citenamefont {et~al.}}]{2016A&A...594A..13P}%
  \BibitemOpen
  \bibfield  {author} {\bibinfo {author} {\bibnamefont {{Planck
  Collaboration}}}, \bibinfo {author} {\bibfnamefont {P.~A.~R.}\ \bibnamefont
  {{Ade}}}, \bibinfo {author} {\bibfnamefont {N.}~\bibnamefont {{Aghanim}}},
  \bibinfo {author} {\bibfnamefont {M.}~\bibnamefont {{Arnaud}}}, \bibinfo
  {author} {\bibfnamefont {M.}~\bibnamefont {{Ashdown}}}, \bibinfo {author}
  {\bibfnamefont {J.}~\bibnamefont {{Aumont}}}, \bibinfo {author}
  {\bibfnamefont {C.}~\bibnamefont {{Baccigalupi}}}, \bibinfo {author}
  {\bibfnamefont {A.~J.}\ \bibnamefont {{Banday}}}, \bibinfo {author}
  {\bibfnamefont {R.~B.}\ \bibnamefont {{Barreiro}}}, \bibinfo {author}
  {\bibfnamefont {J.~G.}\ \bibnamefont {{Bartlett}}},\ and\ \bibinfo {author}
  {\bibnamefont {et~al.}},\ }\bibfield  {title} {\bibinfo {title} {{Planck 2015
  results. XIII. Cosmological parameters}},\ }\href
  {https://doi.org/10.1051/0004-6361/201525830} {\bibfield  {journal} {\bibinfo
   {journal} {A\&A}\ }\textbf {\bibinfo {volume} {594}},\ \bibinfo {eid} {A13}
  (\bibinfo {year} {2016})},\ \Eprint {https://arxiv.org/abs/1502.01589}
  {arXiv:1502.01589 [astro-ph.CO]} \BibitemShut {NoStop}%
\bibitem [{\citenamefont {{LIGO Scientific Collaboration}}\ and\ \citenamefont
  {{Virgo Collaboration}}(2019)}]{LVC_O2_StS}%
  \BibitemOpen
  \bibfield  {author} {\bibinfo {author} {\bibnamefont {{LIGO Scientific
  Collaboration}}}\ and\ \bibinfo {author} {\bibnamefont {{Virgo
  Collaboration}}},\ }\href@noop {} {\bibinfo {title} {A gravitational-wave
  measurement of the hubble constant following the second observing run of
  advanced ligo and virgo}} (\bibinfo {year} {2019}),\ \Eprint
  {https://arxiv.org/abs/1908.06060} {arXiv:1908.06060 [astro-ph.CO]}
  \BibitemShut {NoStop}%
\bibitem [{\citenamefont {{Palmese}}\ \emph {et~al.}(2020)\citenamefont
  {{Palmese}}, \citenamefont {{deVicente}}, \citenamefont {{Pereira}},
  \citenamefont {{Annis}},\ and\ \citenamefont {{DES
  Collaboration}}}]{palmese20_sts}%
  \BibitemOpen
  \bibfield  {author} {\bibinfo {author} {\bibfnamefont {A.}~\bibnamefont
  {{Palmese}}}, \bibinfo {author} {\bibfnamefont {J.}~\bibnamefont
  {{deVicente}}}, \bibinfo {author} {\bibfnamefont {M.~E.~S.}\ \bibnamefont
  {{Pereira}}}, \bibinfo {author} {\bibfnamefont {J.}~\bibnamefont {{Annis}}},\
  and\ \bibinfo {author} {\bibnamefont {{DES Collaboration}}},\ }\bibfield
  {title} {\bibinfo {title} {{A Statistical Standard Siren Measurement of the
  Hubble Constant from the LIGO/Virgo Gravitational Wave Compact Object Merger
  GW190814 and Dark Energy Survey Galaxies}},\ }\href
  {https://doi.org/10.3847/2041-8213/abaeff} {\bibfield  {journal} {\bibinfo
  {journal} {\apjl}\ }\textbf {\bibinfo {volume} {900}},\ \bibinfo {eid} {L33}
  (\bibinfo {year} {2020})},\ \Eprint {https://arxiv.org/abs/2006.14961}
  {arXiv:2006.14961 [astro-ph.CO]} \BibitemShut {NoStop}%
\bibitem [{\citenamefont {Abbott}\ \emph
  {et~al.}(2021{\natexlab{c}})\citenamefont {Abbott}, \citenamefont {Abbott},
  \citenamefont {Abraham}, \citenamefont {Acernese}, \citenamefont {Ackley},
  \citenamefont {Adams}, \citenamefont {Adams}, \citenamefont {Adhikari},
  \citenamefont {Adya}, \citenamefont {Affeldt},\ and\ \citenamefont
  {et~al.}}]{gwtc2}%
  \BibitemOpen
  \bibfield  {author} {\bibinfo {author} {\bibfnamefont {R.}~\bibnamefont
  {Abbott}}, \bibinfo {author} {\bibfnamefont {T.}~\bibnamefont {Abbott}},
  \bibinfo {author} {\bibfnamefont {S.}~\bibnamefont {Abraham}}, \bibinfo
  {author} {\bibfnamefont {F.}~\bibnamefont {Acernese}}, \bibinfo {author}
  {\bibfnamefont {K.}~\bibnamefont {Ackley}}, \bibinfo {author} {\bibfnamefont
  {A.}~\bibnamefont {Adams}}, \bibinfo {author} {\bibfnamefont
  {C.}~\bibnamefont {Adams}}, \bibinfo {author} {\bibfnamefont
  {R.}~\bibnamefont {Adhikari}}, \bibinfo {author} {\bibfnamefont
  {V.}~\bibnamefont {Adya}}, \bibinfo {author} {\bibfnamefont {C.}~\bibnamefont
  {Affeldt}},\ and\ \bibinfo {author} {\bibnamefont {et~al.}},\ }\bibfield
  {title} {\bibinfo {title} {Gwtc-2: Compact binary coalescences observed by
  ligo and virgo during the first half of the third observing run},\ }\bibfield
   {journal} {\bibinfo  {journal} {Physical Review X}\ }\textbf {\bibinfo
  {volume} {11}},\ \href {https://doi.org/10.1103/physrevx.11.021053}
  {10.1103/physrevx.11.021053} (\bibinfo {year}
  {2021}{\natexlab{c}})\BibitemShut {NoStop}%
\bibitem [{\citenamefont {{Madau}}\ and\ \citenamefont
  {{Dickinson}}(2014)}]{2014ARA&A..52..415M}%
  \BibitemOpen
  \bibfield  {author} {\bibinfo {author} {\bibfnamefont {P.}~\bibnamefont
  {{Madau}}}\ and\ \bibinfo {author} {\bibfnamefont {M.}~\bibnamefont
  {{Dickinson}}},\ }\bibfield  {title} {\bibinfo {title} {{Cosmic
  Star-Formation History}},\ }\href
  {https://doi.org/10.1146/annurev-astro-081811-125615} {\bibfield  {journal}
  {\bibinfo  {journal} {ARA\&A}\ }\textbf {\bibinfo {volume} {52}},\ \bibinfo
  {pages} {415} (\bibinfo {year} {2014})},\ \Eprint
  {https://arxiv.org/abs/1403.0007} {arXiv:1403.0007 [astro-ph.CO]}
  \BibitemShut {NoStop}%
\bibitem [{\citenamefont {Belgacem}\ \emph {et~al.}(2018)\citenamefont
  {Belgacem}, \citenamefont {Dirian}, \citenamefont {Foffa},\ and\
  \citenamefont {Maggiore}}]{PhysRevD.98.023510}%
  \BibitemOpen
  \bibfield  {author} {\bibinfo {author} {\bibfnamefont {E.}~\bibnamefont
  {Belgacem}}, \bibinfo {author} {\bibfnamefont {Y.}~\bibnamefont {Dirian}},
  \bibinfo {author} {\bibfnamefont {S.}~\bibnamefont {Foffa}},\ and\ \bibinfo
  {author} {\bibfnamefont {M.}~\bibnamefont {Maggiore}},\ }\bibfield  {title}
  {\bibinfo {title} {Modified gravitational-wave propagation and standard
  sirens},\ }\href {https://doi.org/10.1103/PhysRevD.98.023510} {\bibfield
  {journal} {\bibinfo  {journal} {Phys. Rev. D}\ }\textbf {\bibinfo {volume}
  {98}},\ \bibinfo {pages} {023510} (\bibinfo {year} {2018})}\BibitemShut
  {NoStop}%
\bibitem [{\citenamefont {Corman}\ \emph {et~al.}(2022)\citenamefont {Corman},
  \citenamefont {Ghosh}, \citenamefont {Escamilla-Rivera}, \citenamefont
  {Hendry}, \citenamefont {Marsat},\ and\ \citenamefont
  {Tamanini}}]{Corman:2021avn}%
  \BibitemOpen
  \bibfield  {author} {\bibinfo {author} {\bibfnamefont {M.}~\bibnamefont
  {Corman}}, \bibinfo {author} {\bibfnamefont {A.}~\bibnamefont {Ghosh}},
  \bibinfo {author} {\bibfnamefont {C.}~\bibnamefont {Escamilla-Rivera}},
  \bibinfo {author} {\bibfnamefont {M.~A.}\ \bibnamefont {Hendry}}, \bibinfo
  {author} {\bibfnamefont {S.}~\bibnamefont {Marsat}},\ and\ \bibinfo {author}
  {\bibfnamefont {N.}~\bibnamefont {Tamanini}},\ }\bibfield  {title} {\bibinfo
  {title} {{Constraining cosmological extra dimensions with gravitational wave
  standard sirens: From theory to current and future multimessenger
  observations}},\ }\href {https://doi.org/10.1103/PhysRevD.105.064061}
  {\bibfield  {journal} {\bibinfo  {journal} {Phys. Rev. D}\ }\textbf {\bibinfo
  {volume} {105}},\ \bibinfo {pages} {064061} (\bibinfo {year} {2022})},\
  \Eprint {https://arxiv.org/abs/2109.08748} {arXiv:2109.08748 [gr-qc]}
  \BibitemShut {NoStop}%
\bibitem [{\citenamefont {Lagos}\ \emph {et~al.}(2019)\citenamefont {Lagos},
  \citenamefont {Fishbach}, \citenamefont {Landry},\ and\ \citenamefont
  {Holz}}]{Lagos:2019kds}%
  \BibitemOpen
  \bibfield  {author} {\bibinfo {author} {\bibfnamefont {M.}~\bibnamefont
  {Lagos}}, \bibinfo {author} {\bibfnamefont {M.}~\bibnamefont {Fishbach}},
  \bibinfo {author} {\bibfnamefont {P.}~\bibnamefont {Landry}},\ and\ \bibinfo
  {author} {\bibfnamefont {D.~E.}\ \bibnamefont {Holz}},\ }\bibfield  {title}
  {\bibinfo {title} {{Standard sirens with a running Planck mass}},\ }\href
  {https://doi.org/10.1103/PhysRevD.99.083504} {\bibfield  {journal} {\bibinfo
  {journal} {Phys. Rev. D}\ }\textbf {\bibinfo {volume} {99}},\ \bibinfo
  {pages} {083504} (\bibinfo {year} {2019})},\ \Eprint
  {https://arxiv.org/abs/1901.03321} {arXiv:1901.03321 [astro-ph.CO]}
  \BibitemShut {NoStop}%
\bibitem [{\citenamefont {Finke}\ \emph {et~al.}(2021)\citenamefont {Finke},
  \citenamefont {Foffa}, \citenamefont {Iacovelli}, \citenamefont {Maggiore},\
  and\ \citenamefont {Mancarella}}]{finke2021cosmology}%
  \BibitemOpen
  \bibfield  {author} {\bibinfo {author} {\bibfnamefont {A.}~\bibnamefont
  {Finke}}, \bibinfo {author} {\bibfnamefont {S.}~\bibnamefont {Foffa}},
  \bibinfo {author} {\bibfnamefont {F.}~\bibnamefont {Iacovelli}}, \bibinfo
  {author} {\bibfnamefont {M.}~\bibnamefont {Maggiore}},\ and\ \bibinfo
  {author} {\bibfnamefont {M.}~\bibnamefont {Mancarella}},\ }\href@noop {}
  {\bibinfo {title} {Cosmology with ligo/virgo dark sirens: Hubble parameter
  and modified gravitational wave propagation}} (\bibinfo {year} {2021}),\
  \Eprint {https://arxiv.org/abs/2101.12660} {arXiv:2101.12660 [astro-ph.CO]}
  \BibitemShut {NoStop}%
\bibitem [{\citenamefont {Maga\~na Hernandez}(2023)}]{PhysRevD.107.084033}%
  \BibitemOpen
  \bibfield  {author} {\bibinfo {author} {\bibfnamefont {I.}~\bibnamefont
  {Maga\~na Hernandez}},\ }\bibfield  {title} {\bibinfo {title} {Constraining
  the number of spacetime dimensions from gwtc-3 binary black hole mergers},\
  }\href {https://doi.org/10.1103/PhysRevD.107.084033} {\bibfield  {journal}
  {\bibinfo  {journal} {Phys. Rev. D}\ }\textbf {\bibinfo {volume} {107}},\
  \bibinfo {pages} {084033} (\bibinfo {year} {2023})}\BibitemShut {NoStop}%
\bibitem [{\citenamefont {{Farr}}(2019)}]{2019RNAAS...3...66F}%
  \BibitemOpen
  \bibfield  {author} {\bibinfo {author} {\bibfnamefont {W.~M.}\ \bibnamefont
  {{Farr}}},\ }\bibfield  {title} {\bibinfo {title} {{Accuracy Requirements for
  Empirically Measured Selection Functions}},\ }\href
  {https://doi.org/10.3847/2515-5172/ab1d5f} {\bibfield  {journal} {\bibinfo
  {journal} {Research Notes of the American Astronomical Society}\ }\textbf
  {\bibinfo {volume} {3}},\ \bibinfo {eid} {66} (\bibinfo {year} {2019})},\
  \Eprint {https://arxiv.org/abs/1904.10879} {arXiv:1904.10879 [astro-ph.IM]}
  \BibitemShut {NoStop}%
\end{thebibliography}%

\end{document}